\documentstyle[a4p,12pt,epsfig,cite]{article}

\begin{document}
\newcommand {\epem}     {e$^+$e$^-$}
\newcommand {\durham}     {$k_{\perp}$}
\newcommand {\ecm} { E_{\mathrm c.m.} }
\newcommand {\ejet} { E_{\mathrm jet} }
\newcommand {\eqprime} { E_{\mathrm q}^\prime }
\newcommand {\egprime} { E_{\mathrm g}^\prime }
\newcommand {\eqstar} { E_{\mathrm q}^* }
\newcommand {\eqbarstar} { E_{\mathrm \overline{q}}^* }
\newcommand {\egstar} { E_{\mathrm g}^* }
\newcommand {\qjet} { \kappa_{\mathrm jet} }
\newcommand {\mnchgluon} { \langle n_{\mathrm gluon}^{\mathrm ch.} \rangle }
\newcommand {\mngluon} { \langle n_{\mathrm gluon} \rangle }
\newcommand {\mnparton} { \langle n_{\mathrm gluon}^{\mathrm parton} \rangle }
\newcommand {\ptcut} { p_{\mathrm\perp,\,cut} }
\newcommand {\ptgluon} { p_{\mathrm\perp,\,gluon} }
\newcommand {\ycut} { y_{\mathrm cut} }
\newcommand {\nchgluon} { n_{\mathrm gluon}^{\mathrm ch.} }
\newcommand {\fragfunc} { { 1/N \,({\mathrm d}n_{\mathrm gluon}^{\mathrm ch.}
                            /{\mathrm d}}x_E^*) }
\newcommand {\factwo}  { {  F_{2} } }
\newcommand {\facthree}  { { F_{3} } }
\newcommand {\factwogluon}  { {  F_{\mathrm 2,\,gluon} } }
\newcommand {\facthreegluon}  { { F_{\mathrm 3,\,gluon} } }
\newcommand {\factwoparton}  { {  F_{\mathrm 2,\,gluon}^{\mathrm parton} } }
\newcommand {\facthreeparton}  { { F_{\mathrm 3,\,gluon}^{\mathrm parton} } }
\newcommand {\gincl}  {{ \mathrm g_{incl.} }}
\newcommand {\nff}  {{  n_F }}
\newcommand {\mzee}  {{  m_{\mathrm Z} }}
\newcommand {\gqratio}  { { r_{\mathrm g/q}} }
\newcommand {\rfactwo}   { {F_2^{\mathrm g/q}} } 
\newcommand {\rfacthree} { {F_3^{\mathrm g/q}} } 
\def\gtsim{\raisebox{-3pt}{\rlap{$\,\sim\,$}} \raisebox{3pt}{$\,>\,$}}
\def\ltsim{\raisebox{-3pt}{\rlap{$\,\sim\,$}} \raisebox{3pt}{$\,<\,$}}

\begin{titlepage}
\noindent
\begin{center}  {\large EUROPEAN ORGANIZATION FOR NUCLEAR RESEARCH }
\end{center}

%\bigskip\bigskip\bigskip
\begin{tabbing}
\` CERN-EP-2003-067 \\
\` 09 October 2003 \\
\end{tabbing}

\bigskip\bigskip\bigskip
%\vspace*{-5mm}

\begin{center}{\LARGE\bf
Experimental studies of
unbiased gluon jets \\[-1mm]
from {\epem} annihilations using the jet boost \\[1mm]
algorithm
}
\end{center}

\begin{center}
{\Large
The OPAL Collaboration
}
\end{center}

\begin{center}{\large\bf Abstract}\end{center}
\bigskip
\noindent
We present the first experimental results based on the 
jet boost algorithm,
a technique to select 
unbiased samples of gluon jets in {\epem} annihilations,
i.e.~gluon jets free of biases introduced by 
event selection or jet finding criteria.
Our results are derived from hadronic Z$^0$ decays observed
with the OPAL detector
at the LEP {\epem} collider at CERN.
First,
we test the boost algorithm through studies with Herwig Monte Carlo 
events and find that it provides
accurate measurements of the charged particle multiplicity 
distributions of unbiased gluon jets
for jet energies larger than about 5~GeV,
and of the jet particle energy spectra 
(fragmentation functions)
for jet energies larger than about 14~GeV.
Second, we apply the boost algorithm to our data to derive unbiased
measurements of the gluon jet multiplicity distribution 
for energies between about 5 and 18~GeV,
and of the gluon jet fragmentation function at
14 and 18~GeV.
In conjunction with our earlier results at 40~GeV,
we then test QCD calculations for the energy evolution
of the distributions,
specifically the mean and first
two non-trivial normalized factorial moments of the
multiplicity distribution,
and the fragmentation function.
The theoretical results are found to be in global agreement
with the data,
although the factorial moments are not well described
for jet energies below about 14~GeV.
%From a fit of the next-to-leading-order QCD expression for
%the energy evolution of the gluon jet fragmentation function
%between about 14 and 40~GeV,
%we find a value of the strong coupling strength
%in the $\mathrm\overline{MS}$
%renormalization scheme of
%$\alpha_S$$\,=\,$$\mathrm0.128\pm0.008\,(stat.)\pm 0.015\,(syst.)$,
%We test the energy evolution of the gluon jet fragmentation function
%between about 14 and 40~GeV to
%provide a unique consistency check of QCD.

\vspace*{2cm}
\begin{center}{\large
  (Submitted to Phys. Rev. D)
}\end{center}

\end{titlepage}

\begin{center}{\Large        The OPAL Collaboration
}\end{center}\bigskip
\begin{center}{
%begin authorlist PLEASE DO NOT DELETE THIS COMMENT
G.\thinspace Abbiendi$^{  2}$,
C.\thinspace Ainsley$^{  5}$,
P.F.\thinspace {\AA}kesson$^{  3,  y}$,
G.\thinspace Alexander$^{ 22}$,
J.\thinspace Allison$^{ 16}$,
P.\thinspace Amaral$^{  9}$, 
G.\thinspace Anagnostou$^{  1}$,
K.J.\thinspace Anderson$^{  9}$,
S.\thinspace Arcelli$^{  2}$,
S.\thinspace Asai$^{ 23}$
% ...
%end authorlist PLEASE DO NOT DELETE THIS COMMENT
}\end{center}\bigskip
\bigskip
%begin institutes
$^{  1}$School of Physics and Astronomy, University of Birmingham,
Birmingham B15 2TT, UK
\newline
$^{  2}$Dipartimento di Fisica dell' Universit\`a di Bologna and INFN,
I-40126 Bologna, Italy
\newline
$^{  3}$Physikalisches Institut, Universit\"at Bonn,
D-53115 Bonn, Germany
\newline
% ...
$^{  5}$Cavendish Laboratory, Cambridge CB3 0HE, UK
\newline
% ...
$^{  9}$Enrico Fermi Institute and Department of Physics,
University of Chicago, Chicago IL 60637, USA
\newline
% ...
$^{ 16}$Department of Physics, Schuster Laboratory, The University,
Manchester M13 9PL, UK
\newline
% ...
$^{ 22}$Department of Physics and Astronomy, Tel Aviv University,
Tel Aviv 69978, Israel
\newline
$^{ 23}$International Centre for Elementary Particle Physics and
Department of Physics, University of Tokyo, Tokyo 113-0033, and
Kobe University, Kobe 657-8501, Japan
\newline
% ...
$^{  q}$ now at IPHE Universit\'e de Lausanne, CH-1015 Lausanne, Switzerland
\newline
% ...
$^{  y}$ now at CERN
\newline
% ...
%end notes

\clearpage\newpage

\section{Introduction}
\label{sec-introduction}

Gluon jets were first observed in 1979~\cite{bib-petra},
at the PETRA {\epem} collider at DESY.
Certain features of the jets were quickly measured,
such as their production angular distributions,
leading,
for example,
to a determination of the gluon spin~\cite{bib-spin}.
In contrast,
it has proved difficult to obtain meaningful information
about internal characteristics of gluon jets.
The difficulty arises because gluon jets
are usually produced in conjunction with other jets
or the beam remnants at accelerators,
making their identification ambiguous.
Gluon jets in {\epem} annihilations are usually studied
using three-jet $\mathrm q\overline{q}g$ events,
for example,
where q denotes a quark jet,
$\mathrm\overline{q}$ an antiquark jet,
and g a gluon jet.
At hadron colliders,
gluon jets are studied using events with two
energetic gluon jets
produced in conjunction with the beam remnants
and less energetic jets.
In either case,
the gluon jets are identified
%isolated from the other jets or from the beam remnants
using jet finding algorithms such as
the {\durham}~\cite{bib-durham}
or cone~\cite{bib-cone} jet finder,
which assign particles in an event to the jets.
The jet finding algorithms employ resolution criteria.
Different jet algorithms or choices of the
resolution scales yield different assignments
of particles to the jets.
This produces the ambiguities mentioned above.
Many studies employ
fixed values for the resolution scales,
leading to truncation of higher order radiation from the jets
and thus to further ambiguity.
Because of these intrinsic ambiguities,
jets defined in this manner are called ``biased.''

Theoretical descriptions of gluon jets
are usually based on a different approach.
The theoretical approach assumes the production of 
a pair of gluons in an overall color singlet,
i.e.~gg events from a point source.
There are neither beam remnants nor other jets.
The gg system is divided into hemispheres 
in a frame in which the two gluons are back-to-back
(they move in opposite directions),
using the plane
perpendicular to the direction of the separating gluons.
The particles in a hemisphere define a jet.
%The energy scale of the jet is given by the
%total particle energy in the hemisphere.
Since all particles in the event arise from one of the
two original gluons,
there is no ambiguity about which particles to
assign to the gluon jets.\footnote{Note that if the event
is boosted along the gg event axis,
the jet energies and multiplicities change.
The relationship between an unbiased jet's energy and its mean particle
multiplicity is universal, however,
independent of this boost or
of the invariant mass of the gg system.
The same comment applies to the $\mathrm q\overline{q}$ color
singlet systems discussed in Sect.~\ref{sec-boost}.
}
Furthermore,
there are no jet resolution criteria
and thus no truncation of higher order radiation,
i.e.\ all events in the sample are used.
The properties of the jets depend
on a single scale:
the jet energy.
Jets defined in this manner are called ``unbiased,''
in contrast to biased jets,
whose properties depend on the jet resolution scales as well.
Many theoretical results 
have been presented for unbiased gluon jets,
based on Quantum Chromodynamics (QCD),
the gauge theory of strong interactions.
Because most experimental studies are performed using 
biased jets,
tests of the theory have often been indirect.

So far,
only three methods have been used
to measure gluon jet properties in a manner consistent 
with the theoretical
prescription outlined in the previous paragraph,
%i.e.\ to measure properties of unbiased gluon jets,
avoiding the ambiguities associated with biased jets.
First,
radiative $\Upsilon$ decays,
$\Upsilon$$\,\rightarrow\,$$\gamma$gg$\,\rightarrow\,$$\gamma
+\mathit{hadrons}$,
have been studied~\cite{bib-cleo92,bib-cleo97}.
The gg system in these events corresponds to
the event class of the theoretical approach,
described above.
Second,
rare events from hadronic 
Z$^0$$\,\rightarrow\,$$\mathrm q\overline{q}$ 
decays have been 
selected~\cite{bib-opalhemisphere-1,bib-opalhemisphere-2,bib-opalhemisphere-3},
in which the q and $\mathrm\overline{q}$ jets 
are approximately colinear:
the event hemisphere ``$\gincl$'' against which the 
q and $\mathrm\overline{q}$ recoil
corresponds almost exactly to 
an unbiased gluon jet as shown in~\cite{bib-jwg94}.
Third,
the theoretical formalism of~\cite{bib-eden99}
has been applied~\cite{bib-opaleden}
to extract properties of unbiased gluon jets indirectly,
by subtracting results obtained from two-jet
$\mathrm q\overline{q}$ events
from those obtained from three-jet
$\mathrm q\overline{q}g$ events
(see~\cite{bib-eden99,bib-opaleden} for more details).

The first and second of the above techniques 
provide an explicit association of particles 
in an event to the gluon jets.
This allows many characteristics of the jets,
e.g.~the distributions of particle multiplicity and energy, 
to be studied.
The jet energies associated with these two techniques
are limited, however,
to $\ejet$$\,\sim\,$5 and 40~GeV,
respectively.
The third technique,
based on comparing results from
$\mathrm q\overline{q}$ and
$\mathrm q\overline{q}g$ events,
yields measurements over a range of jet energies,
from about 5 to 15~GeV.
This method does not associate particles in
an event with the gluon jet,
however,
yielding only the mean particle 
multiplicity of the jets, $\mngluon$.

In~\cite{bib-eden},
an additional method to determine
properties of unbiased gluon jets is proposed:
the so-called jet boost algorithm.
The jet boost algorithm is described
in Sect.~\ref{sec-boost}.
So far,
no experimental results have been presented
based on this technique.
Like the third method mentioned above,
the jet boost algorithm provides results over a range of jet energies.
Like the first and second methods,
it specifies which particles in an event 
to associate with the gluon jet.
The jet boost method therefore combines features of the other approaches,
offering a means to measure a variety of properties
of unbiased gluon jets as a function of energy.

In this paper,
we present the first experimental study to use
the jet boost algorithm.
The study is based on hadronic decays of the Z$^0$ boson.
The data were collected with the OPAL detector at the
LEP {\epem} storage ring at CERN.
We measure the charged particle multiplicity distribution
and the particle energy spectrum (fragmentation function) of the jets
for a variety of jet energies.
The results are compared to QCD calculations to provide new
and unique tests of that theory.

\section{The jet boost algorithm}
\label{sec-boost}

\begin{figure}[t]
 \begin{center}
 \begin{tabular}{cc}
      \epsfxsize=5cm
      \epsffile{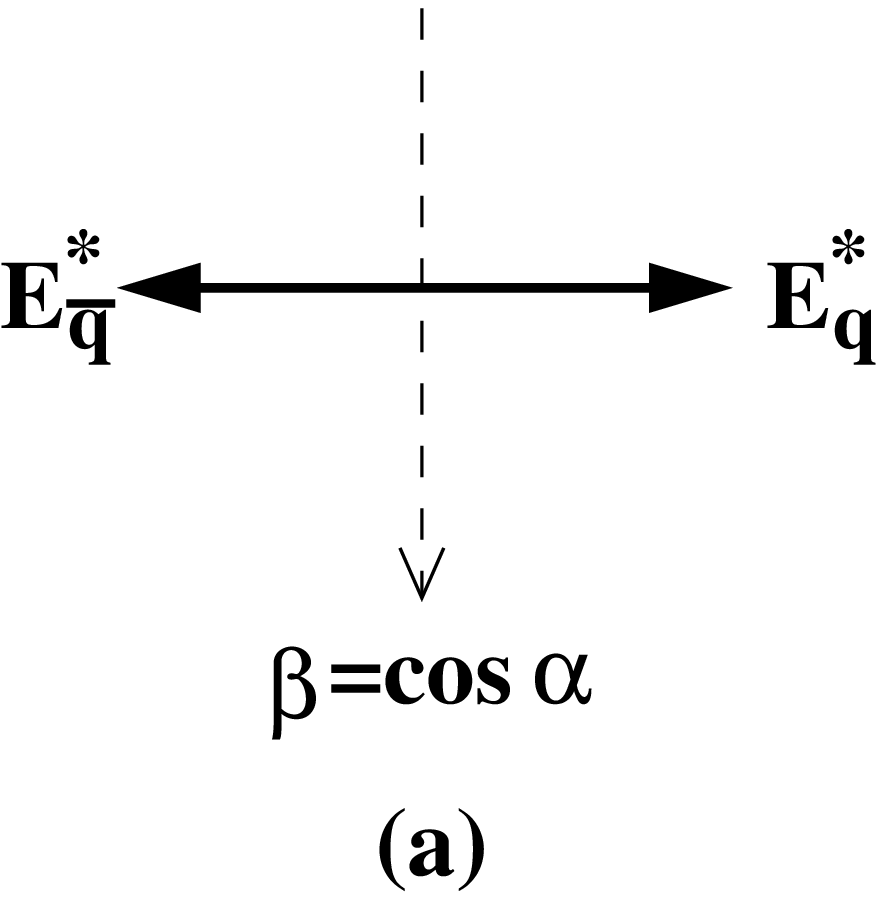} & \hspace*{1.5cm}
      \epsfxsize=5cm
      \epsffile{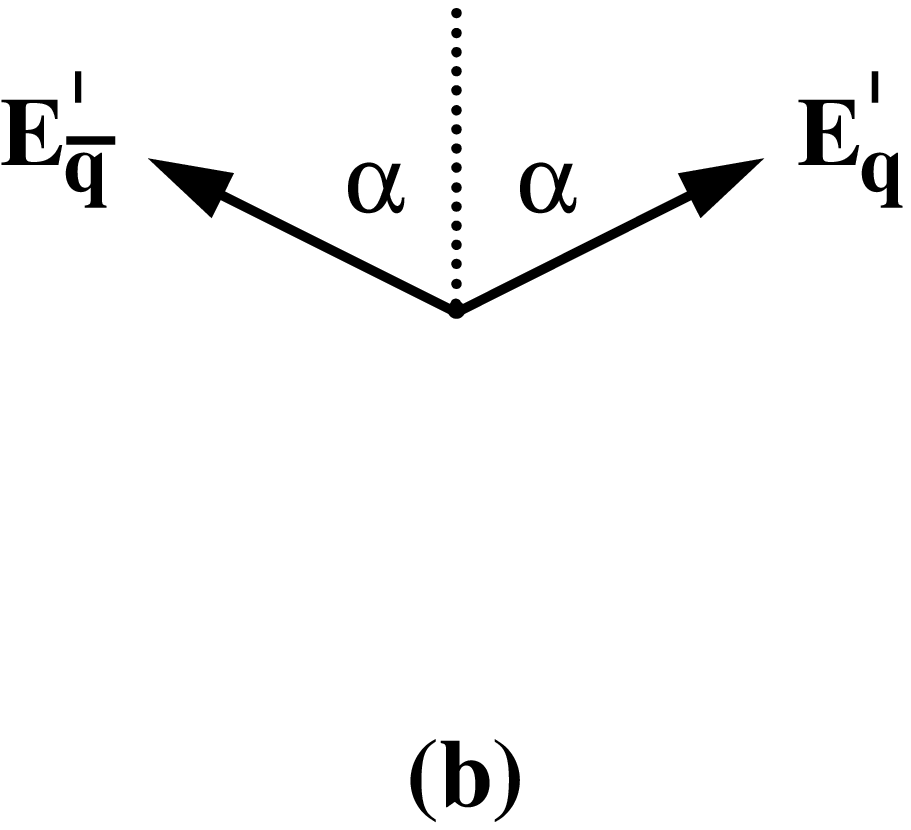}
 \end{tabular}
 \end{center}
\caption{
Schematic illustration of a $\mathrm q\overline{q}$
color dipole viewed 
(a)~in a frame in which the q and $\mathrm\overline{q}$
are back-to-back,
%(not necessarily the c.m. frame), 
and (b)~in a frame boosted in a direction bisecting the dipole.
The quark and antiquark jets are labelled by their energies:
$E_{\mathrm q}^*$ and $E_{\mathrm\overline{q}}^*$
in the back-to-back frame
and $E_{\mathrm q}^\prime$ and $E_{\mathrm\overline{q}}^\prime$
in the boosted frame.
%The two frames are related by the Lorentz boost factor $\beta$,
%where $\beta$$\,=\,$$\cos\alpha$ with $\theta$$\,=\,$$2\alpha$
%the angle between the 
%Unbiased quark jets are defined by hemispheres in the 
%back-to-back frame,
%as indicated schematically by the dashed line in~(a).
%In the boosted frame,
%the unbiased jets correspond to cones 
%of half angle $\alpha$ around the jet axes,
%where $\beta$$\,=\,$$\cos\alpha$
%is the Lorentz boost factor relating the back-to-back and
%boosted frames as indicated schematically by the dotted lines in~(b).
}
\label{fig-qqdipole}
\end{figure}

The jet boost algorithm 
(henceforth referred to as the ``boost algorithm'' or ``boost method'')
is motivated by the color dipole model of QCD~\cite{bib-dipole}.
Thus consider a quark-antiquark system created 
from a color singlet source,
e.g.\ {\epem}$\,\rightarrow\,$$\mathrm q\overline{q}$ events.
%The q and $\mathrm\overline{q}$ move back-to-back.
Because the q and $\mathrm\overline{q}$ carry opposite color charges,
they form a dipole.
Unbiased quark jets are defined by dividing the event in half 
in a frame in which the q and $\mathrm \overline{q}$ move back-to-back,
using the plane perpendicular to the direction of
the separating q and $\mathrm\overline{q}$
(see Fig.~\ref{fig-qqdipole}a).
This is analogous to the definition of
unbiased gluon jets presented in the introduction.
Note that the back-to-back frame is not necessarily the 
center-of-momentum (c.m.) frame
of the dipole.
The energy scales of the jets, 
$\eqstar$ and $\eqbarstar$,
are given by the hemisphere energies in the back-to-back frame.
If a Lorentz boost is performed along the hemisphere boundary
assuming the q and $\mathrm\overline{q}$ are massless,
the dipole appears as shown in Fig.~\ref{fig-qqdipole}b.
In Appendix~\ref{sec-boost1} it is shown that
the Lorentz $\beta$ factor relating the 
back-to-back and boosted frames
is $\beta$$\,=\,$$\cos\alpha$,
where $\alpha$$\,=\,$$\theta/2$ with $\theta$
the angle between the
q and $\mathrm\overline{q}$ in the boosted frame.
Furthermore, it is shown that the jet energies in the boosted frame,
$E_i^\prime$ (with $i$$\,=\,$q or $\mathrm\overline{q}$), 
are related to the jet energies in the
back-to-back frame, $E_i^*$, by
\begin{equation}
  E_i^* = E_i^\prime \sin{\frac{\theta}{2}} \;\;\;\; .
  \label{eq-ejetstar}
\end{equation}

%Note that because of QCD coherence,
%particle emission are ordered by angle such that the
%opening angle between 
%the angles of emission of particles (quarks or gluons) 
%by the quark or antiquark are constrained to 
%gluon emission is constrained 
%each particle (quark or gluon) emitted by the jet is constrained to
%a smaller emission polar angle 
%appear at a smaller polar angle with respect to
%the jet axis than particles
%created by the preceding emission
%(see~e.g.~\cite{bib-angorder}).
%Therefore,
%in the boosted frame,
%the jets are constrained
%to cones of half angle $\alpha$ around the jet axis
%as indicated by the dotted lines in Fig.~\ref{fig-qqdipole}b.

%The jet boost algorithm~\cite{bib-eden}
%(henceforth referred to as the ``boost algorithm'' or ``boost method'')
%is motivated by the color dipole model of QCD~\cite{bib-dipole}.
%In a quark-antiquark system created from a color singlet source,
%e.g.\ {\epem}$\,\rightarrow\,$$\mathrm q\overline{q}$ events,
%the q and $\mathrm\overline{q}$ carry opposite color charges
%and so form a dipole.

\begin{figure}[t]
 \begin{center}
 \begin{tabular}{ccc}
      \epsfxsize=5cm
      \epsffile{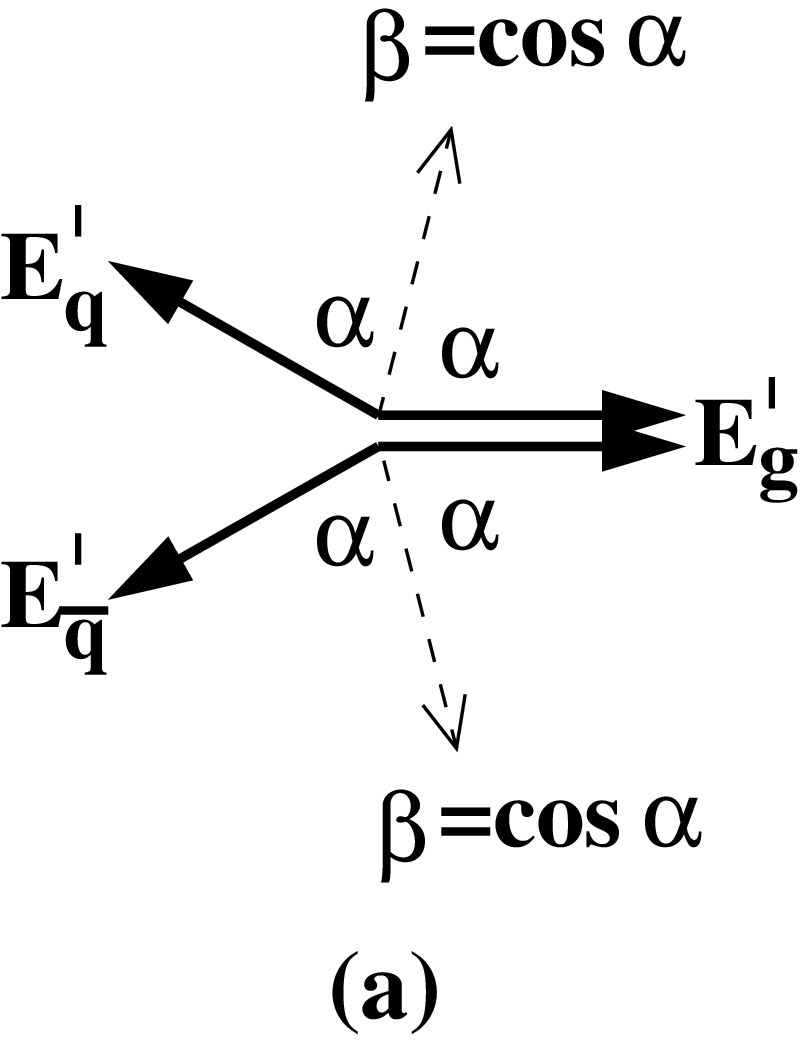} & \hspace*{.5cm}
      \epsfxsize=5cm
      \epsffile{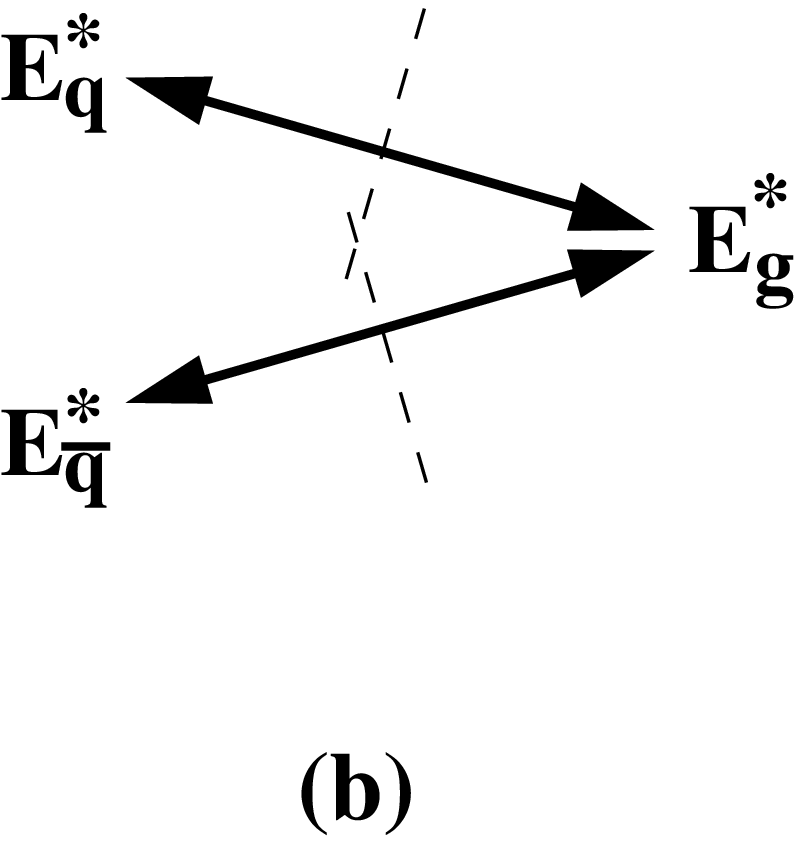} & \hspace*{.5cm}
      \epsfxsize=5cm
      \epsffile{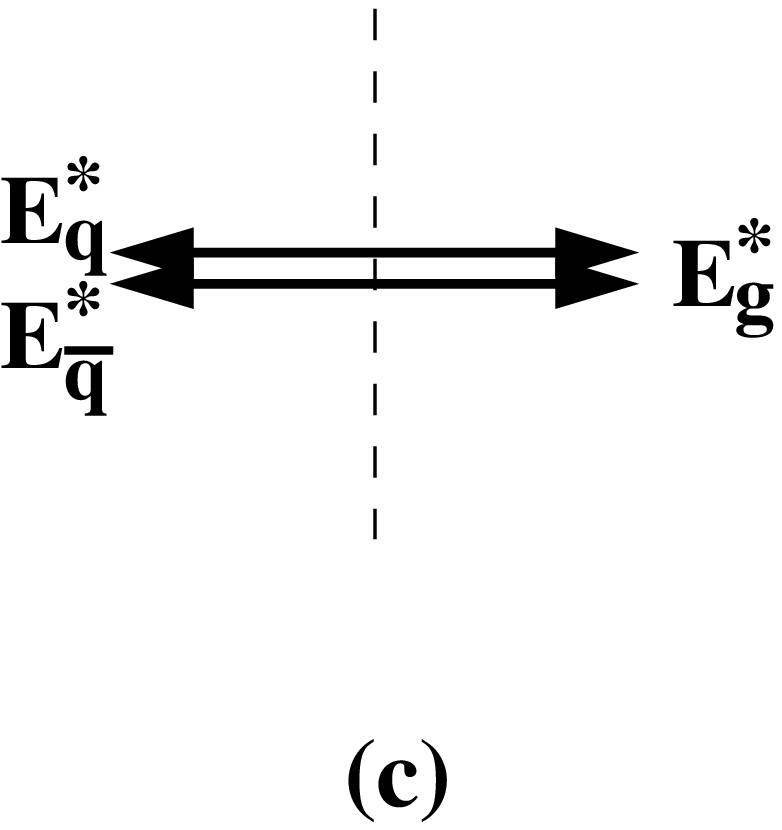}
 \end{tabular}
 \end{center}
\caption{
(a)~A symmetric three-jet $\mathrm q\overline{q}g$ event
in which the angle $\theta$$\,=\,$$2\alpha$
between the quark and gluon jets is the same as the angle
between the antiquark and gluon jets.
In the QCD dipole model, the $\mathrm q\overline{q}g$
event consists of two independent color dipoles.
(b)~Each of the dipoles can be independently boosted to 
a back-to-back frame.
(c)~The dipoles in the back-to-back frames can be combined to yield
an event with the color structure of a gluon-gluon event
in a color singlet.
Note that the combined quark-antiquark jet system in
{\epem}$\,\rightarrow\,$$\mathrm q\overline{q}g$ events has 
the color structure of a gluon jet.
}
\label{fig-qgdipole}
\end{figure}

In {\epem}$\,\rightarrow\,$$\mathrm q\overline{q}g$ events,
the color charge of the gluon can be decomposed into two parts:
one equal and opposite to the color charge of the quark
and the other equal and opposite to the charge of the antiquark.
A $\mathrm q\overline{q}g$ event therefore consists of
two independent dipoles,
one defined by the q and g and the other by
the $\mathrm \overline{q}$ and~g.
In a frame in which the angle 
$\theta$$\,=\,$$2\alpha$ between the q and g is the same as
the angle between the $\mathrm\overline{q}$ and~g, 
yielding a symmetric event as in~Fig.~\ref{fig-qgdipole}a,
each dipole can be independently
boosted to a back-to-back frame along the bisector of the dipole
using the boost factor $\beta$$\,=\,$$\cos\alpha$
mentioned above,
again assuming the partons are massless
(see Fig.~\ref{fig-qgdipole}b).
The two dipoles in the back-to-back frames can then be combined
to yield an event with the color structure of
a gg event in a color singlet,
i.e.~two back-to-back gluon jet hemispheres
(see Fig.~\ref{fig-qgdipole}c),
since the combined quark-antiquark system has
%{\epem}$\,\rightarrow\,$$\mathrm q\overline{q}g$ events has 
the color structure of the gluon jet as mentioned above.
This corresponds to the production of unbiased gluon jets
as discussed in the introduction.
%Therefore,
%in the frame of the symmetric $\mathrm q\overline{q}g$ events
%(Fig.~\ref{fig-qgdipole}a),
%an unbiased gluon jet can be defined by the particles in a cone 
%of half angle $\alpha$ around the gluon jet axis,
%as follows from the correspondence between
%Figs.~\ref{fig-qqdipole} and~\ref{fig-qgdipole}
%(see the dashed lines in Fig.~\ref{fig-qgdipole}a).
In the frame of the symmetric event
(Fig.~\ref{fig-qgdipole}a),
the unbiased gluon jet is defined by the particles in a cone 
of half angle $\alpha$ around the gluon jet axis~\cite{bib-eden}.
The energy of the unbiased gluon jet,
$\egstar$
(Fig.~\ref{fig-qgdipole}c),
is related to the energy of the gluon jet in the
symmetric $\mathrm q\overline{q}g$ event, $\egprime$
(Fig.~\ref{fig-qgdipole}a),
by eq.~(\ref{eq-ejetstar}) with $i$$\,=\,$g,
as follows from the correspondence between 
Figs.~\ref{fig-qqdipole} and~\ref{fig-qgdipole}.

%Taking the coherent nature of gluon radiation into account,
%it has been shown~\cite{}
%that the appropriate energy
%scale of quark jets in 
%{\epem}$\,\rightarrow\,$$\mathrm q\overline{q}g$ events is
%\begin{equation}
%  \qjet = \ejet \sin{\frac{\theta_{\mathrm qg}}{2}}
%  \label{eq-qscale}
%\end{equation}
%with $\theta_{\mathrm qg}$ the angle between
%the gluon and quark.
%For example,
%if the gluon is colinear with the quark 
%so that $\theta_{\mathrm qg}$$\,\approx\,$0,
%(\ref{eq-qscale}) yields $\qjet$$\,\approx\,$0.
%This corresponds to the fact that
%the quark's color charge is neutralized by the
%corresponding anticolor of the gluon
%for $\theta_{\mathrm qg}$$\,\approx\,$0,
%so that the qg dipole cannot radiate.
%The energy scale of gluon jets in 
%{\epem}$\,\rightarrow\,$$\mathrm q\overline{q}g$ events
%is in general more difficult to specify,
%because the gluon jet is connected
%to both the qg and $\mathrm\overline{q}$g dipoles.
%In symmetric events with 
%$\theta_{\mathrm qg}$$\,=\,$$\theta_{\mathrm\overline{q}g}$
%as in Fig.~\ref{}c,
%the scales associated with the two dipoles are the same,
%however.
%Therefore,
%for these symmetric events,
%$\qjet$ as defined in~(\ref{eq-qscale}) also
%gives the energy scale of gluon jets.

Three-jet $\mathrm q\overline{q}g$ events 
from {\epem} annihilations are
usually identified using a jet finding algorithm.
Some of the most common jet finders
are based on a transverse momentum cutoff, $\ptcut$,
to resolve the jets.
%``two-jet'' $\mathrm q\overline{q}$
%from ``three-jet'' $\mathrm q\overline{q}g$ events.
Examples of such algorithms are the
{\durham}, Cambridge~\cite{bib-cambridge}
and Luclus~\cite{bib-luclus} jet finders.
The value of $\ptcut$ 
(sometimes referred to as the virtuality scale~\cite{bib-eden})
specifies the maximum 
transverse momentum of radiated particles within a jet.
As a necessary but not sufficient condition to avoid biasing the jets,
$\ptcut$
should be adjusted separately for each event so that exactly
three jets are reconstructed.
In contrast,
a fixed value of $\ptcut$
truncates higher order radiation in the jet.
%and introduces a bias.
%as mentioned in the introduction.
For gluon jets identified in this manner,
any radiation (``sub-jet'')
emitted within the jet must necessarily
have a smaller transverse momentum than the gluon jet itself,
otherwise the roles of the ``sub-jet'' and ``gluon jet''
would be reversed.
Thus the transverse momentum of the gluon jet,
$\ptgluon$,
defines an effective cutoff for sub-jet radiation,
i.e.~$\ptcut$$\,=\,$$\ptgluon$.
Note that the definition of transverse momentum is ambiguous
in events with hard, acolinear gluon radiation
(for a discussion, see e.g.~\cite{bib-eden99}).
In the color dipole model,
the transverse momentum of a gluon jet
in a $\mathrm q\overline{q}g$ event is
defined by~\cite{bib-eden99}
\begin{equation}
  \ptgluon = \frac{1}{2}\,
    \sqrt {\frac{ s_{\mathrm qg}s_{\mathrm \overline{q}g} }{ s } } \;\;\;\; ,
  \label{eq-ptlu}
\end{equation}
where $s_{ij}$ ($i,j$$\,=\,$$\mathrm q,\, \overline{q},\,g$) is
the invariant mass squared of the $ij$ pair,
and $s$$\,=\,$$\ecm^2$ with
$\ecm$ the event energy in the c.m.\ frame.
Thus eq.~(\ref{eq-ptlu}) defines the virtuality scale
of gluon jets in the $\mathrm q\overline{q}g$ events.
An experimental demonstration that $\ptgluon$ 
is an appropriate scale for gluon jets 
in $\mathrm q\overline{q}g$ events 
is presented in~\cite{bib-opaleden}.

For a gluon jet to be unbiased,
its properties should be independent
of the jet resolution scale(s).
%as stated in the introduction.
%i.e.\ its properties should depend on the energy scale alone
%and not the virtuality scale as well,
In~\cite{bib-eden} it is noted that independence
from the resolution scales implies that the 
energy and virtuality scales are the same:
\begin{equation}
  \egstar = \ptgluon \;\;\;\; .
  \label{eq-onescale}
\end{equation}
%i.e.\ an unbiased jet depends on a single scale only,
%see~\cite{bib-eden}.
The boost algorithm prescription for identifying an unbiased
gluon jet
%a three-jet {\epem}$\,\rightarrow\,$$\mathrm q\overline{q}g$ event
is then as follows~\cite{bib-eden}.
Three-jet events are defined using
a transverse momentum based jet algorithm.
%For our standard analysis,
%we use the {\durham} jet finder.
The resolution parameter of the algorithm
is adjusted for every event so that exactly
three jets are reconstructed.
After identification of the gluon jet using 
standard experimental techniques (see
e.g.\ Sect.~\ref{sec-gluon}),
the event is boosted to the symmetric frame in which the
angle between the gluon and quark jets is the
same as the angle between the gluon and antiquark jets,
as in Fig.~\ref{fig-qgdipole}a.
The algebra of this boost
is uniquely specified by the requirement
of eq.~(\ref{eq-onescale})
(see Appendix~\ref{sec-boost2}).
In the symmetric frame,
the unbiased gluon jet is defined by all particles
in a cone of half angle $\alpha$$\,=\,$$\theta/2$ around the
gluon jet direction,
where $\theta$ is the angle between the gluon jet
and the other two jets
(cf.~Fig.~\ref{fig-qgdipole}a and the discussion above).
The energy of the unbiased jet, $\egstar$,
is given by eqs.~(\ref{eq-ptlu}) and~(\ref{eq-onescale}).

\section{Detector and data sample}
\label{sec-detector}

The OPAL detector is described in
detail elsewhere~\mbox{\cite{bib-detector,bib-si}}.
OPAL operated from 1989 to 2000.
The analysis presented here is based on the tracking
system and electromagnetic calorimeter.
The tracking system
consisted of a silicon microvertex detector,
an inner vertex chamber,
a large volume jet chamber,
and specialized chambers at the outer radius of the 
jet chamber to improve the measurements in the
$z$-direction.\footnote{Our right handed
coordinate system is defined so that
$z$~is parallel to the e$^-$ beam axis,
$x$ points towards the center of the LEP ring,
$r$~is the coordinate normal to the beam axis,
$\phi$~is the azimuthal angle around the beam axis with
respect to $x$,
and $\theta$ is the polar angle \mbox{with respect to~$z$.}}
The tracking system covered the region
$|\cos\theta|$$\,<\,$0.98 and
was enclosed by a solenoidal magnet coil
with an axial field of~0.435~T.
Electromagnetic energy was measured by a
lead-glass calorimeter located outside the magnet coil,
which also covered $|\cos\theta|$$\,<\,$0.98.

The present analysis is based on a sample of
about 3.13 million hadronic annihilation events,
corresponding to the OPAL sample collected
within 3~GeV of the Z$^0$ peak 
($\mzee$) from 1993 to 2000.
This sample includes readout 
of both the $r$--$\phi$ and $z$ 
coordinates of the silicon strip microvertex detector~\cite{bib-si}.
The procedures for identifying hadronic annihilation
events are described in~\cite{bib-opaltkmh}.

We employ the tracks of charged particles
reconstructed in the tracking chambers
and clusters of energy
deposited in the electromagnetic calorimeter.
Tracks are required to have at least 20 measured
points (of 159 possible) in the jet chamber,
or at least 50\% of the number of points expected 
based on the track's polar angle,
whichever is larger.
In addition, the tracks are required 
to have a momentum component perpendicular 
to the beam axis greater than 0.05~GeV/$c$,
to lie in the region $|\cos\theta|$$\,<\,$0.96,
to point to the origin to within 5~cm in the $r$--$\phi$ plane
and 30~cm in the $z$ direction,
and to yield a reasonable $\chi^2$ per
degree-of-freedom for the track fit in the $r$--$\phi$ plane.
Electromagnetic clusters are required to have an energy greater 
than 0.10~GeV if they are in the barrel section 
of the detector ($|\cos\theta|$$\,<\,$0.82)
or 0.25~GeV if they are in the endcap section 
(0.82$\,<\,$$|\cos\theta|$$\,<\,$0.98).
A matching algorithm~\cite{bib-mt} is used to
reduce double counting of energy in cases where 
charged tracks point towards electromagnetic clusters.
Specifically,
if a charged track points towards a cluster,
the cluster's energy is re-defined by subtracting 
the energy that is expected to be deposited in
the calorimeter by the track.
If the energy of the cluster is smaller than 
this expected energy,
the cluster is not used.
In this way,
the energies of the clusters are primarily associated
with neutral particles.

Each accepted track and cluster
is considered to be a particle.
Tracks are assigned the pion mass.
Clusters are assigned zero mass since they originate
mostly from photons.

To eliminate residual background and events
in which a significant number of particles is lost
near the beam direction,
the number of accepted charged tracks in an event
is required to be at least five and the 
thrust axis of the event,
calculated using the particles,
is required to satisfy
$|\cos (\theta_{\mathrm{thrust}})|$$\,<\,$0.90,
where $\theta_{\mathrm{thrust}}$ is the
angle between the thrust and beam axes.
The number of events which pass these cuts is
about 2.77~million.
The residual background to this sample
from all sources is estimated to be less 
than~1\%
%~\cite{bib-opaltkmh} 
and is neglected.

\section{Gluon jet selection}
\label{sec-gluon}

We apply the {\durham} jet finder to the
sample of events described in Sect.~\ref{sec-detector}.
The resolution scale, $y_{cut}$,
is adjusted separately for each event so that exactly
three jets are reconstructed.
Both charged and neutral particles are used for the
definition of the jets.
The jets are assigned energies using the technique
of calculated energies with massless kinematics
(see for example~\cite{bib-opalstring}).
Jet energies determined in this manner are more accurate than
visible jet energies,
with the latter defined by a sum over the reconstructed
energies of the particles assigned to the~jet.
We employ massless kinematics because the
boost algorithm assumes massless jets
(see Sect.~\ref{sec-boost} and the Appendices).
The jets are ordered such that jet~1 has the largest energy
and jet~3 the smallest energy.
%Due to finite detector capabilities,
%ordering by visible jet energies is not necessarily
%the same as ordering by calculated energies.

Due to the gluon radiation spectrum
in {\epem}$\,\rightarrow\,$$\rm q\overline{q}g$ events,
jet~1 is likely to be a quark 
\mbox{(q or $\mathrm\overline{q}$)} jet.
%if the gluon jet energy is relatively small.
%To make use of this spectrum to identify 
%gluon jets with good efficiency,
%we implement a gluon jet identification procedure
%which depends on the gluon jet's energy.
We therefore assume jet~1 is always a quark jet.
We then use the technique of displaced secondary vertices
to identify the other quark jet.
Displaced secondary vertices are associated with 
heavy quark decay,
especially that of the b quark.
At LEP, b quarks are produced almost exclusively
at the electroweak vertex:
thus a jet containing a b hadron is almost always a quark jet.
To reconstruct secondary vertices in jets,
we use the method described in~\cite{bib-qg95a}.
For jets with a secondary vertex,
the signed decay length, $L$,
is calculated with respect to the primary vertex,
along with its uncertainty,~$\sigma_L$.
%The sign of $L$ is determined by summing the 3-momenta
%of the tracks fitted to the secondary vertex;
%$L$$\,>\,$0 if the secondary vertex is displaced from the primary
%vertex in the same hemisphere as this momentum sum,
%and $L$$\,<\,$0 otherwise.
To be tagged as a quark jet,
a jet is required to contain a successfully reconstructed 
secondary vertex
with $L/\sigma_L>\;$3.0.
We select events for which 
exactly one of the lower energy jets is
tagged as a quark jet.
%We require exactly one of the lower energy jets to be tagged
%as a quark jet.
The remaining lower energy jet in these events 
is identified as the gluon jet.

We next examine the selected events as a function of the
energy $\egstar$ (see eq.~(\ref{eq-onescale}))
of the identified gluon jet.
$\egstar$ is calculated using the
jet 4-momenta in the laboratory frame
and the Lorentz invariant expression eq.~(\ref{eq-ptlu}).
We require $\egstar$ to be at least 5.0~GeV 
so that the jet is well defined.
For 5.0$\,\leq\,$$\egstar$$\,<\,$9.5~GeV,
the estimated gluon jet purity is
about 80\% or larger once the final selection
cuts have been applied
(see below).
For values of $\egstar$ above this,
the purity is lower
because the assumption that jet~1 is a quark jet becomes
less accurate as the gluon jet energy increases.
Therefore,
for $\egstar$$\,\geq\,$9.5~GeV,
we impose additional requirements
on the two identified quark jets.
A quark jet in an event with
9.5$\,\leq\,$$\egstar$$\,<\,$16.0~GeV
is required to contain a successfully reconstructed secondary
vertex with $L/\sigma_L$$\,>\,$3.0 if it is either jet 1 or~2,
or $L/\sigma_L$$\,>\,$5.0 if it is jet~3.
These cuts account for the fact that the $L/\sigma_L$
distributions of jets depend upon the jet energy.
For events with 16.0$\,\leq\,$$\egstar$$\,<\,$20.0~GeV,
a quark jet is required to contain a secondary vertex
with $L/\sigma_L$$\,>\,$5.0
irrespective of whether it is jet 1, 2 or~3.
We retain events in which the two identified quark jets 
(as defined in the previous paragraph)
satisfy these more stringent requirements.
We do not consider gluon jets with $\egstar$$\,\geq\,$20.0~GeV
because of the low event statistics.

The resulting $\mathrm q\overline{q}g$ sample
contains many events with soft or nearly colinear jets.
To eliminate these events,
we impose cuts on the jet
energies and angles with respect to the other jets.
Besides the requirement $\egstar$$\,\geq\,$5.0~GeV
for gluon jets,
mentioned above,
we determine the following scale for quark jets in the 
laboratory frame:
\begin{equation}
   \qjet = \ejet\,\sin\left(
         \frac{\theta_{\mathrm min.}}{2} \right)
     \;\;\;\; ,
  \label{eq-scale}
\end{equation}
%where $E_{\mathrm jet}$ is the energy of the jet,
with $\theta_{\mathrm min.}$ the smaller of the angles between
the jet under consideration and the other two jets.
The scale eq.~(\ref{eq-scale})
was proposed in~\cite{bib-dok88}
(see also~\cite{bib-kappascale}).
Note the similarity between eqs.~(\ref{eq-ejetstar})
and~(\ref{eq-scale}).
We require the quark jets to satisfy
$\qjet$$\,\geq\,$8.0~GeV.
After applying all cuts, 
the number of selected events is~$25\,396$.

The purity of this sample is evaluated
using simulated events generated with the Herwig
Monte Carlo event generator, version 6.2~\cite{bib-herwig}.
Herwig is chosen because it is known to provide
a better description of gluon jets in {\epem} annihilations
than the available alternatives
(see e.g.~\cite{bib-opalhemisphere-1}).
The Monte Carlo events are examined at the ``detector level.''
The detector level includes initial-state photon radiation, 
simulation of the OPAL detector~\cite{bib-gopal},
and the same analysis procedures as are applied to the data.
The detector level Herwig sample in our study contains
six million inclusive Z$^0$ events.
%Six million Herwig events was processed through
%the detector simulation and used as the detector level sample in our study.
The parameter values we use for Herwig
are documented in~\cite{bib-opalrapgap}.
We determine the directions of the primary
quark and antiquark from the Z$^0$ 
decay after the parton shower has terminated.
The reconstructed jet closest to 
the direction of an evolved primary quark or antiquark
is considered to be a quark jet.
The distinct jet closest to the evolved primary
quark or antiquark not associated with this first
jet is considered to be the other quark jet.
The remaining jet is the gluon jet.
Using this method,
the overall purity of the final gluon jet sample
is found to be $85.1\pm 0.2\,\mathrm{(stat.)}$\%.

The data are binned in seven intervals of $\egstar$.
The bin edges are chosen so that the mean gluon jet energy
for most bins corresponds to an energy
at which unbiased quark jet multiplicity data are available
for comparison
(see Sect.~\ref{sec-ratios}).
Table~\ref{tab-purities} summarizes the bin definition,
number of gluon jets,
mean jet energy $\langle\egstar\rangle$
and estimated gluon jet purity,
for each bin.
The systematic uncertainties attributed to the $\langle\egstar\rangle$
and purity results are discussed in Sect.~\ref{sec-systematic}.

\begin{table}[t]
 \begin{center}
  \begin{tabular}{|c|ccc|}
\hline
    Bin in $\egstar$ (GeV) & Number of jets 
  & $\langle\egstar\rangle$ (GeV) & Purity (\%) \\
\hline
    5.0--5.5    & 4022 &  $5.25\pm 0.01\pm 0.01$ & $88.8\pm 0.4\pm 1.4$ \\
    5.5--6.5    & 6652 &  $5.98\pm 0.01\pm 0.01$ & $87.3\pm 0.3\pm 1.6$ \\
    6.5--7.5    & 5017 &  $6.98\pm 0.01\pm 0.01$ & $84.2\pm 0.4\pm 2.3$ \\
    7.5--9.5    & 7390 &  $8.43\pm 0.01\pm 0.01$ & $79.2\pm 0.3\pm 2.2$ \\
    9.5--13.0   & 1713 & $10.92\pm 0.02\pm 0.04$ & $94.5\pm 0.3\pm 3.6$ \\
    13.0--16.0  &  485 & $14.24\pm 0.04\pm 0.05$ & $86.1\pm 0.9\pm 4.2$ \\
    16.0--20.0  &  117 & $17.72\pm 0.11\pm 0.21$ & $73.9\pm 2.5\pm 8.9$ \\
\hline
  5.0--20.0 & $25\,396$ &  $7.32\pm 0.01\pm 0.07$ & $85.1\pm 0.2\pm 2.6$ \\
\hline
  \end{tabular}
\caption{Bins in the unbiased jet energy $\egstar$,
 and the corresponding number of jets, mean energies,
 and estimated purities,
 for the gluon jets in our final event sample.
 The last row gives the results for the entire sample.
 For the $\langle\egstar\rangle$ and purity results,
 the first uncertainty is statistical and the second systematic.
}
\label{tab-purities}
 \end{center}
\end{table}

The boost algorithm 
(Sect.~\ref{sec-boost})
is applied to the selected
$\mathrm q\overline{q}g$ events.
Henceforth,
by ``gluon jet,'' we refer to gluon jets treated
according to this prescription.

Because we rely on b quark tagging to identify gluon jets,
the events we study are enriched in heavy quark jets.
This is in apparent contradiction with the assumption
of the boost algorithm that the jets are massless.
The Herwig Monte Carlo predicts that about 80\% of
the events in the final sample are b events.
In Sect.~\ref{sec-test},
we show that this reliance on b events does not affect
the applicability of the method
(see Fig.~\ref{fig-btest} below).
Also note that the properties of hard, acolinear gluon jets
do not depend on the event flavor according to QCD,
as has been experimentally demonstrated in 
e.g.~\cite{bib-qg95b}.

\section{Experimental distributions}

We study the charged particle multiplicity distributions
of the identified gluon jets,
$\nchgluon$.
The multiplicity distributions are presented in terms
of their fractional probabilities,
$P(\nchgluon)$,
and are thus normalized to have unit area.
We also study the fragmentation functions of the jets.
The fragmentation function $\fragfunc$ is defined by
the inclusive distribution of scaled charged particle
energies $x_E^*$$\,=\,$$E^*/\egstar$
in the back-to-back frames of the qg and $\mathrm\overline{q}g$ dipoles
(see Fig.~\ref{fig-qgdipole}b).
The fragmentation functions are
normalized to the number of events $N$ in the respective
bins of $\egstar$ (see Table~\ref{tab-purities}).
To determine the particle energies $E^*$,
particles assigned to the gluon jet in the
symmetric frame (Fig.~\ref{fig-qgdipole}a)
are boosted to the back-to-back frames of the dipoles
using the boost factor $\beta$$\,=\,$$\cos\alpha$
mentioned in Sect.~\ref{sec-boost}
(see also Appendix~\ref{sec-boost1}).
In the data,
it is not possible to know the dipole with which 
a particle should be associated.
Therefore,
we tried both possibilities.
We found that the same results are obtained
irrespective of whether the particles are boosted to the
frame of the qg or the $\mathrm\overline{q}g$ dipole.

%The multiplicity distributions are normalized to have unit area.
%The fragmentation functions are
%normalized to the number of events $N$ in the respective bins of $\egstar$.

We also examine the mean
and first two non-trivial normalized factorial moments 
of the $\nchgluon$ distribution,
denoted $\mnchgluon$, $\factwogluon$ and $\facthreegluon$,
respectively.
Normalized factorial moments~\cite{bib-facmom}
are defined by
\begin{equation}
   F_\ell = \frac{ \langle n(n-1)\cdots (n-\ell+1) \rangle  }
                        { \langle n \rangle^\ell }
    \;\;\;\; ,
\end{equation}
with $n$$\,=\,$$\nchgluon$ and $\ell$ a positive integer.
Note that $\factwo$ is directly related to the dispersion 
of a distribution
while $\facthree$ is related to both the skew and dispersion
(see e.g.~\cite{bib-opalhemisphere-2}).
Thus normalized factorial moments provide information about
the shape of a distribution,
or equivalently
about event-to-event fluctuations from the mean.
We study normalized factorial moments because
QCD predictions for the shape of multiplicity distributions
are usually presented in that form
(for a review,
see~\cite{bib-eddi}).

\section{Test of the boost algorithm}
\label{sec-test}

Before describing our results,
we present a test of the boost algorithm
using events generated with 
the Herwig Monte Carlo event generator.
With simulated events,
it is possible to compare
gluon jets from {\epem} hadronic Z$^0$ decays
as used in the experiment
with unbiased gluon jets from
color singlet gg events
as used in theoretical calculations.
%The later provide a standard for unbiased gluon jets as
%discussed in the introduction.
%By comparing results from the two samples,
%the validity of the boost algorithm 
%can therefore be established.
%The Herwig Monte Carlo is used for this test
%because it contains an event generator and hadronization
%model for both event samples.
%\footnote{In contrast,
%the hadronization model used e.g.\ by Ariadne 
%does not include a mechanism to hadronize a purely
%system ???}

The Monte Carlo events are examined at the ``hadron level.''
The hadron level does not include 
initial-state radiation or detector simulation and utilizes
all charged and neutral particles with lifetimes
greater than 3$\,\times\,$$10^{-10}$~s,
which are treated as stable.
For the inclusive Z$^0$ hadronic events,
we generated a sample with 10~million events.
For the gg event samples,
10~million events were generated at each energy
(see below).

\begin{figure}[p]
 \begin{center}
   \epsfxsize=16.cm
   \epsffile{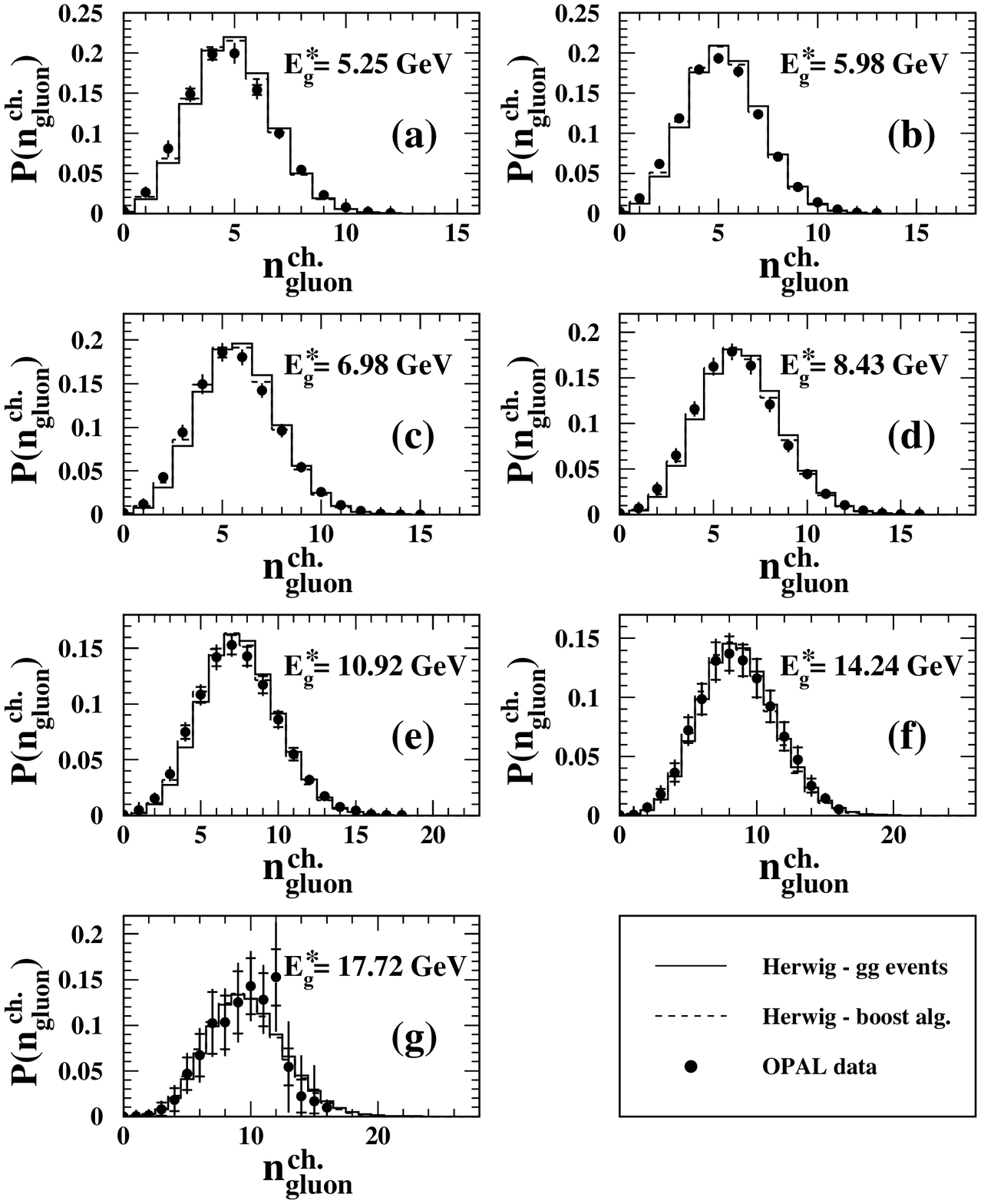}
 \end{center}
\vspace*{-6mm}
\caption{
Charged particle multiplicity distributions of gluon jets,
$\nchgluon$,
for different jet energies~$\egstar$.
The data have been corrected for detector acceptance
and resolution,
for event selection,
and for gluon jet impurity.
The total uncertainties are shown by the vertical lines, 
with the statistical component delimited by small horizontal lines.
The data are presented in comparison to predictions of 
the Herwig Monte Carlo event generator
at the hadron level.
Two different sets of Monte Carlo results are included:
one based on {\epem} events treated using the boost algorithm,
and one based on hemispheres of gg events.
}
\label{fig-nchtest}
\end{figure}

The dashed histograms in 
Fig.~\ref{fig-nchtest} show the prediction of Herwig
for the $\nchgluon$ distributions of Z$^0$ events,
obtained using the boost algorithm.
The results are shown for the seven bins of energy $\egstar$
defined in Table~\ref{tab-purities}.
The events are selected using the procedures
described in Sect.~\ref{sec-gluon} for the data,
except that the quark jet identification is performed
using Monte Carlo information as explained in
Sect.~\ref{sec-gluon}.
%rather than e.g.~secondary vertices.
The solid histograms
show the corresponding results for hemispheres of gg events.
The energies of the gg hemispheres are chosen to equal
the mean energies of the jets obtained from the boost algorithm,
for each bin.
The solid points with uncertainties in Fig.~\ref{fig-nchtest}
show our corrected data:
these are discussed in Sect.~\ref{sec-multiplicity}.

\begin{figure}[p]
 \begin{center}
   \epsfxsize=13.1cm
   \epsffile{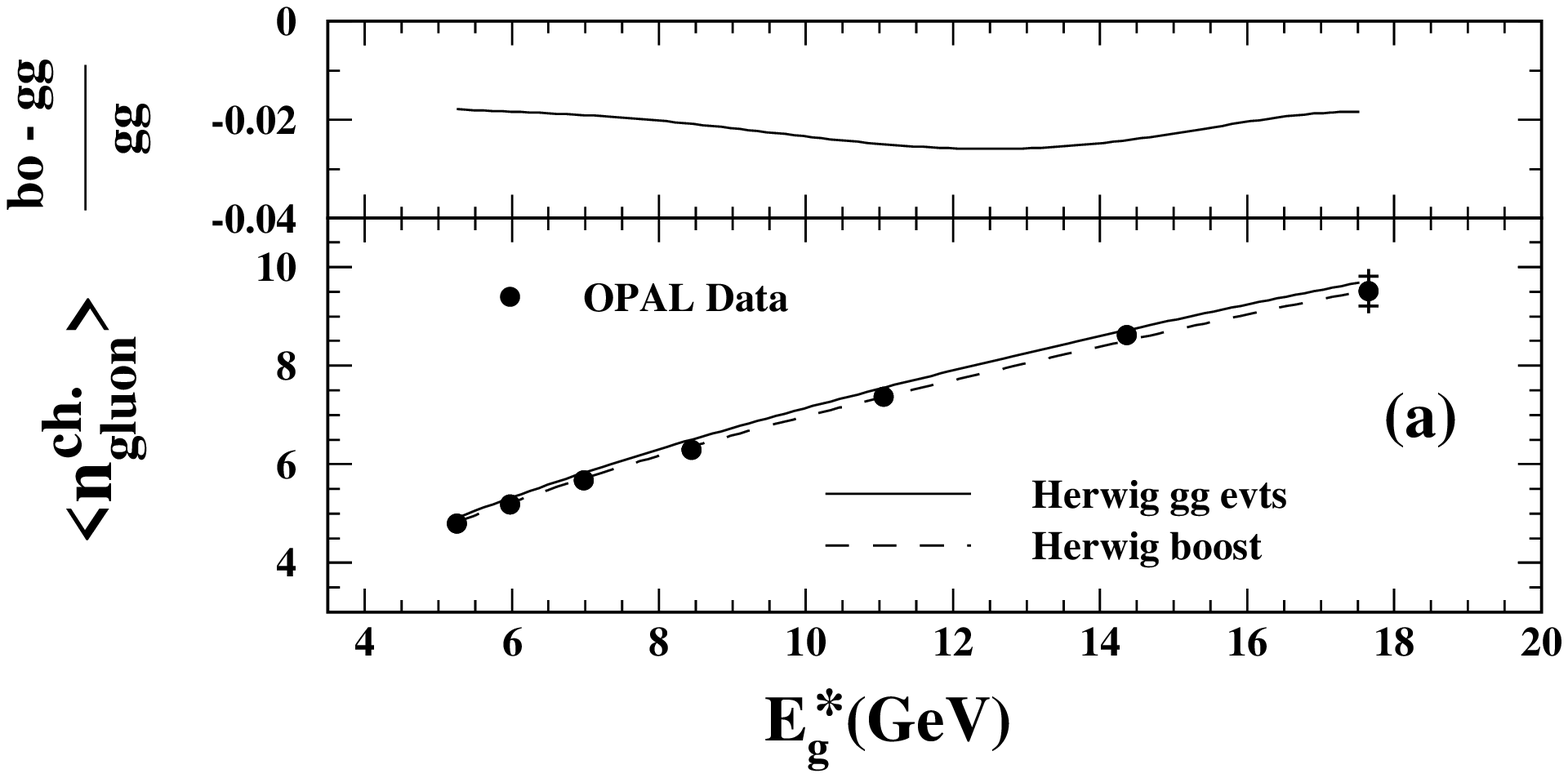} \\
   \epsfxsize=13.1cm
   \epsffile{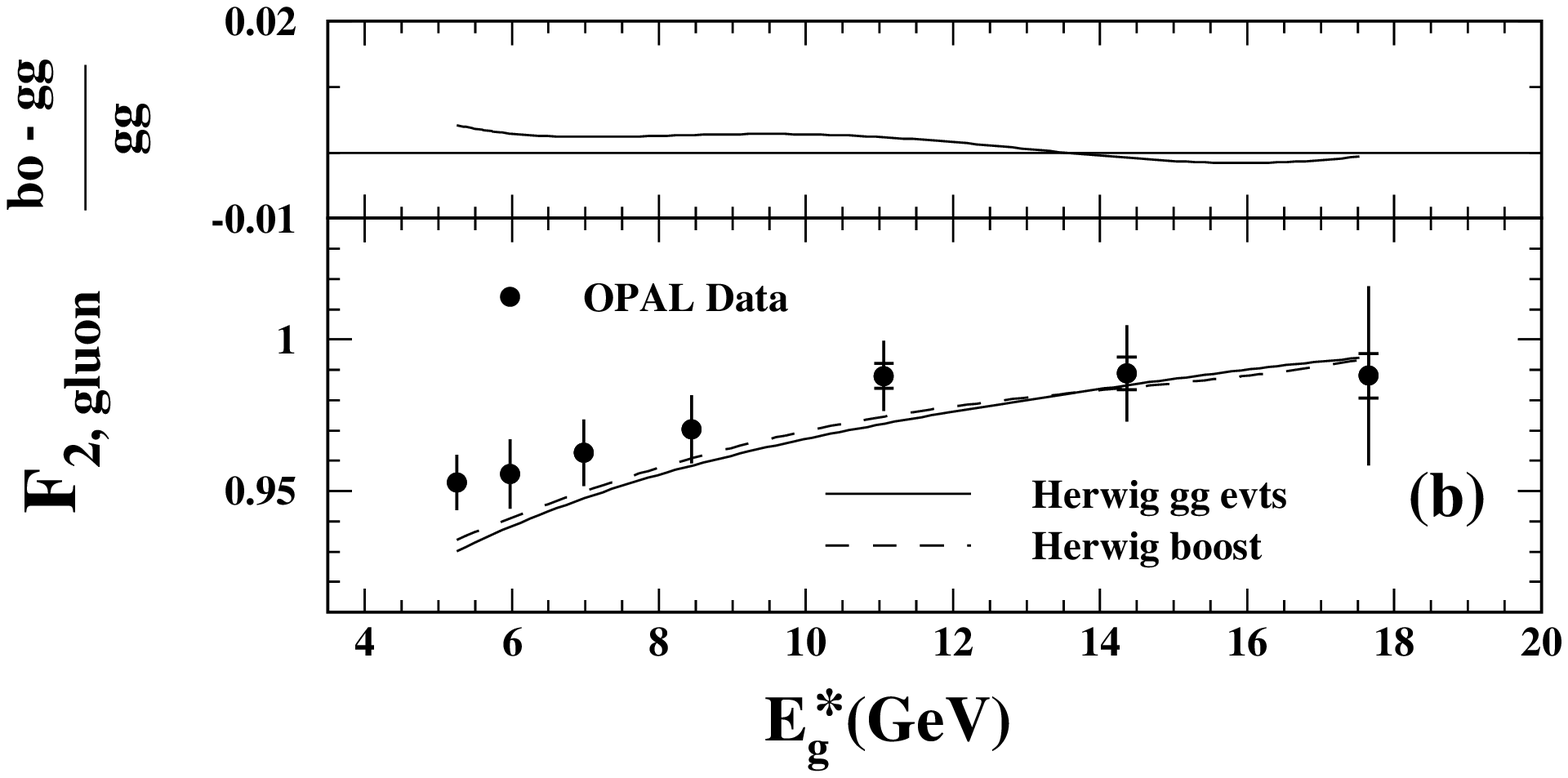} \\
   \epsfxsize=13.1cm
   \epsffile{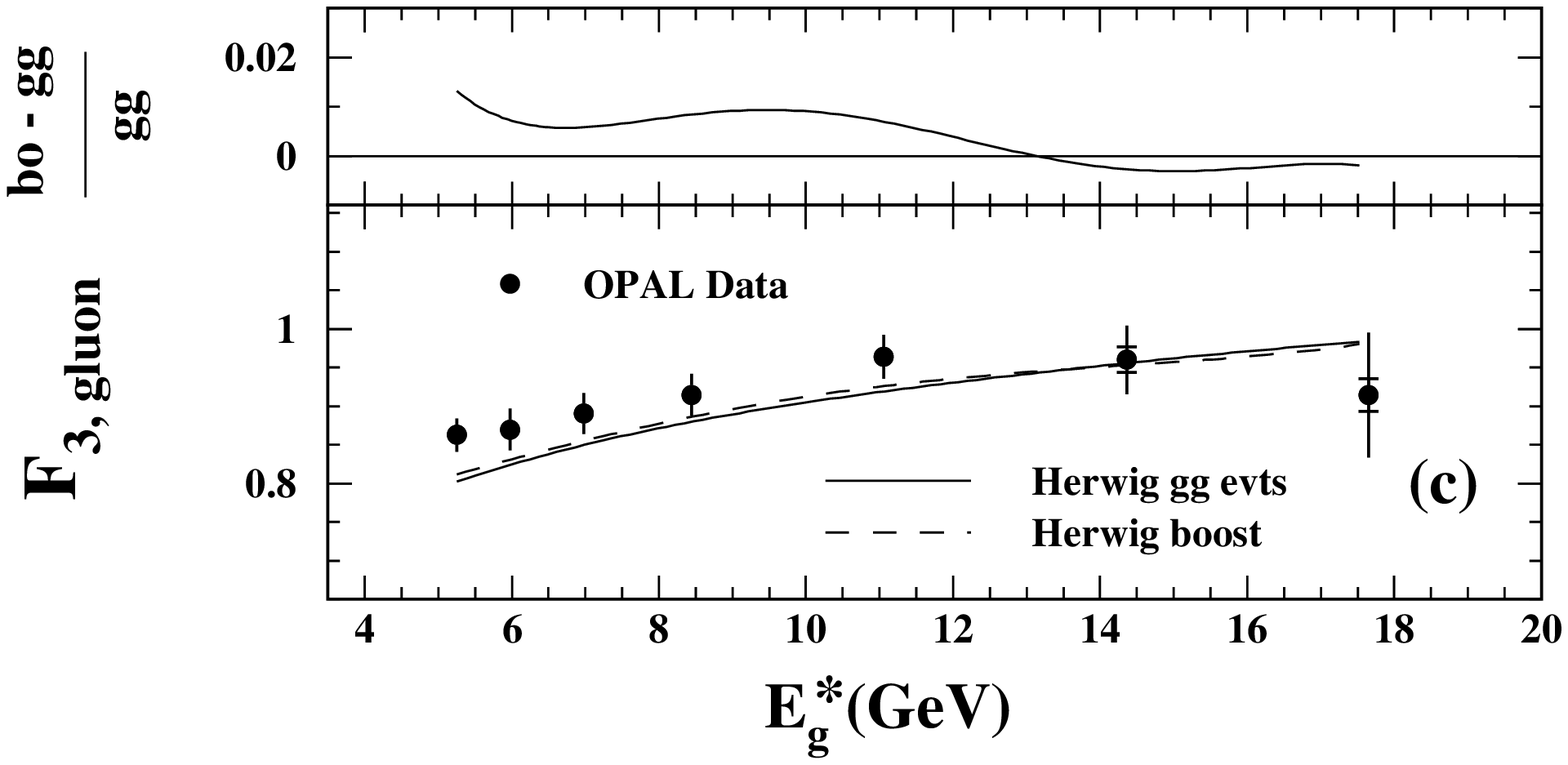}
 \end{center}
\vspace{-6mm}
\caption{
(a)~The mean charged particle multiplicity value of
gluon jets, $\mnchgluon$,
as a function of the gluon jet energy~$\egstar$.
(b,c)~The corresponding results for the two lowest 
non-trivial normalized factorial moments,
$\factwogluon$ and $\facthreegluon$.
The data have been corrected for detector acceptance
and resolution,
for event selection,
and for gluon jet impurity.
The total uncertainties are shown by the vertical lines, 
with the statistical component delimited by small horizontal lines.
The data are presented in comparison to predictions of 
the Herwig Monte Carlo event generator at the hadron level.
The small figures above each distribution show the 
fractional differences between the results of
Herwig found using the boost (``bo'') 
and gg event hemisphere (``gg'') methods.
}
\label{fig-moments}
\end{figure}

The analogous results for the mean value $\mnchgluon$
and the normalized factorial moments
$\factwogluon$ and $\facthreegluon$ 
are presented in Fig.~\ref{fig-moments}.
The small figures above the distributions in Fig.~\ref{fig-moments}
show the fractional differences between the results of the 
boost and gg hemisphere methods.
Note that the statistical uncertainties of these differences are
much smaller than the differences themselves,
as is also true for the other difference plots between
the boost and gg hemisphere methods presented below.

%The statistical uncertainties are too small
%to be visible.
%Fig.~\ref{fig-nchdeviations} shows the deviations of the
%hemisphere and {\epem} results,
%defined by subtracting the latter from the former.

From Fig.~\ref{fig-nchtest} it is seen that
the results of the boost algorithm correspond well
with those of the gg hemispheres.
Nonetheless,
a small shift towards lower $\nchgluon$ 
is present in the distributions from the boost method,
as is most clearly visible from the difference plot
in Fig.~\ref{fig-moments}a.
From this plot,
the shift is seen to be about 2\%,
independent of the energy.
This difference of 2\%
is comparable to the experimental uncertainties
(see Sect.~\ref{sec-multiplicity})
and no correction is made for it.
From the difference plots in Figs.~\ref{fig-moments}b and~c,
it is seen that the results for $\factwogluon$ and $\facthreegluon$ 
from the boost method
agree to better than about 1\% with those of gg hemispheres,
i.e.~the shapes of the $\nchgluon$
distributions found using the two methods
are very similar.
We conclude that the boost algorithm
provides an accurate means to measure unbiased gluon jet
multiplicity,
at least for jet energies larger than 5~GeV.

\begin{figure}[p]
 \begin{center}
   \begin{tabular}{cc}
      \epsfxsize=7.8cm
      \epsffile{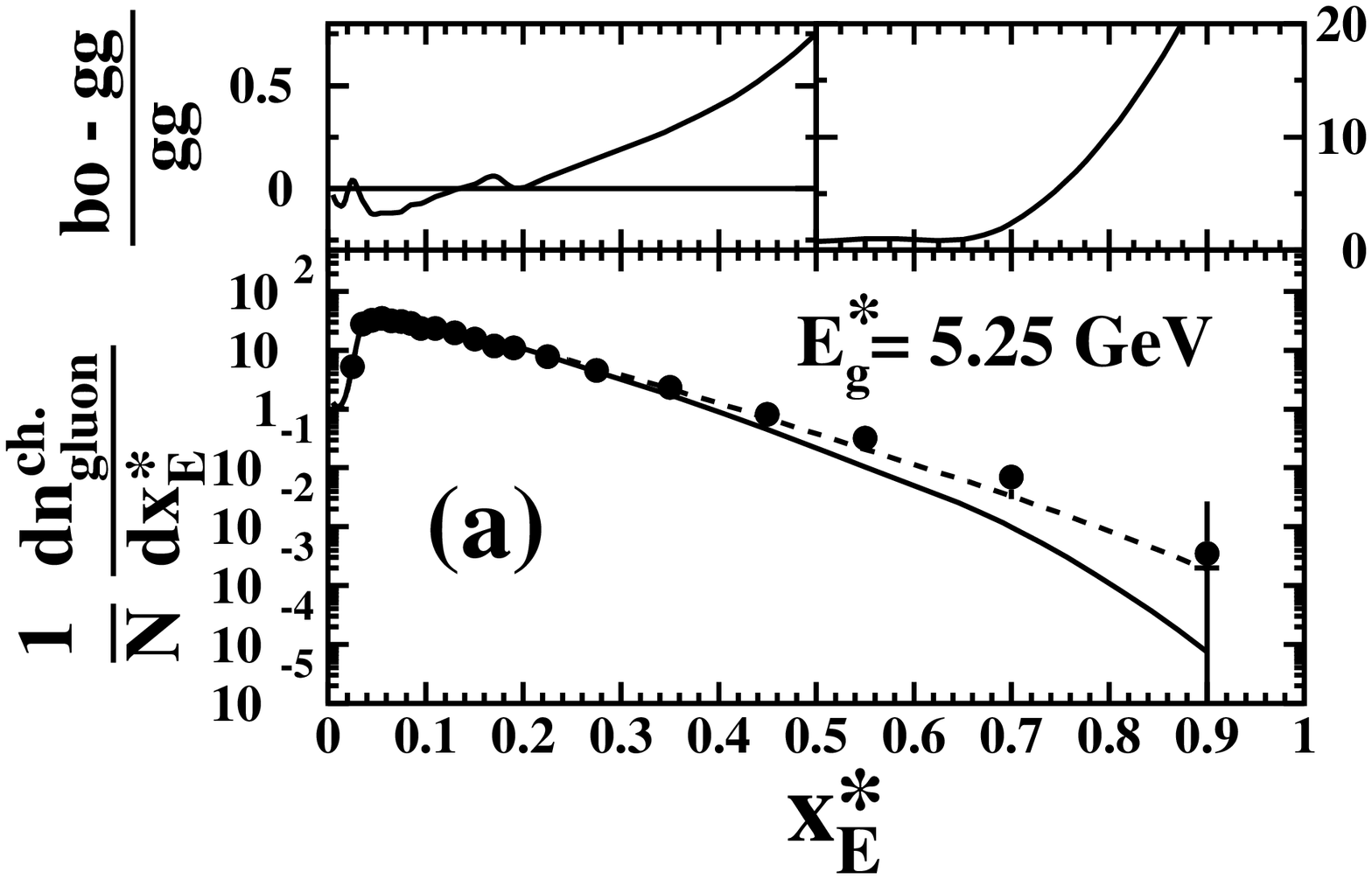} &
      \epsfxsize=7.8cm
      \epsffile{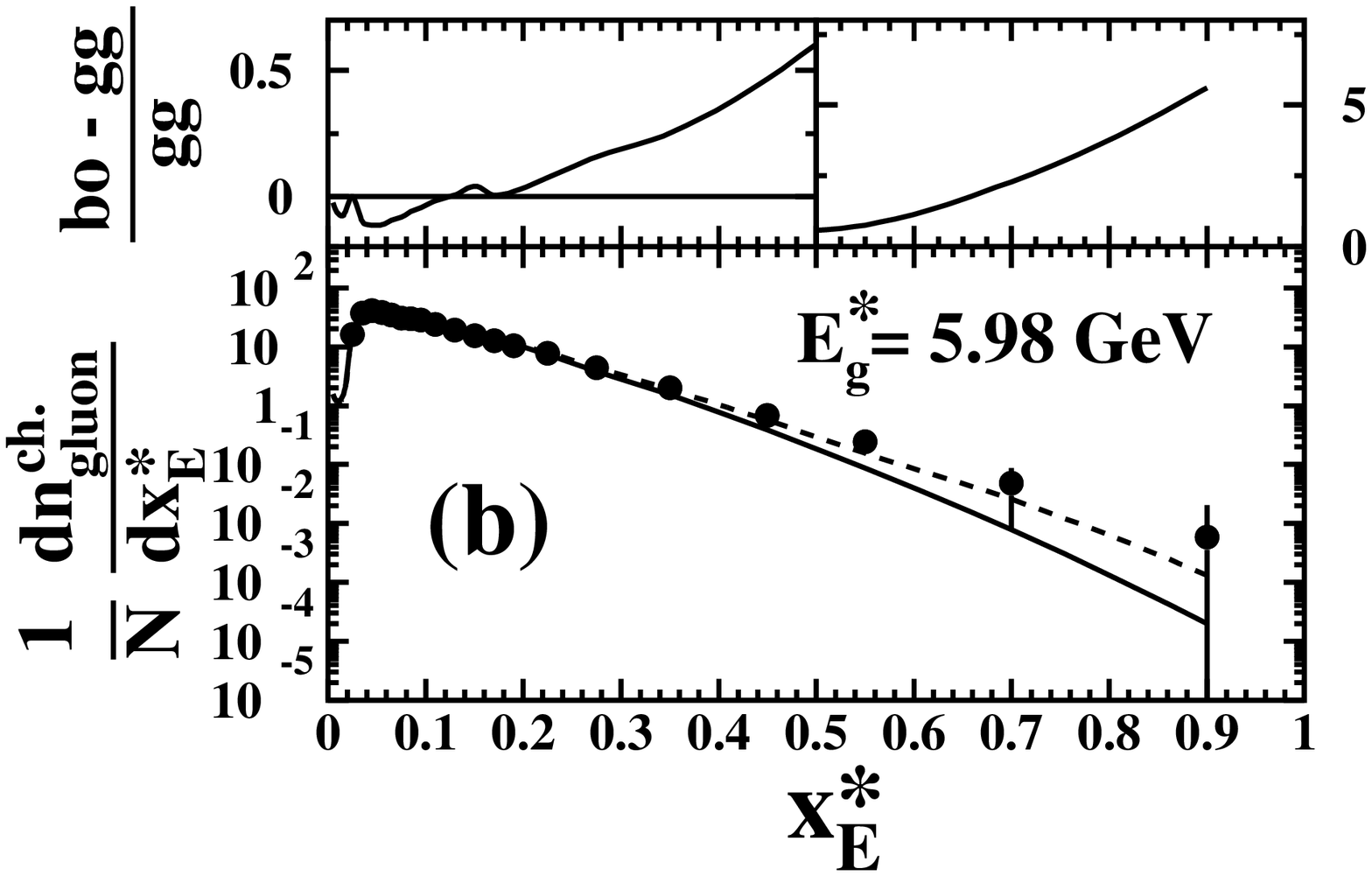} \\[-1mm]
      \epsfxsize=7.8cm
      \epsffile{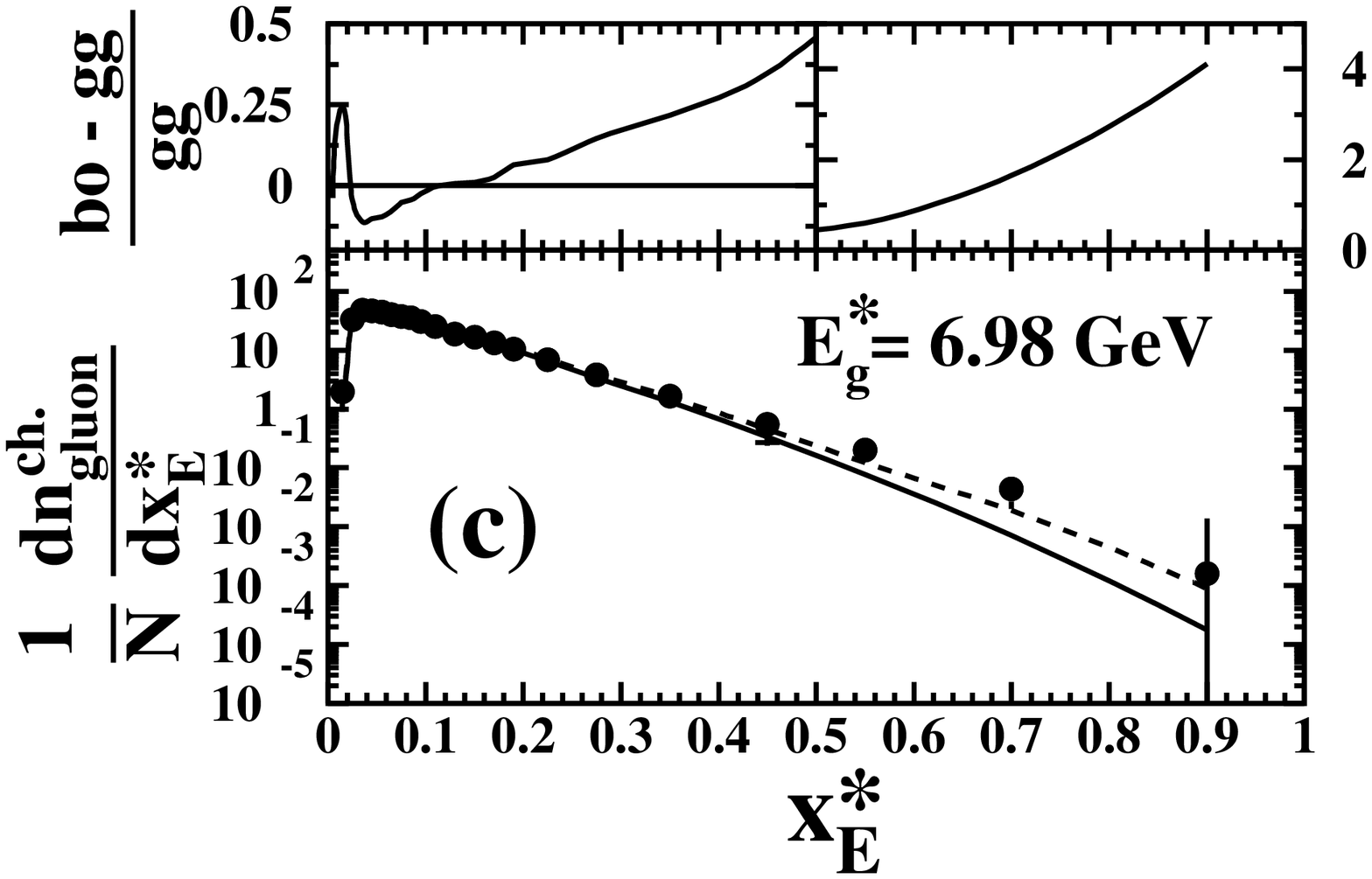} &
      \epsfxsize=7.8cm
      \epsffile{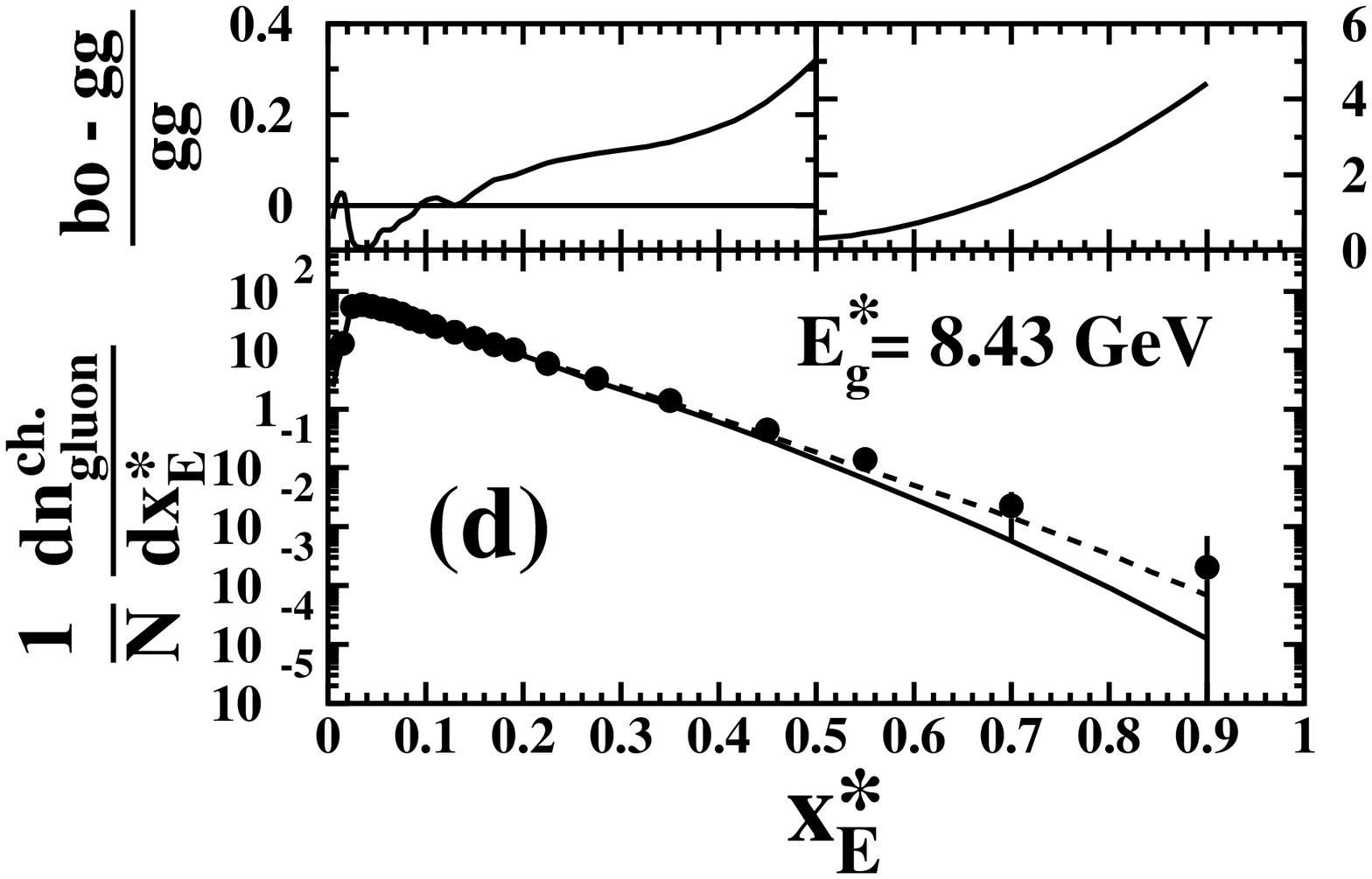} \\[-1mm]
      \epsfxsize=7.8cm
      \epsffile{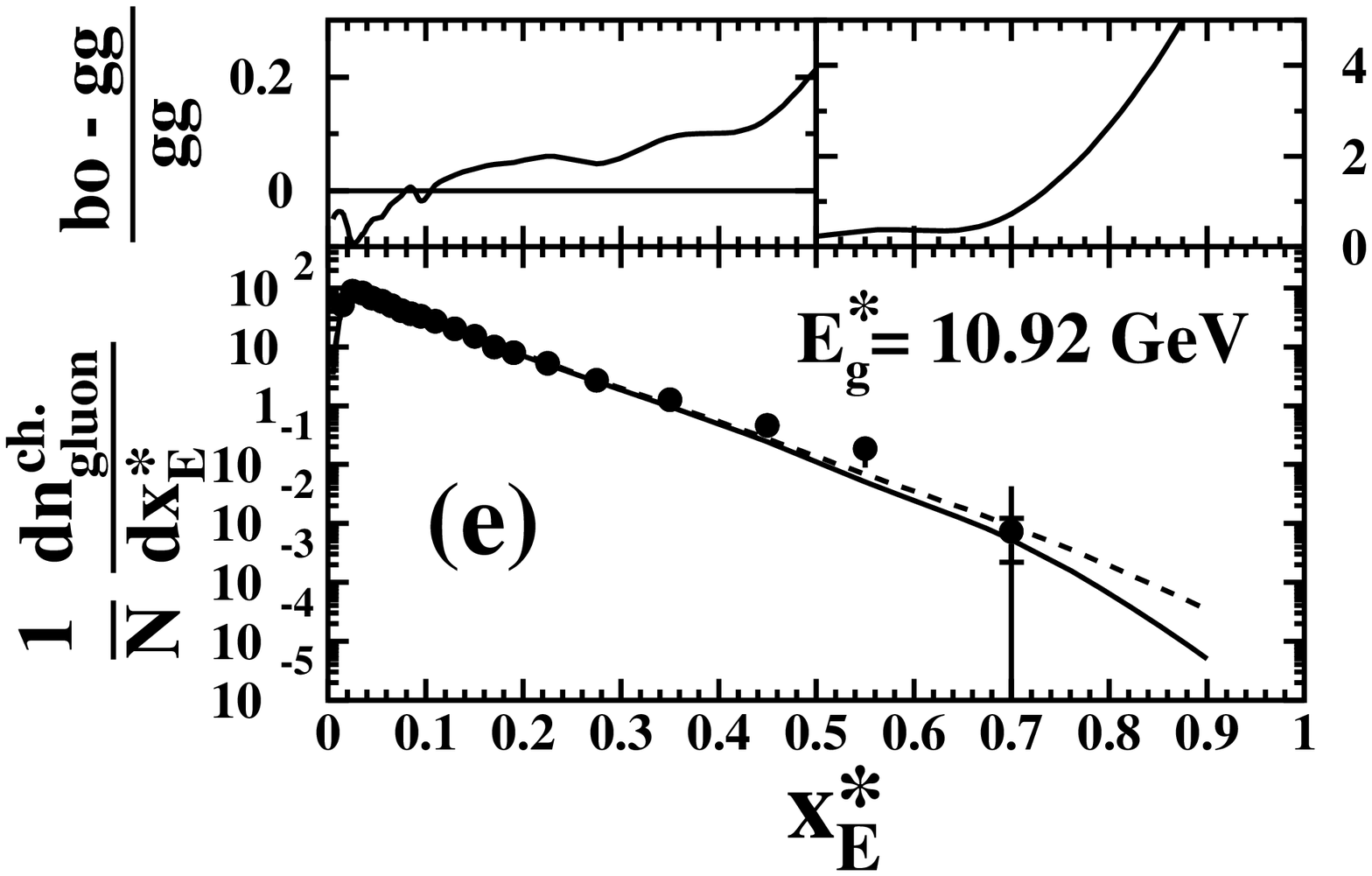} &
      \epsfxsize=7.8cm
      \epsffile{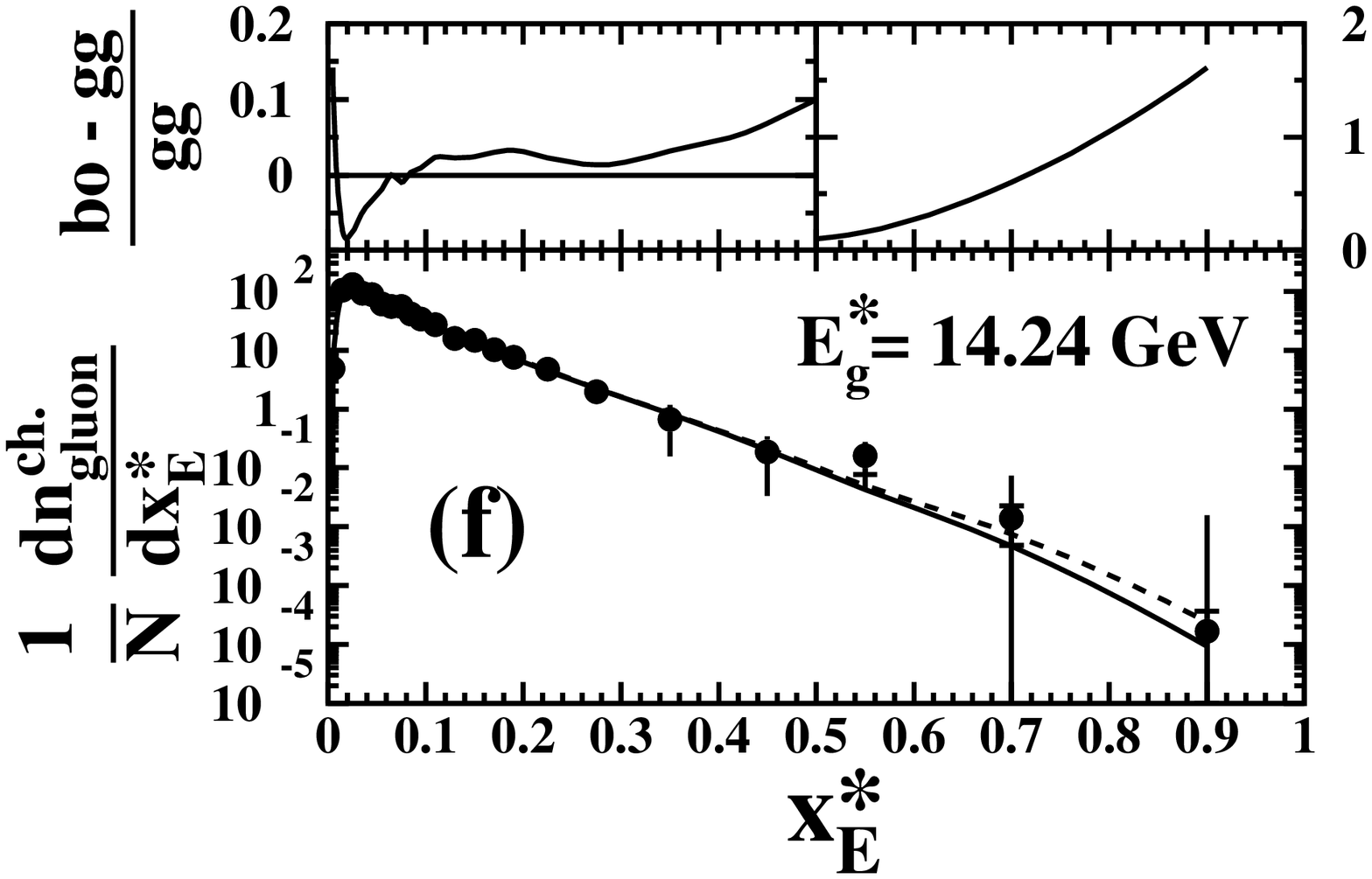} \\[-1mm]
      \epsfxsize=7.7cm
      \epsffile{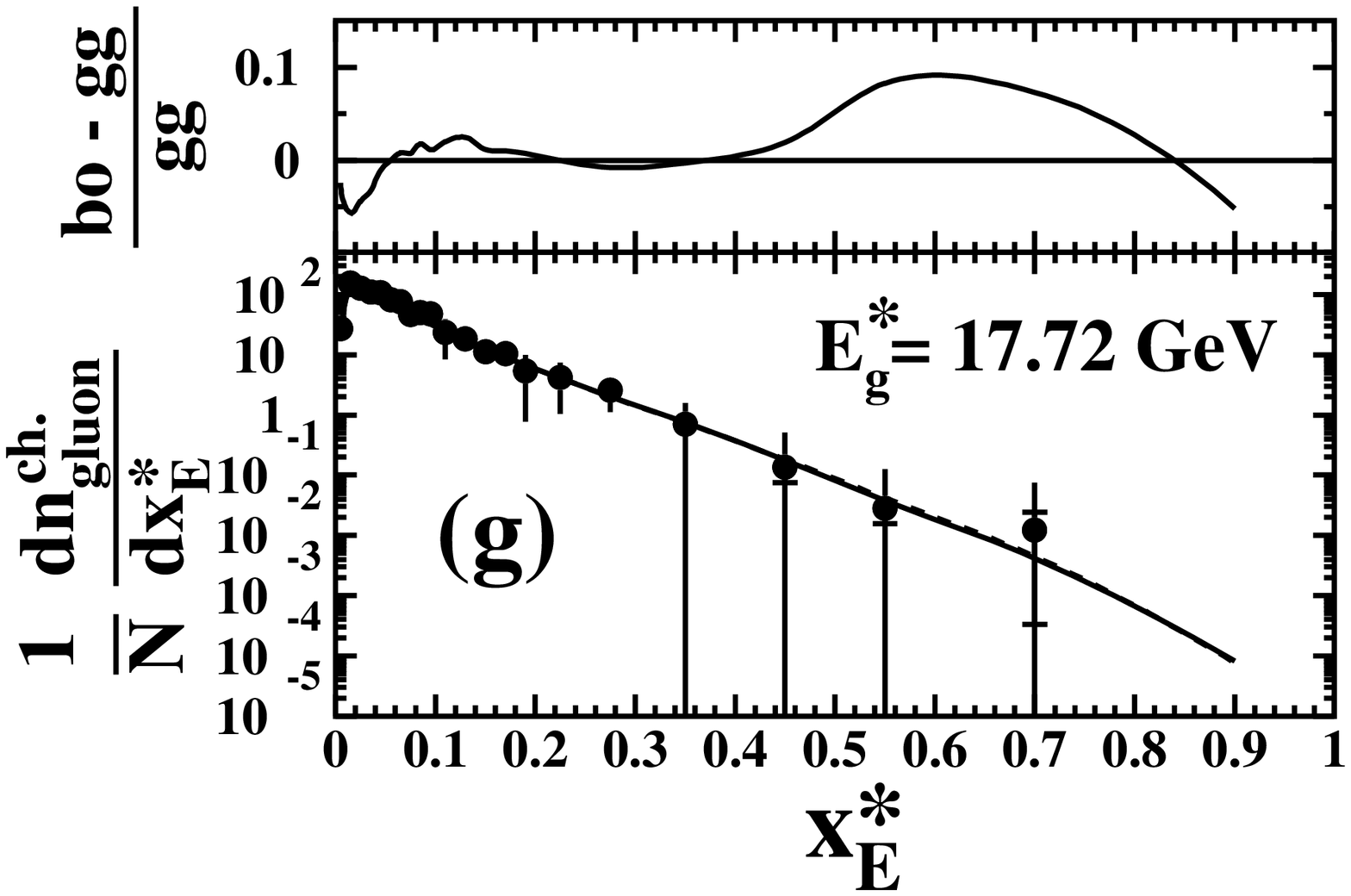} &
      \epsfxsize=5.7cm
      \epsffile[-30 -30 170 103]{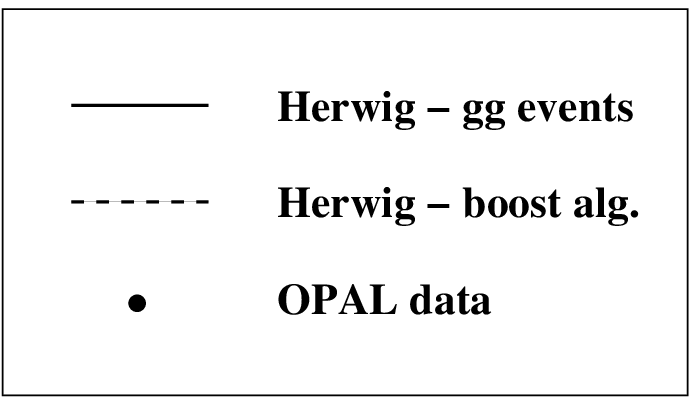} \\
   \end{tabular}
 \end{center}
\vspace*{-7mm}
\caption{
Charged particle fragmentation functions of gluon jets,
$\fragfunc$,
for different jet energy values~$\egstar$.
The data have been corrected for detector acceptance
and resolution,
for event selection,
and for gluon jet impurity.
The total uncertainties are shown by the vertical lines, 
with the statistical component delimited by small horizontal lines.
The data are presented in comparison to predictions of 
the Herwig Monte Carlo event generator at the hadron level.
The small figures above each distribution show the 
fractional differences between the results of
Herwig found using the boost (``bo'') 
and gg event hemisphere (``gg'') methods.
}
\label{fig-fftest}
\end{figure}

An analogous study of the gluon jet
fragmentation function is presented in 
\mbox{Fig.~\ref{fig-fftest}}.
For $\egstar$$\,\gtsim\,$11~GeV (Figs.~\ref{fig-fftest}e--g),
the results of the boost and gg hemisphere methods
are seen to be in reasonable agreement,
i.e.~the solid and dashed curves are quite similar.
For smaller energies (Figs.~\ref{fig-fftest}a--d),
the boost algorithm predicts a significant
excess of particles with large $x_E^*$ values
compared to the gg events, however.
The reason the boost method more accurately describes the
properties of gg events as the jet energy increases is
that the assumption of massless gluon jets
(Sect.~\ref{sec-boost})
becomes more accurate for larger jet energies.
We verified using Monte Carlo events with
$\ecm$$\,>\,$$\mzee$ that the agreement between the two methods
is even better for $\egstar$ values above those in our study.
%than are accessible in our study.

The difference plots in the top portions of 
\mbox{Figs.~\ref{fig-fftest}a--g}
show the fractional differences between the results of the
boost and gg hemisphere methods.
The difference plots for Figs.~\ref{fig-fftest}a--f
are presented on two scales,
one for 0.0$\,\leq\,$$x_E^*$$\,\leq\,$0.50 and
the other for 0.50$\,\leq\,$$x_E^*$$\,\leq\,$1.00,
to improve their visibility.
For $\egstar$$\,\leq\,$10.92~GeV (Figs.~\ref{fig-fftest}a-e),
the results of the boost algorithm are seen to deviate
from those of the gg hemispheres by up to about 20\% or more,
even for $x_E^*$$\,\ltsim\,$0.50 
where the experimental uncertainties are relatively small
(see Sect.~\ref{sec-ff} for a discussion of the data).
For $\egstar$$\,=\,$14.24 and 17.72~GeV
(Figs.~\ref{fig-fftest}f and~g),
the deviations for $x_E^*$$\,\ltsim\,$0.50 
are at most about 10\%
and in most $x_E^*$ bins much less.
In our study of the gluon jet fragmentation function (Sect.~\ref{sec-ff}),
we therefore restrict our attention to the jet samples with 
$\egstar$$\,=\,$14.24 and 17.72~GeV.

\begin{figure}[t]
 \begin{center}
      \epsfxsize=17cm
      \epsffile{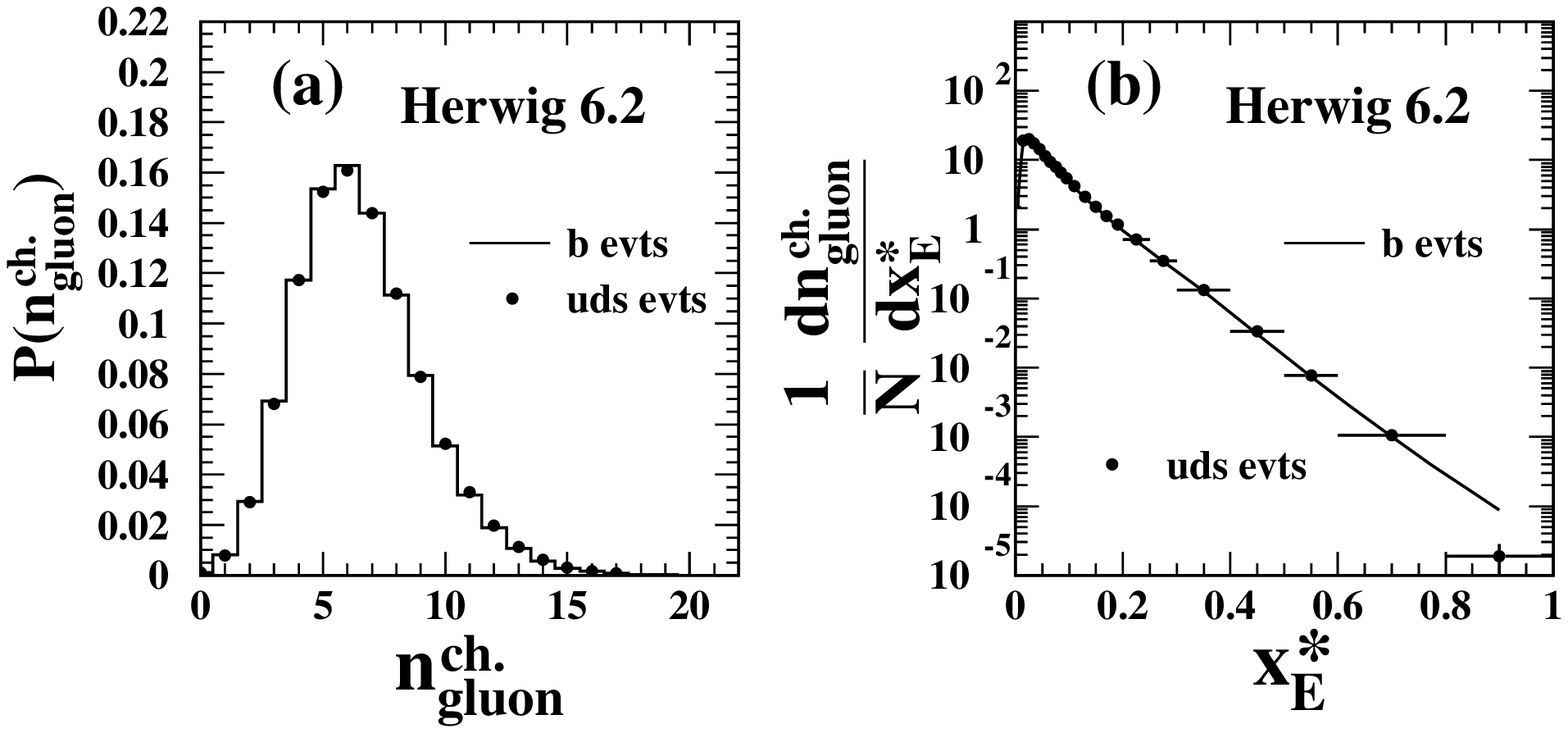}
 \end{center}
\caption{
Hadron level results from the Herwig Monte Carlo for
the (a)~$\nchgluon$ and (b)~$\fragfunc$ distributions,
for uds and b flavor events.
The results in (a)~are given for jet energies
5$\,\leq\,$$\egstar$$\,\leq\,$20~GeV,
corresponding to the range for which we find
the boost method to be applicable for the $\nchgluon$ distribution.
Analogously,
the results in (b)~are given for
13$\,\leq\,$$\egstar$$\,\leq\,$20~GeV,
corresponding to the more limited range for which
we find the 
boost method to be applicable for the $\fragfunc$ distribution.
}
\label{fig-btest}
\end{figure}

\begin{figure}[t]
 \begin{center}
      \epsfxsize=17cm
      \epsffile{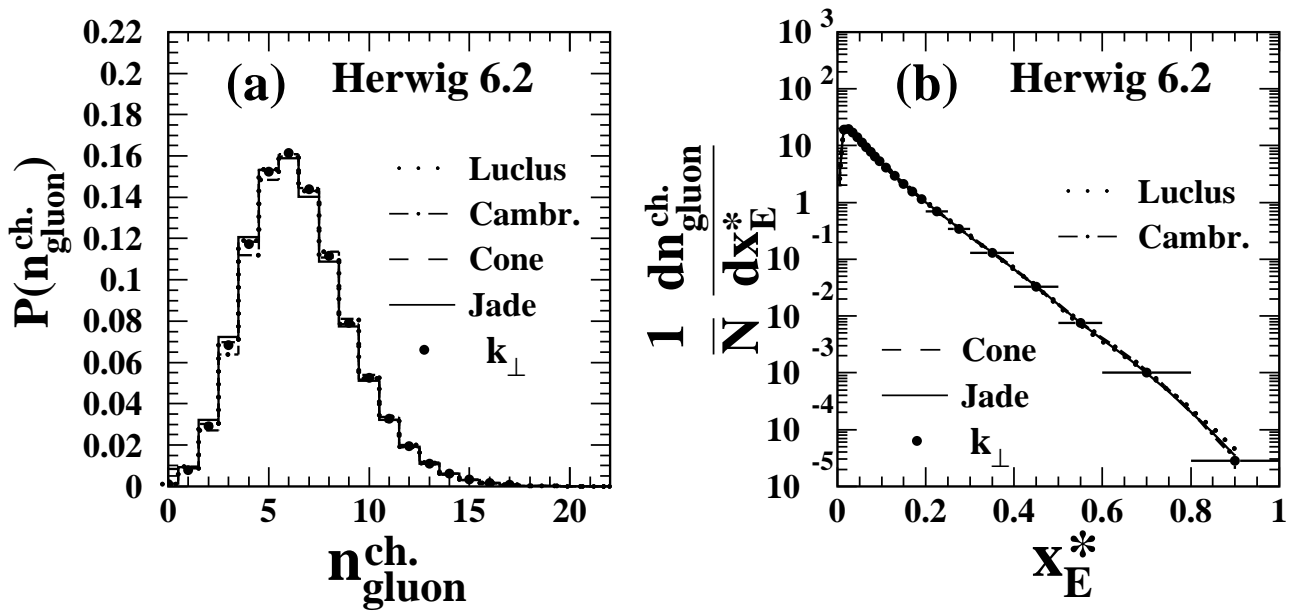}
 \end{center}
\caption{
Hadron level results from the Herwig Monte Carlo for
the (a)~$\nchgluon$ and (b)~$\fragfunc$ distributions,
for different choices of the jet finding algorithm
used for the initial definition of gluon jets.
The results in (a)~are given for jet energies
5$\,\leq\,$$\egstar$$\,\leq\,$20~GeV,
corresponding to the range for which we find
the boost method to be applicable for the $\nchgluon$ distribution.
Analogously,
the results in (b)~are given for
13$\,\leq\,$$\egstar$$\,\leq\,$20~GeV,
corresponding to the more limited range for which
we find the 
boost method to be applicable for the $\fragfunc$ distribution.
}
\label{fig-jetfinders}
\end{figure}

Fig.~\ref{fig-btest} shows the results we obtain 
from applying the boost algorithm to
uds and b flavor events from Herwig.
For simplicity,
the results for the $\nchgluon$ distribution 
(Fig.~\ref{fig-btest}a) include all jet energies,
5$\,\leq\,$$\egstar$$\,\leq\,$20~GeV.
The results for the $\fragfunc$ distribution
(Fig.~\ref{fig-btest}b) are restricted to the
two highest energy bins
(13$\,\leq\,$$\egstar$$\,\leq\,$20~GeV)
for the reason stated in the previous paragraph.
With the exception of the highest bin in 
Fig.~\ref{fig-btest}b ($x_E^*$$\,\geq\,$0.80),
it is seen that the uds and b events yield essentially
identical results for the gluon jet properties.
This establishes that our reliance on b events
to identify gluon jets (Sect.~\ref{sec-gluon})
does not introduce a significant bias,
i.e.~the theoretical assumption of massless jets is not
an important consideration for the quark jets.
We also tested the massless parton assumption of the boost algorithm
by repeating the comparisons of the gg and boost results
shown in Figs.~\ref{fig-nchtest}--\ref{fig-fftest}
after scaling the charged particle 3-momenta so that the magnitude
of a particle's 3-momentum equaled its energy,
and found that our conclusions were unchanged.

It is interesting to establish the degree to which
gluon jet properties
determined using the boost method are independent of the
jet algorithm chosen for the initial definition of the jets.
Fig.~\ref{fig-jetfinders}a shows the Herwig prediction for the 
$\nchgluon$ distribution,
for jets defined 
using the Luclus, Cambridge, cone and Jade~\cite{bib-jade} jet finders,
in addition to the {\durham} jet finder used for
our standard analysis.
%The results are based on Herwig Monte Carlo events at the hadron level.
%For simplicity,
%the results for all jet energies 
%(5$\,\leq\,$$\egstar$$\,\leq\,$20~GeV) are combined.
Note that the Jade algorithm uses the invariant mass between jets
as a resolution criterion.
The cone jet finder uses the total particle energy within a cone.
These two jet finders
--~unlike the other three~--
are therefore not based on the transverse momentum $\ptcut$
between jets and so do not correspond to the framework
of the dipole model or boost algorithm
(see Sect.~\ref{sec-boost}).
The five jet algorithms are seen to yield essentially identical results,
demonstrating the independence of the boost method from the
jet finder choice.
The fact that the cone and Jade jet finders yield essentially the same 
results as the 
three $\ptcut$ based algorithms demonstrates the robustness of the 
boost algorithm in this respect.
The corresponding results for the gluon jet fragmentation function 
are shown in Fig.~\ref{fig-jetfinders}b.
%In this case we restrict the study to
%jets in the two highest energy bins
%(13$\,\leq\,$$\egstar$$\,\leq\,$20~GeV)
%for the reason stated in the previous paragraph.
Again,
all five jet algorithms are seen to
yield essentially identical results.

\section{Correction procedure}

We correct the data to the hadron level (Sect.~\ref{sec-test})
and for gluon jet impurity.
This allows our data to be compared more directly to the
results of other studies and to theoretical calculations
(Sect.~\ref{sec-results}).

The multiplicity distributions are corrected in two steps.
In the first step,
the data are corrected for particle acceptance,
resolution, and secondary electromagnetic and hadronic interactions
using an unfolding matrix.
The matrix is constructed using detector level Monte Carlo events
(Sect.~\ref{sec-gluon})
subjected to the same analysis procedures as the data.
The matrix relates the value of $\nchgluon$ at
the detector level to the corresponding value
before the same event is processed by the detector simulation.
%Therefore the matrix corrects the data to the hadron level with
%the exception that initial-state radiation and the experimental
%event acceptance are included.
In the second step,
the data are corrected for event acceptance and selection,
initial-state radiation and gluon jet impurity
using bin-by-bin factors.
The factors are constructed by taking the ratio of
hadron to detector level Monte Carlo predictions.
The method of bin-by-bin corrections is
described in~\cite{bib-opal1990}.
The matrices and bin-by-bin factors are determined
using Herwig.
The matrices indicate that about 80\% of the events
exhibit a migration of one $\nchgluon$ bin or less
between the detector and hadron levels.
About 50\% of the events have the same value of $\nchgluon$
at the two levels.
The overall size of the corrections,
including the bin-by-bin factors,
varies from about 10 to~30\%.

The fragmentation functions are corrected 
using the bin-by-bin method, also based on Herwig.
A matrix procedure is not used for the fragmentation
functions because they include more than
one entry per event.
The typical size of the corrections is~15\%.

\section{Systematic uncertainties}
\label{sec-systematic}

To evaluate systematic uncertainties for the corrected data, 
we repeated the analysis with the changes given in the list below.
The differences between the standard results
and those found using each of these conditions
were used to define symmetric systematic uncertainties.
The systematic uncertainties were added in quadrature to define
the total systematic uncertainties.
The systematic uncertainty evaluated for each bin was
averaged with the results from its two neighbors to reduce the
effect of bin-to-bin fluctuations.
The single neighbor was used for bins at the ends of
the distributions.

The applied changes are:
\begin{enumerate}
  \item  The Ariadne Monte Carlo~\cite{bib-ariadne},
    version~4.11,
    and the Jetset Monte Carlo~\cite{bib-jetset}, version 7.4,
    were used to correct the data, 
    rather than Herwig.
    Samples of six million Ariadne and Jetset events at the
    detector level (Sect.~\ref{sec-gluon}) were used for this purpose.
    The parameter values used for these two models are given
    in~\cite{bib-opalrapgap} and~\cite{bib-qg95b}, respectively.
  \item  Charged tracks alone were used for the data and
    Monte Carlo samples with detector simulation,
    rather than charged tracks plus electromagnetic clusters
    (note: in the standard analysis,
    electromagnetic clusters are used in the definition
    of the jets, see Sect.~\ref{sec-gluon}).
  \item  The particle selection was further varied, 
    first by restricting charged tracks and electromagnetic clusters
    to the central region of the detector,
    $|\cos\theta|<0.70$,
    rather than $|\cos\theta|<0.96$ for the charged tracks and
    $|\cos\theta|<0.98$ for the clusters, 
    and second by increasing the minimum transverse momentum 
    of charged tracks with respect to the beam axis
    from 0.05~GeV/$c$ to 0.15~GeV/$c$.
  \item The quark jet tagging requirements were changed
    by requiring the decay length of the lower energy quark
    jet to satisfy $L/\sigma_L$$\,>\,$2.0
    for 5.0$\,\leq\,$$\egstar$$\,<\,$9.5~GeV,
    rather than $L/\sigma_L$$\,>\,$3.0,
    and at the same time by requiring the decay lengths of jets 1 and~2 to 
    satisfy $L/\sigma_L$$\,>\,$2.0 for 9.5$\,\leq\,$$\egstar$$\,<\,$16.0~GeV,
    again rather than $L/\sigma_L$$\,>\,$3.0.
    This resulted in $35\,607$~events
    with an estimated purity of $\mathrm 80.5\pm 0.1\,(stat.)$\%.
    As an additional check on the quark jet selection
    we increased the minimum $\qjet$ value of quark jets 
    (see eq.~(\ref{eq-scale}))
    from 8 to 10~GeV,
    with the $L/\sigma_{L}$ requirements at their standard values.
    This resulted in $23\,128$~events with
    an estimated purity of $\mathrm 85.7\pm 0.2\,(stat.)$\%.
\end{enumerate}
For the first item,
the largest of the described differences
with respect to the standard result
was assigned as the systematic uncertainty,
and similarly for the third and fourth items.

The largest contribution to the
systematic uncertainties generally arose
from using Ariadne or Jetset to correct the data.
The second largest contribution generally
arose from using charged particles alone or from 
restricting particles to $|\cos\theta|<0.70$.

Systematic uncertainties were also evaluated for the
gluon jet purities listed in Table~\ref{tab-purities}.
These uncertainties were derived by repeating the analysis
using each of the systematic variations given in the above list,
except for item~4 since this check is specifically
designed to alter the purities.
The results are given in Table~\ref{tab-purities}.
Similarly,
the systematic uncertainties listed in Table~\ref{tab-purities}
for the mean gluon jet energies $\langle\egstar\rangle$
were derived using the systematic variations in the above list,
except for item~1 
since data at the detector level do not depend on the Monte Carlo.

%Note we do not consider different choices of the jet finder
%(Cambridge, cone, etc.) 
%in our evaluation of systematic uncertainties.
%This is because all five jet finders in our study
%yield essentially identical results,
%see Fig.~\ref{fig-jetfinders}.

\section{Results}
\label{sec-results}

\subsection{Multiplicity distributions}
\label{sec-multiplicity}

The corrected multiplicity distributions are
shown by the solid points with uncertainties in Fig.~\ref{fig-nchtest}.
The vertical lines show the total uncertainties,
with statistical and systematic terms added in quadrature.
Statistical uncertainties were evaluated for the corrected data
using 50 independent samples of Monte Carlo events at
the hadron level,
each with about the same event statistics as the data
(this comment applies to all the corrected measurements
presented in this paper).
The statistical components of the uncertainties are
delimited by small horizontal lines
(for some points the statistical uncertainties
are too small to be visible).
These data are listed in 
Tables~\ref{tab-nchfirst}--\ref{tab-nchlast}.
The corresponding results for 
$\mnchgluon$, $\factwogluon$ and $\facthreegluon$ 
are presented in Fig.~\ref{fig-moments}
and Table~\ref{tab-meanfacs}.

\begin{figure}[p]
 \begin{center}
   \epsfxsize=12.9cm
   \epsffile{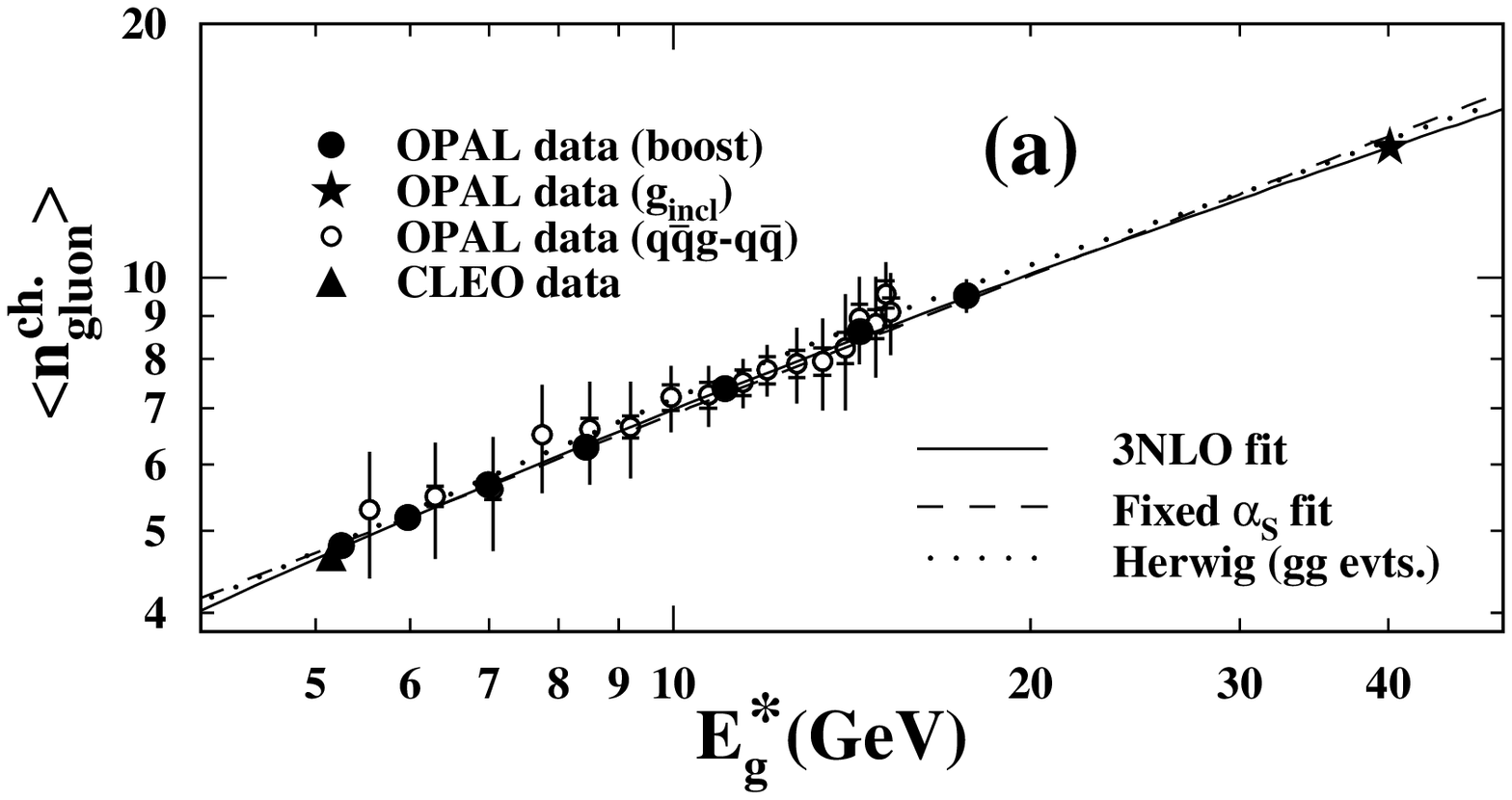} \\
   \epsfxsize=13.cm
   \epsffile{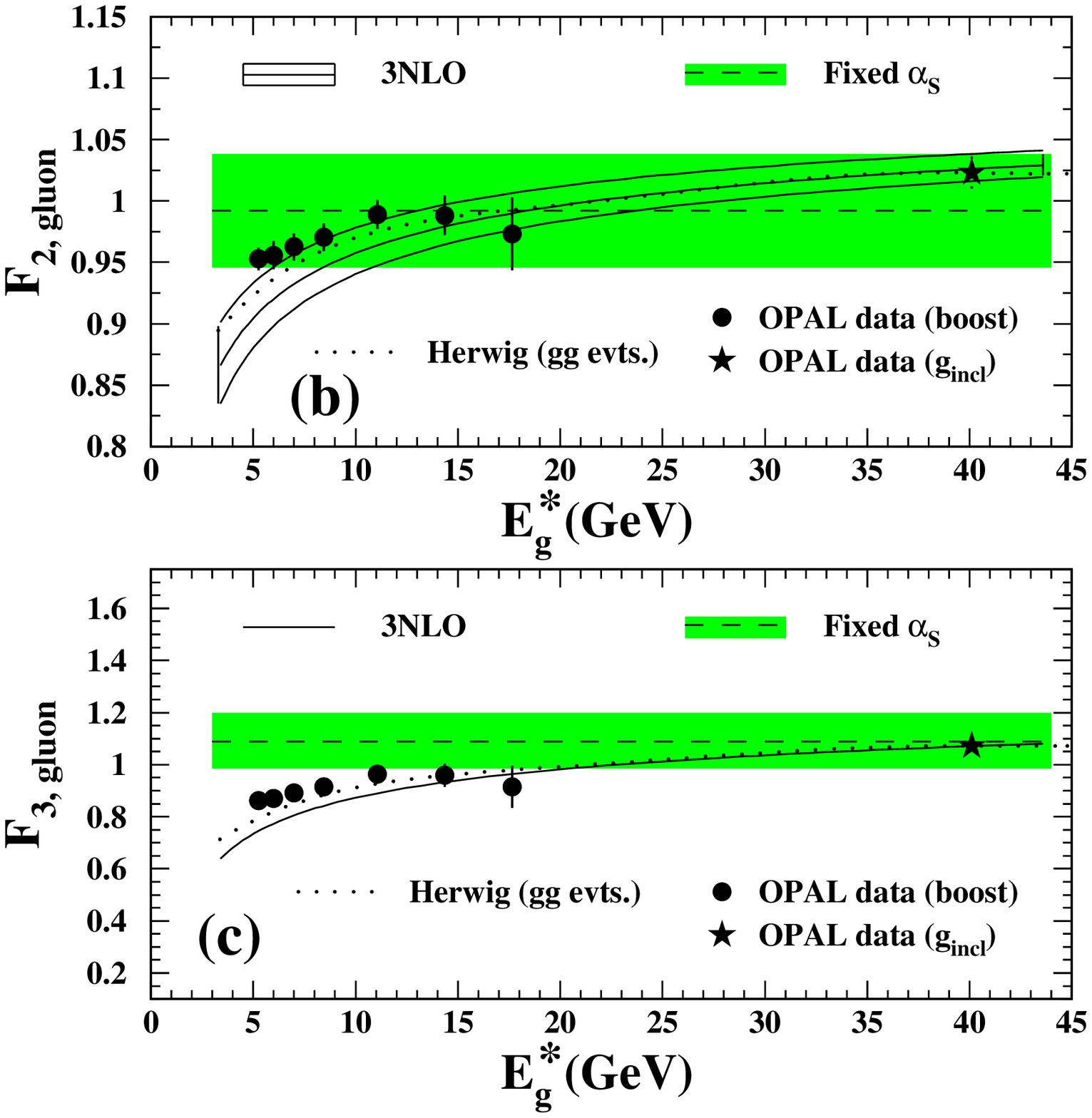} 
 \end{center}
\vspace*{-6mm}
\caption{
(a)~The mean charged particle multiplicity value of
gluon jets, $\mnchgluon$,
as a function of the gluon jet energy~$\egstar$.
The data have been corrected for detector acceptance
and resolution,
for event selection,
and for gluon jet impurity.
The total uncertainties are shown by the vertical lines, 
with the statistical component delimited by small horizontal lines.
(b,c)~The corresponding results for the two lowest 
non-trivial normalized factorial moments,
$\factwogluon$ and $\facthreegluon$.
The data are presented in comparison to the result of QCD analytic
calculations,
and to the Herwig Monte Carlo at the hadron level.
}
\label{fig-corrected-mnch}
\end{figure}

In Fig.~\ref{fig-corrected-mnch} we again present the
corrected results for $\mnchgluon$, $\factwogluon$ and $\facthreegluon$,
this time including our direct measurements
at $\egstar$$\,=\,$40.1~GeV~\cite{bib-opalhemisphere-2,bib-opalhemisphere-3}
based on 
{\epem}$\,\rightarrow\,$$\mathrm q\overline{q}\gincl$ events.
Fig.~\ref{fig-corrected-mnch}a also includes
a direct measurement of $\mnchgluon$
from the CLEO Collaboration at $\egstar$$\,=\,$5.2~GeV~\cite{bib-cleo92},
based on radiative $\Upsilon(3S)$ decays.
The open points in Fig.~\ref{fig-corrected-mnch}a
show our earlier results~\cite{bib-opaleden}
based on 
subtracting multiplicities in $\mathrm q\overline{q}$
and $\mathrm q\overline{q}g$ events~\cite{bib-eden99}.
The results from the present study are seen to be consistent
with these latter data,
and are considerably more precise.
Our results are also consistent with the CLEO measurement.

%There are no other previous measurements of 
%$\factwogluon$ and $\facthreegluon$ based on unbiased gluon jets.

\begin{table}[p]
\begin{center}
\scalebox{.94}{
\begin{tabular}{|c|c|c|c|}
  \hline
  $\nchgluon$ & $P(\nchgluon)$, $\egstar$$\,=\,$5.25~GeV &
    $P(\nchgluon)$, $\egstar$$\,=\,$5.98~GeV  &
    $P(\nchgluon)$, $\egstar$$\,=\,$6.98~GeV \\
  \hline
    0  & $0.0036 \pm 0.0009  \pm 0.0036$   
           & $0.0025\pm0.0005\pm0.0025$
               & $0.0019\pm 0.0004\pm 0.0019$ \\
    1  & $0.0266 \pm 0.0021  \pm 0.0068 $
           & $0.0190\pm0.0014\pm0.0040$
               & $0.0117\pm 0.0011\pm 0.0069$\\
    2  & $0.0808 \pm 0.0046  \pm 0.0093$       
           & $0.0618\pm0.0026\pm0.0048$
               & $0.0427\pm 0.0022\pm 0.0057$ \\
    3  & $0.1490 \pm 0.0067  \pm 0.0078$       
           & $0.1186\pm0.0046\pm0.0044$
               & $0.0943\pm 0.0046\pm 0.0073$ \\
    4  & $0.1982 \pm 0.0063  \pm 0.0058$       
           & $0.1794\pm0.0057\pm0.0029$ 
               & $0.1494\pm 0.0052\pm 0.0099$ \\
    5  & $0.199  \pm 0.006   \pm 0.012$        
           & $0.1935\pm0.0044\pm0.0050$
               & $0.1857\pm 0.0058\pm 0.0088$ \\
    6  & $0.154  \pm 0.006   \pm 0.010 $       
           & $0.1769\pm0.0040\pm0.0061$
               & $0.1803\pm 0.0050\pm 0.0078$ \\
    7  & $0.1000 \pm 0.0046  \pm 0.0059$       
           & $0.1236\pm0.0043\pm0.0042$
               & $0.1422\pm 0.0048\pm 0.0069$ \\
    8  & $0.0545 \pm 0.0029  \pm 0.0049$       
           & $0.0710\pm0.0029\pm0.0024$
               & $0.0961\pm 0.0044\pm 0.0064$ \\
    9  & $0.0230 \pm 0.0021  \pm 0.0012$       
           & $0.0330\pm0.0020\pm0.0019$
               & $0.0543\pm 0.0029\pm 0.0053$ \\
   10  & $0.0079 \pm 0.0012  \pm 0.0012$       
           & $0.0139\pm0.0011\pm0.0018$
               & $0.0256\pm 0.0020\pm 0.0035$ \\
   11  & $0.00232 \pm 0.00046  \pm 0.00071$    
           & $0.0050\pm0.0007\pm0.0014$
               & $0.0106\pm 0.0011\pm 0.0020$ \\
   12  & $0.00041 \pm 0.00028 \pm 0.00041$     
           & $0.00129\pm0.00038\pm0.00077$
               & $0.0039\pm 0.0007\pm 0.0010$ \\
   13  & ---    
           & $0.00034\pm0.00016\pm0.00031$
               & $0.00094\pm 0.00043\pm 0.00055$ \\
   14  & ---    
           & ---
               & $0.00017\pm 0.00017\pm 0.00017$ \\
   15  & ---    
           & ---
               & $0.00017\pm 0.00006\pm 0.00017$ \\
  \hline
\end{tabular}
} % scalebox
\end{center}
  \caption{The charged particle multiplicity distribution of gluon jets,
$\nchgluon$, for $\egstar$$\,=\,$5.25, 5.98 and 6.98~GeV.
The data have been corrected for detector acceptance and
resolution,
for event selection,
and for gluon jet impurity.
The first uncertainty is statistical and the 
second systematic.}
  \label{tab-nchfirst}
%\end{table}
%
%\begin{table}[t]
\begin{center}
\begin{tabular}{|c|c|c|}
  \hline
  $\nchgluon$ & $P(\nchgluon)$, $\egstar$$\,=\,$8.43~GeV &
    $P(\nchgluon)$, $\egstar$$\,=\,$10.92~GeV  \\
  \hline
    0  & $0.0012\pm0.0003\pm0.0012$     & $0.0008\pm0.0004\pm0.0008$ \\
    1  & $0.0069\pm0.0010\pm0.0066$     & $0.0047\pm0.0012\pm0.0047$ \\
    2  & $0.0280\pm0.0019\pm0.0075$     & $0.0152\pm0.0028\pm0.0047$  \\
    3  & $0.0646\pm0.0029\pm0.0078$     & $0.0373\pm0.0043\pm0.0052$ \\
    4  & $0.1157\pm0.0035\pm0.0075$     & $0.0749\pm0.0060\pm0.0062$ \\
    5  & $0.1621\pm0.0052\pm0.0085$     & $0.1083\pm0.0071\pm0.0063$ \\
    6  & $0.1783\pm0.0043\pm0.0080$     & $0.1419\pm0.0080\pm0.0082$   \\
    7  & $0.1632\pm0.0045\pm0.0086$     & $0.1530\pm0.0087\pm0.0061$  \\
    8  & $0.1207\pm0.0039\pm0.0073$     & $0.1428\pm0.0085\pm0.0063$  \\
    9  & $0.0756\pm0.0031\pm0.0064$     & $0.1173\pm0.0075\pm0.0066$  \\
   10  & $0.0443\pm0.0021\pm0.0053$     & $0.0862\pm0.0070\pm0.0053$  \\
   11  & $0.0226\pm0.0010\pm0.0031$     & $0.0550\pm0.0055\pm0.0033$  \\
   12  & $0.0103\pm0.0008\pm0.0020$     & $0.0320\pm0.0034\pm0.0032$  \\
   13  & $0.0044\pm0.0006\pm0.0012$     & $0.0172\pm0.0031\pm0.0030$  \\
   14  & $0.00141\pm0.00043\pm0.00071$  & $0.0074\pm0.0019\pm0.0016$  \\
   15  & $0.00048\pm0.00022\pm0.00032$  & $0.0042\pm0.0013\pm0.0011$  \\
   16  & $0.00019\pm0.00007\pm0.00015$  & $0.0011\pm0.0005\pm0.0011$  \\
   17 & ---                             & $0.0004\pm0.0004\pm0.0004$ \\
   18 & ---                             & $0.00018\pm0.00018\pm0.00018$ \\
  \hline
\end{tabular}
\end{center}
  \caption{The charged particle multiplicity distribution of gluon jets,
$\nchgluon$, for $\egstar$$\,=\,$8.43 and 10.92~GeV.
The data have been corrected for detector acceptance and
resolution,
for event selection,
and for gluon jet impurity.
The first uncertainty is statistical and the 
second systematic.}
  \label{tab-nchnext}
\end{table}

\begin{table}[p]
\begin{center}
\begin{tabular}{|c|c|c|}
  \hline
  $\nchgluon$ & $P(\nchgluon)$, $\egstar$$\,=\,$14.24~GeV &
    $P(\nchgluon)$, $\egstar$$\,=\,$17.72~GeV  \\
  \hline
    0  & $0.0001\pm0.0001\pm0.0001$ & $0.00004\pm0.00001\pm0.00004$  \\
    1  & $0.0005\pm0.0005\pm0.0005$ & $0.0004\pm0.0004\pm0.0004$ \\
    2  & $0.0068\pm0.0028\pm0.0036$ & $0.0013\pm0.0013\pm0.0013$  \\
    3  & $0.0179\pm0.0047\pm0.0062$ & $0.0080\pm0.0072\pm0.0057$ \\
    4  & $0.0363\pm0.0078\pm0.0079$ & $0.018\pm0.013\pm0.011$ \\
    5  & $0.072\pm0.011\pm0.008$    & $0.047\pm0.018\pm0.014$ \\
    6  & $0.099\pm0.013\pm0.006$    & $0.067\pm0.023\pm0.019$   \\
    7  & $0.131\pm0.016\pm0.010$    & $0.102\pm0.033\pm0.016$ \\
    8  & $0.137\pm0.014\pm0.010$    & $0.103\pm0.029\pm0.023$  \\
    9  & $0.132\pm0.013\pm0.007$    & $0.125\pm0.034\pm0.027$  \\
   10  & $0.116\pm0.016\pm0.006$    & $0.143\pm0.031\pm0.023$  \\
   11  & $0.0924\pm0.013\pm0.006$   & $0.128\pm0.029\pm0.024$  \\
   12  & $0.067\pm0.012\pm0.006$    & $0.153\pm0.031\pm0.051$  \\
   13  & $0.047\pm0.010\pm0.006$    & $0.054\pm0.020\pm0.046$  \\
   14  & $0.025\pm0.006\pm0.007$    & $0.022\pm0.018\pm0.022$  \\
   15  & $0.01447\pm0.0020\pm0.0040$& $0.017\pm0.013\pm0.017$  \\
   16  & $0.0051\pm0.0019\pm0.0039$ & $0.010\pm0.010\pm0.009$  \\
  \hline
\end{tabular}
\end{center}
  \caption{The charged particle multiplicity distribution of gluon jets,
$\nchgluon$, for $\egstar$$\,=\,$14.24 and 17.72~GeV.
The data have been corrected for detector acceptance and
resolution,
for event selection,
and for gluon jet impurity.
The first uncertainty is statistical and the 
second systematic.}
  \label{tab-nchlast}
%\end{table}
%
%\begin{table}[t]
\begin{center}
\begin{tabular}{|c|ccc|}
  \hline
   $\egstar$ & $\mnchgluon$ & $\factwogluon$ & $\facthreegluon$ \\
  \hline
    5.25  & $4.803\pm0.030\pm0.047$ & $0.9528\pm0.0030\pm0.0087$ & $0.863\pm0.008\pm0.020$ \\
    5.98  & $5.190\pm0.030\pm0.062$ & $0.956\pm0.002\pm0.011$ & $0.870\pm0.006\pm0.027$ \\
    6.98  & $5.677\pm0.030\pm0.074$ & $0.963\pm0.002\pm0.011$ & $0.891\pm0.006\pm0.026$ \\
    8.43  & $6.291\pm0.030\pm0.090$ & $0.970\pm0.002\pm0.011$ & $0.915\pm0.005\pm0.027$ \\
   10.92  & $7.378\pm0.062\pm0.077$ & $0.989\pm0.004\pm0.011$ & $0.964\pm0.012\pm0.026$ \\
   14.24  & $8.62\pm0.13\pm0.10$ & $0.988\pm0.005\pm0.015$ & $0.960\pm0.016\pm0.041$ \\
   17.72  & $9.52\pm0.30\pm0.33$ & $0.973\pm0.007\pm0.029$ & $0.914\pm0.021\pm0.078$ \\
  \hline
\end{tabular}
\end{center}
  \caption{The mean, $\mnchgluon$,
and first two non-trivial normalized factorial moments,
$\factwogluon$ and $\facthreegluon$,
of the charged particle multiplicity distribution of gluon jets.
The data have been corrected for detector acceptance and resolution,
for event selection,
and for gluon jet impurity.
The first uncertainty is statistical and the 
second systematic.}
  \label{tab-meanfacs}
\end{table}

%From Fig.~\ref{{fig-nchtest} it is seen that the data
%are, in general,
%well described by the Herwig Monte Carlo,
%both using the boost method and gg event hemispheres.

From Fig.~\ref{fig-moments}a
(or Fig.~\ref{fig-corrected-mnch}a)
it is seen that the
energy evolution of $\mnchgluon$ is well described by Herwig.
The Herwig predictions for the higher moments
$\factwogluon$ and $\facthreegluon$ 
are also in reasonable agreement with the data,
as seen from Figs.~\ref{fig-moments}b and~c
(or Figs.~\ref{fig-corrected-mnch}b and~c),
although the Monte Carlo curves lie somewhat below the
measurements for jet energies smaller than about 12~GeV.

In the following,
we present fits of QCD expressions to the 
$\mnchgluon$, $\factwogluon$ and $\facthreegluon$ data.
The theoretical expressions are at the ``parton level.''
The corresponding distributions are denoted
$\mnparton$, $\factwoparton$ and $\facthreeparton$.
The parton level is based on quarks and gluons 
present at the end of the perturbative shower.
The theoretical results are compared to the charged
particle hadron level data\footnote{The issue of 
how this comparison differs from one
based on both charged and neutral particles at the hadron level
is addressed in Sect.~\ref{sec-3nlo}.},
without hadronization corrections.
By hadronization correction,
we mean the ratio of the parton to hadron level 
predictions from a QCD Monte Carlo program, e.g.~Herwig.
We do not apply hadronization corrections
because they are model dependent.
%to introduce as little Monte Carlo dependence into our
%results as possible.
The fitted parameters 
%we obtain from the theoretical expressions
therefore incorporate effects from hadronization,
in addition to possible effects from approximations in the
QCD expressions themselves.
Our strategy is to compare
the parameter values obtained from different distributions
(where appropriate) to see whether they are generally similar
despite hadronization,
and thereby to test the global consistency of the formalism
in a qualitative way.

\subsubsection{3NLO perturbative expressions}
\label{sec-3nlo}

A QCD analytic calculation of the energy evolution of
$\mnparton$,
valid to the next-to-next-to-next-to-leading order (3NLO)
of perturbation theory,
is presented in~\cite{bib-dremingary}.
So far,
only two tests of this expression have been performed.
The first test~\cite{bib-dremingary}
is based on two data points only:
the $\gincl$ and $\Upsilon(3S)$-derived
results shown in Fig.~\ref{fig-corrected-mnch}a.
The second test~\cite{bib-opaleden}
is based on these same two data points and
the less direct measurements
shown by the open symbols in Fig.~\ref{fig-corrected-mnch}a.
3NLO analytic results for 
$\factwoparton$ and $\facthreeparton$ are presented in~\cite{bib-f2f3-3nlo}.
So far,
there have been no experimental tests of 
the energy evolution of these expressions.

The solid curve in Fig.~\ref{fig-corrected-mnch}a
shows the result of a two parameter $\chi^2$
fit of the 3NLO expression for $\mnparton$
to the $\mnchgluon$ data.
The fit is performed assuming $\nff$$\,=\,$5,
where $\nff$ is the number of active quark
flavors in the perturbative stage of an event.
Essentially identical curves are obtained if $\nff$$\,=\,$3
or 4 (see below) is used instead.
The fitted data are the seven measurements of 
$\mnchgluon$ from the present study
(see Table~\ref{tab-meanfacs})
and the $\gincl$ and $\Upsilon(3S)$
results shown in Fig.~\ref{fig-corrected-mnch}a.
The fits are performed using statistical uncertainties
only to determine the~$\chi^2$.
The fitted parameters are the QCD scale parameter $\Lambda$
and an overall normalization constant~$K$
(see~\cite{bib-dremingary}).
Note that $\Lambda$ is strongly correlated 
with $\mathrm\Lambda_{\overline{MS}}$~\cite{bib-lambda-msbar}
but is not necessarily the same.
Note also that there is an ambiguity in the appropriate value to
use for $\nff$ because c and b quarks are rarely produced
in the perturbative evolution of jets at LEP.
The fitted parameter values and corresponding $\chi^2$ results
are listed in the top portion of Table~\ref{tab-fits}.
The results are given for $\nff$$\,=\,$3, 4 and~5.
%Since the 3NLO expression is fitted to the hadron level data,
%these parameters incorporate effects of hadronization.
The systematic uncertainties attributed to 
the parameters are 
defined by adding the following contributions in quadrature:
(1)~the uncertainty of the fitted parameters returned by the
fitting routine when the total uncertainties of the data are used
to perform the fit,
rather than the statistical uncertainties only
(note: point-to-point systematic uncertainties
are treated as uncorrelated);
(2)~the difference between the results of the standard fit
and those found by fitting only the $\mnchgluon$ data 
of Table~\ref{tab-meanfacs}
(i.e.~excluding the $\gincl$ and $\Upsilon(3S)$ measurements).
%We note that fits of the 3NLO expression to the data using either 
%of these systematic variations yield curves which are 
%almost indistinguishable from the solid curve
%in Fig.~\ref{fig-corrected-mnch}a.

%The width of the open band centered on the 3NLO curve in
%Fig.~\ref{fig-corrected-mnch}a (barely visible)
%indicates the effects of increasing or decreasing by 
%one standard deviation of the total 
%uncertainty of the 3NLO result,
%defined by adding the systematic and statistical terms in quadrature.
%The systematic terms are defined by the difference between the
%curve found using the standard fit and those found using
%the two systematic variations listed above.

\begin{table}[t]
 \begin{center}
  \begin{tabular}{|cc|ccc|}
  \hline
& $\nff$ & $\Lambda$ (GeV) & $K$ & $\chi^2$/d.o.f. \\
  \hline
& 3 & $0.470\pm0.027\pm0.050$ & $0.1366\pm0.0047\pm0.0084$ & 6.2/7 \\
$\mnchgluon $
& 4 & $0.385\pm0.024\pm0.046$ & $0.1164\pm0.0042\pm0.0080$ & 5.6/7 \\
& 5 & $0.296\pm0.019\pm0.038$ & $0.0986\pm0.0037\pm0.0073$ & 5.2/7 \\
  \hline
          & 3 & $0.166\pm0.012\pm0.049$  & --- & 8.7/2 \\
$\factwogluon$ 
          & 4 & $0.143\pm0.009\pm0.042$  & --- & 8.2/2 \\
          & 5 & $0.114\pm0.009\pm0.032$  & --- & 8.2/2 \\
  \hline
            & 3 & $0.051\pm0.004\pm0.016$  & --- & 8.4/2 \\
$\facthreegluon$ 
            & 4 & $0.040\pm0.003\pm0.013$  & --- & 8.4/2 \\
            & 5 & $0.029\pm0.002\pm0.011$  & --- & 8.1/2 \\
  \hline
  \end{tabular}
\caption{Results of fits of the 3NLO expressions
for $\mnchgluon$~\cite{bib-dremingary},
$\factwogluon$ and~$\facthreegluon$~\cite{bib-f2f3-3nlo}
to our data.
The $\chi^2$ values are based on the statistical 
uncertainties of the data points.
The first uncertainty is statistical and
the second systematic.
}
\label{tab-fits}
 \end{center}
\end{table}

From Fig.~\ref{fig-corrected-mnch}a and Table~\ref{tab-fits},
it is seen that the 3NLO expression
provides a good description of the $\mnchgluon$ measurements,
i.e.~$\chi^2$/d.o.f.$\,=\,$0.74 for $\nff$$\,=\,$5,
with slightly higher $\chi^2$ for $\nff$$\,=\,$3 and~4.
The result 
$\Lambda$$\,=\,$$0.296\pm 0.037\,$(stat.+syst.)~GeV
we find
%\footnote{Note: ``stat.+syst.'' means the statistical and systematic
%terms have been added in quadrature.}
for $\nff$$\,=\,$5
is much more similar to the corresponding quark jet result,
$\Lambda$$\,=\,$$0.190\pm 0.032\,$(stat.)~GeV~\cite{bib-opaleden},
than to the value 
\mbox{$\Lambda$$\,=\,$$0.60\pm0.06\,$(stat.)~GeV}
found previously~\cite{bib-opaleden}
%(see also~\cite{bib-dremingary})
(``stat.+syst.'' means the statistical and systematic
terms have been added in quadrature).
Our data therefore provide a much improved demonstration
of the consistency
of the 3NLO expressions for the scale dependence 
of unbiased quark and gluon jet multiplicities
than previously available.

%given that the values of $\Lambda$ in the expressions for
%the quark and gluon jets are expected to be the same
%(see, for example~\cite{bib-dremingary}).
%At the scale of the Z$^0$ mass,
%the results $\Lambda$$\,=\,$$0.296\pm 0.037$ (gluons)
%and $\Lambda$$\,=\,$$0.190\pm0.032$ (quarks)
%correspond to 
%$\alpha_S$$\,=\,$$0.123\pm0.002$
%and $0.115\pm0.003$,
%respectively,
%which agree to within about~7\%.\footnote{We relate $\Lambda$
%to $\alpha_S\,(\mzee)$ using the two-loop formula given,
%for example, by (75) in~\cite{bib-dremingary-physrep}.}
%Using the previously available results for $\Lambda$ from 
%unbiased gluon jets~\cite{bib-opaleden,bib-dremingary},
%mentioned above,
%the level of this agreement is only about~21\%.
%Since $\Lambda$ in the 3NLO calculations does not
%correspond to $\mathrm\Lambda_{\overline{MS}}$,
%note that these values of $\alpha_S\,(\mzee)$ cannot be compared directly
%to the world average value in~\cite{bib-pdg}.
%Nonetheless,
%the results found in the fits of the gluon and quark jet data
%are consistent with the expectation $\alpha_S\,(\mzee)$$\,\approx\,$0.10.

The solid curves in Figs.~\ref{fig-corrected-mnch}b and~c
show the corresponding results of fits of the 
3NLO expressions for $\factwoparton$ and $\facthreeparton$ to the data.
We note that the hadronization corrections predicted for
$\factwogluon$ and $\facthreegluon$ (from Herwig)
exhibit a significant dependence on energy,
especially for $\egstar$$\,\ltsim\,$12~GeV.
The hadronization correction predicted for $\factwogluon$ changes
by about 19\% 
for 5$\,\leq\,$$\egstar$$\,\leq\,$12~GeV,
for example (from 0.70 to 0.83),
compared to about 12\%
for 12$\,\leq\,$$\egstar$$\,\leq\,$40~GeV
(from 0.83 to 0.93).
For $\facthreegluon$,
the results are 71\% (from 0.34 to 0.58) 
and 43\% (from 0.58 to 0.83), respectively.
(In comparison,
the hadronization correction predicted for the $\mnchgluon$
distribution in Fig.~\ref{fig-corrected-mnch}a changes by
only about 10\% and 6\% over these intervals,
corresponding to corrections of 0.39, 0.35 and 0.33 at
5, 12 and 40~GeV.)
Therefore,
the fits of the 3NLO expressions 
for $\factwoparton$ and $\facthreeparton$ 
shown in Figs.~\ref{fig-corrected-mnch}b and~c
are restricted to the three data points
with $\egstar$$\,>\,$12~GeV,
i.e.~the data at 14.24, 17.72 and 40.1~GeV.
The results of the fits are listed in the central and
bottom portions of Table~\ref{tab-fits}.
Since $\factwoparton$ and $\facthreeparton$
are normalized moments,
they are independent of an overall normalization factor,
i.e.~$\Lambda$ is the only free parameter.
The $\chi^2$/d.o.f. of these fits 
are seen to be quite large (Table~\ref{tab-fits}):
this is because the statistical uncertainties are relatively
small and the $\factwogluon$ and $\facthreegluon$
measurements at 17.72~GeV
are low compared to the corresponding
data at 14.24~GeV and 40.1~GeV
(see Figs.~\ref{fig-corrected-mnch}b and~c).
Note, however,
that the $\factwogluon$ and $\facthreegluon$
measurements at 17.72~GeV have large systematic uncertainties
and that the fitted curves in 
Figs.~\ref{fig-corrected-mnch}b and~c
describe the energy evolution of $\factwogluon$ and $\facthreegluon$
from 14 to 40~GeV quite well if the total uncertainties
of the measurements are considered.
In contrast,
the 3NLO curves lie below the data at smaller energies.

The systematic uncertainties attributed to the $\Lambda$
values found from fitting the $\factwogluon$ and $\facthreegluon$ data
(Table~\ref{tab-fits})
are defined by adding the following contributions in quadrature:
(1)~the uncertainty of the fitted parameters returned by the
fitting routine when the total uncertainties of the data are used
to perform the fit,
rather than the statistical uncertainties only;
(2)~the difference between the standard results and those found by
repeating the fits including the data at 10.92~GeV;
(3)~the difference between the standard results and those found by
repeating the fits excluding the $\gincl$ measurements at 40.1~GeV.
The open band in Fig.~\ref{fig-corrected-mnch}b shows the
uncertainty of the 3NLO curve,
defined by increasing or decreasing $\Lambda$
by one standard deviation of its total uncertainty
as determined using the $\factwogluon$ data.
The corresponding one standard deviation band for the
$\facthreegluon$ curve is too small to be visible.

The fitted results for $\Lambda$ from $\factwogluon$ and $\facthreegluon$,
viz.~0.114 and 0.029~GeV (for $\nff$$\,=\,$5),
differ from each other and also from the result
$\Lambda$$\,=\,$0.296~GeV found from the fit to the $\mnchgluon$ data
(Table~\ref{tab-fits}).
These differences may be a consequence of the different 
energy dependence of the hadronization
corrections for the three distributions,
predicted to be more substantial for $\facthreegluon$
than for $\factwogluon$,
and for $\factwogluon$ than for $\mnchgluon$,
as discussed above.
For purposes of comparison,
it is interesting to express these $\Lambda$ results
in terms of the coupling strength at the Z$^0$ pole,
$\alpha_S\,(\mzee)$.\footnote{We relate $\Lambda$
to $\alpha_S\,(\mzee)$ using the two-loop formula given,
for example, by (75) in~\cite{bib-dremingary-physrep}.}
The fitted $\Lambda$ results for 
$\mnchgluon$, $\factwogluon$ and $\facthreegluon$ correspond to
$\alpha_S\,(\mzee)$$\,=\,$$0.123\pm0.002$,
$0.107\pm0.003$ and $0.090\pm0.002$, respectively,
where the statistical and systematic uncertainties
have been added in quadrature.
Since $\Lambda$ in the 3NLO calculations does not
correspond to $\mathrm\Lambda_{\overline{MS}}$,
these values of $\alpha_S\,(\mzee)$ cannot be compared directly
to the world average
$\alpha_S\,(\mzee)$$\,=\,$$0.117\pm0.002$~\cite{bib-pdg}.
Furthermore they cannot be compared directly to each other
since the effects of hadronization are different for the
$\mnchgluon$, $\factwogluon$ and $\facthreegluon$ distributions
as noted above.
Nonetheless,
the three $\alpha_S$ results
are globally similar to each other and to the world average value,
i.e.~they are more similar to $\alpha_S$$\,\sim\,$0.1
than to e.g. $\alpha_S$$\,\sim\,$0.01 or~1.0.
It is notable that the 
3NLO results for $\factwoparton$ and $\facthreeparton$ found
using these qualitatively sensible values ($\alpha_S$$\,\sim\,$0.1)
are much more similar to the experimental measurements
in Figs.~\ref{fig-corrected-mnch}b and~c
than to the leading order QCD predictions
of 4/3 and 9/4, respectively~\cite{bib-f2f3-3nlo}.
In this general sense, 
the 3NLO calculations provide a qualitatively
consistent and successful description of 
the gluon jet multiplicity data,
at least for $\egstar$$\,\gtsim\,$14~GeV.

%In this sense,
%the 3NLO calculations provide a qualitatively
%successful description of the gluon jet multiplicity data,
%at least for $\egstar$$\,\gtsim\,$14~GeV.

%Also, since the distributions incorporate the effect of hadronization
%which are different for each distribution as noted above.
%gluon jet multiplicity distribution
%are generally successful, 
%at least for $\egstar$$\,\gtsim\,$14~GeV.
%The central results of
%are derived assuming $\nff$$\,=\,$5,
%with $\Lambda$$\,=\,$0.311~GeV taken from 
%the fit to the $\mnchgluon$ data,
%see Table~\ref{tab-fits} and the discussion above.
%The widths of the bands correspond to the maximum 
%deviations with respect to the standard result if
%$\nff$$\,=\,$4 with $\Lambda$$\,=\,$0.401~GeV,
%or $\nff$$\,=\,$3 with $\Lambda$$\,=\,$0.489~GeV,
%are used instead (Table~\ref{tab-fits}).

We note that most of the multiplicity in high energy jets 
is generated by hard, virtual gluons,
common to both charged and neutral particles
at the hadron level.
As a consequence,
the shapes of the multiplicity distributions
of neutral and charged hadrons are expected to be very similar,
so that it makes no difference if parton level expressions
are compared to charged particle data only (as is done here)
or to data including neutral hadrons as well.
The shapes of the multiplicity distributions of charged
and neutral particles at the hadron level can differ, however, 
because of resonance decays which introduce correlations,
e.g.\ in~$\pi^0$$\,\rightarrow\,$$\gamma\gamma$
decays which produce most of the stable neutral particles at
the hadron level.
Using hadron level Herwig events,
we verified that the fitted results for $\Lambda$
from the $\mngluon$, $\factwogluon$ and $\facthreegluon$ distributions
are almost identical if neutral particles at the hadron level
are used to define the multiplicity distributions,
rather than charged particles,
as long as the $\pi^0$ is declared stable.
The results for the normalization constant $K$ 
in the 3NLO expression for $\mnparton$ differ in the fits
for neutral and charged hadrons,
however,
because the mean numbers of charged and neutral hadrons
are not the same.

\subsubsection{Fixed {\boldmath$\alpha_S$} expressions}
\label{sec-fixed}

Analytic expressions for $\mnparton$, 
$\factwoparton$ and $\facthreeparton$
have also been derived assuming a fixed value of 
$\alpha_S$~\cite{bib-dremin-hwa-n,bib-dremin-hwa-f2f3}.
By assuming $\alpha_S$ is fixed,
the QCD evolution equations for multiplicity can be solved exactly,
without recourse to a perturbative approximation
(for a recent review, see~\cite{bib-dremingary-physrep}).
The solutions based on fixed $\alpha_S$ therefore more completely
incorporate such higher order effects as energy conservation
than do the 3NLO calculations.
On the other hand,
the fixed $\alpha_S$ results do not account 
for the change in $\alpha_S$ with scale.

The dashed curve in Fig.~\ref{fig-corrected-mnch}a
shows the result of a fit of the fixed $\alpha_S$ expression
for gluon jet multiplicity~\cite{bib-dremin-hwa-n},
\begin{equation}
  \mnparton = \left(\frac{\egstar}{Q_0}\right)^{\gamma} \;\;\;\; ,
\end{equation}
to the $\mnchgluon$ data.
The fitted data are the seven measurements of $\mnchgluon$
in Table~\ref{tab-meanfacs}
and the $\gincl$ and $\Upsilon(3S)$
results in Fig.~\ref{fig-corrected-mnch}a.
The fitted parameters are $\gamma$ and~$Q_0$.
$Q_0$ is a cutoff for soft gluon radiation
while $\gamma$ is the so-called
anomalous dimension of QCD,
which takes into account
perturbative corrections to the coupling strength.
The results are 
$\gamma$$\,=\,$$\mathrm 0.548\pm 0.009\, (stat.)\pm 0.028 \,(syst,)$
and
$Q_0$$\,=\,$$\mathrm 0.295\pm 0.017\, (stat.)\pm 0.053 \,(syst.)$~GeV,
where the systematic uncertainties are evaluated as explained for
the 3NLO fit to $\mnchgluon$ in Sect.~\ref{sec-3nlo}.
The $\chi^2/$d.o.f.\ is 21/7,
larger than the result found using
the 3NLO expression (see Table~\ref{tab-fits}).
The fixed $\alpha_S$ calculation provides a reasonable description
of the data within the total uncertainties of the measurements, however.

%If the total uncertainties of the data are used to perform the fit,
%rather than the statistical uncertainties only,
%the $\chi^2/$d.o.f.\ for the fixed $\alpha_S$ solution is 9.7/7,
%demonstrating its general consistency
%with the measurements.

Assuming a specific value for $\nff$,
i.e.~$\nff$$\,=\,$3, 4 or~5,
our result for $\gamma$ can be used to derive
values for $\alpha_S$ and $\gqratio$,
where $\gqratio$ is the ratio between the mean particle
multiplicities of gluon and quark jets:
\begin{equation}
  \gqratio = \frac{\langle n_{\mathrm gluon} \rangle }{\langle n_{\mathrm quark} \rangle}
    \;\;\;\; ,
  \label{eq-gqratio}
\end{equation}
(see e.g.~(120) and~(121) in~\cite{bib-dremingary-physrep}
and the ensuing text).
The results
%values of $\alpha_S$ and $\gqratio$
are given in Table~\ref{tab-fixed-alphas}.
%These results
%are seen to be essentially independent of~$\nff$.
Note that since $\alpha_S$ is constant in this formalism, 
as is $\gqratio$,
there is an ambiguity in the energy scale
of these results.
This ambiguity may partly explain the large value
$\alpha_S$$\,\sim\,$0.3 we obtain for the coupling strength.
In addition,
the assumption that $\alpha_S$ is constant
is not entirely realistic for the energy range of our study.
For these reasons,
the results for $\alpha_S$ in Table~\ref{tab-fixed-alphas}
are not very meaningful.
They are included for completeness only.
In contrast,
the results for $\gqratio$ are found to be only weakly
dependent on the energy scale 
and on the corresponding variation in $\alpha_S$~\cite{bib-dremin-hwa-n}
and thus have more significance.
For the multiplicity ratio,
we obtain $\gqratio$$\,\approx\,$1.7.
This result is discussed further in Sect.~\ref{sec-ratios}.

\begin{table}[t]
 \begin{center}
  \begin{tabular}{|c|cc|}
  \hline
$\nff$ & $\alpha_S$ & $\gqratio$ \\
  \hline
3 & $0.293\pm0.016\pm0.035$ & $1.718\pm0.014\pm0.040$  \\
4 & $0.297\pm0.017\pm0.036$ & $1.697\pm0.014\pm0.041$  \\
5 & $0.301\pm0.017\pm0.036$ & $1.679\pm0.014\pm0.042$  \\
  \hline
  \end{tabular}
\caption{Results for $\alpha_S$ and $\gqratio$
from fitting the fixed $\alpha_S$ expression
for $\mnchgluon$~\cite{bib-dremin-hwa-n}
to our data.
Note that the energy scale associated with these
results is ambiguous (see text).
The first uncertainty is statistical and
the second systematic.
}
\label{tab-fixed-alphas}
 \end{center}
\end{table}

The fixed $\alpha_S$ expressions for 
$\factwogluon$ and $\facthreegluon$~\cite{bib-dremin-hwa-f2f3}
depend on $\nff$ and $\gamma$.
Because these expressions are complicated,
we do not fit them to data
but instead evaluate them using the result
for $\gamma$ found by fitting the $\mnchgluon$ measurements
(cf.~the dashed curve Fig.~\ref{fig-corrected-mnch}a).
The results,
evaluated for $\nff$$\,=\,$5,
are shown by the dashed lines in 
Figs.~\ref{fig-corrected-mnch}b and~c.
Almost identical results are obtained for
$\nff$$\,=\,$3 or~4.
The shaded regions indicate the total uncertainties,
defined by repeating the study 
after increasing or decreasing $\gamma$
by its total uncertainty (see above).
The fixed $\alpha_S$ prediction for $\factwogluon$ 
(Fig.~\ref{fig-corrected-mnch}b)
is seen to accommodate the data
within its fairly large uncertainty.
The corresponding result for $\facthreegluon$ 
(Fig.~\ref{fig-corrected-mnch}c)
agrees with the data point at 40.1~GeV but lies
above the measurements at lower energies.
%The overall descriptions of $\factwogluon$ and $\facthreegluon$ by the
%fixed $\alpha_S$ calculation are quite reasonable
%considering that these descriptions are not based on fits.

\subsubsection{Comparison to quark jets}
\label{sec-ratios}

It is interesting to compare the results
of Table~\ref{tab-meanfacs}
to corresponding measurements for quark jets.
This allows further tests of QCD calculations.

The particle multiplicity of
unbiased quark jets has been measured at many scales.
For our study,
we choose results from the ARGUS~\cite{bib-argus},
JADE~\cite{bib-jade-nch} and HRS~\cite{bib-hrs}
experiments at c.m.\ energies of 10.5, 12.0 and 29.0~GeV,
and from the TASSO~\cite{bib-tasso} experiment
at 14.0, 22.0 and 34.5~GeV.
We select these data because the quark jet energies,
given by half the c.m.\ values,
correspond to the mean energies $\langle\egstar\rangle$
of our gluon jets,
with the exception of the sample with 
$\langle\egstar\rangle$$\,=\,$8.43~GeV
(see Table~\ref{tab-purities}).

\begin{figure}[p]
 \begin{center}
   \epsfxsize=14.2cm
   \epsffile{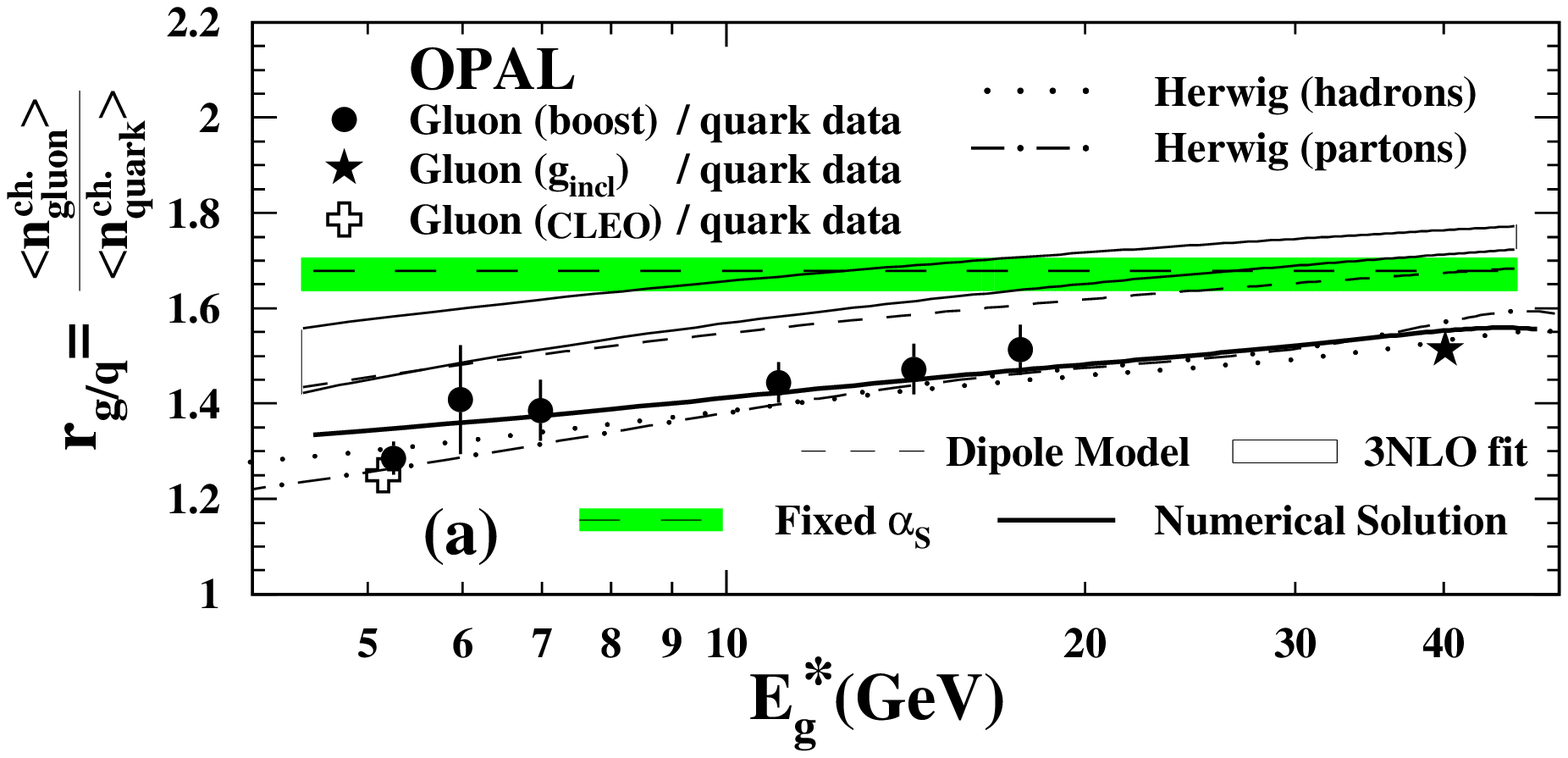} \\
   \epsfxsize=14.2cm
   \epsffile{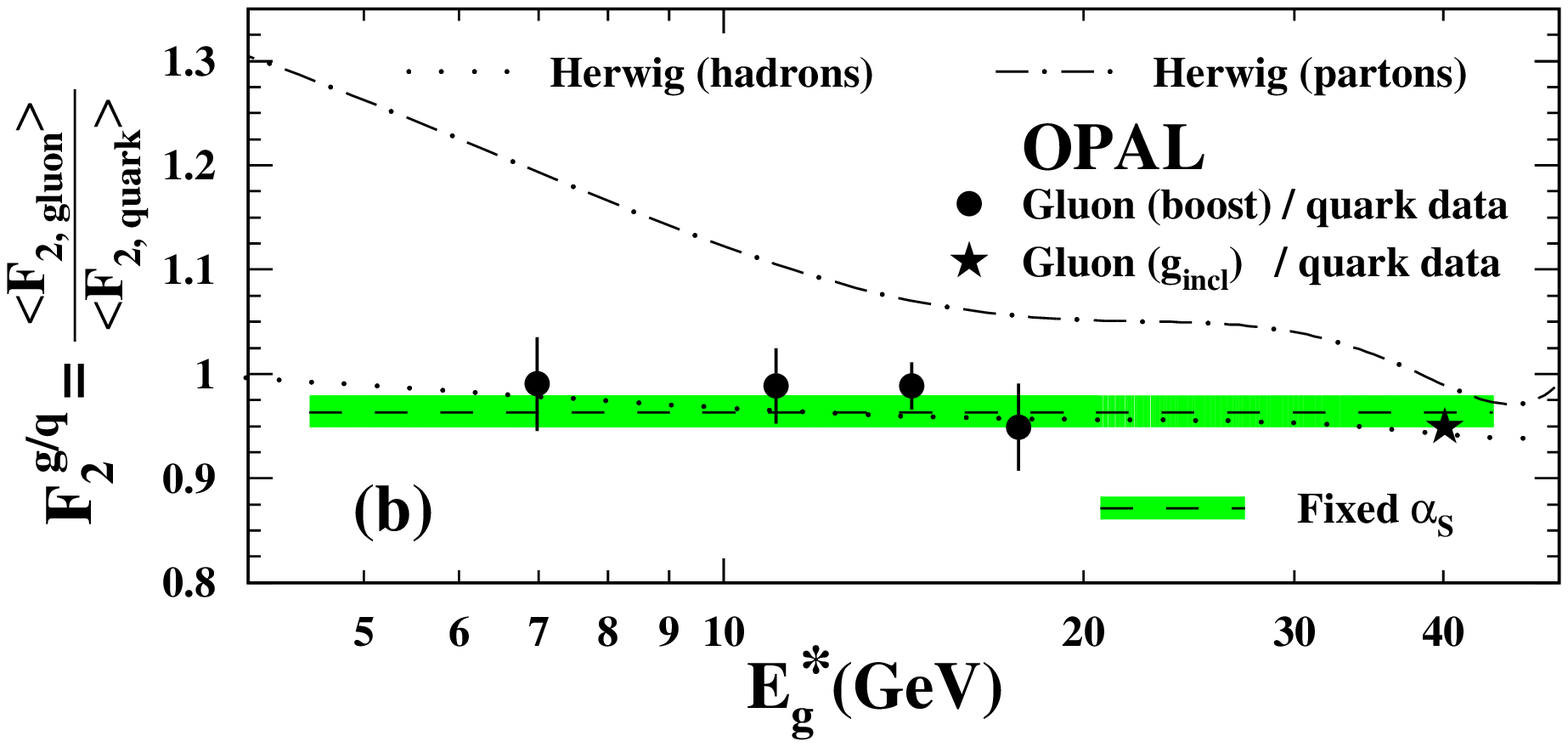} \\
   \epsfxsize=14.2cm
   \epsffile{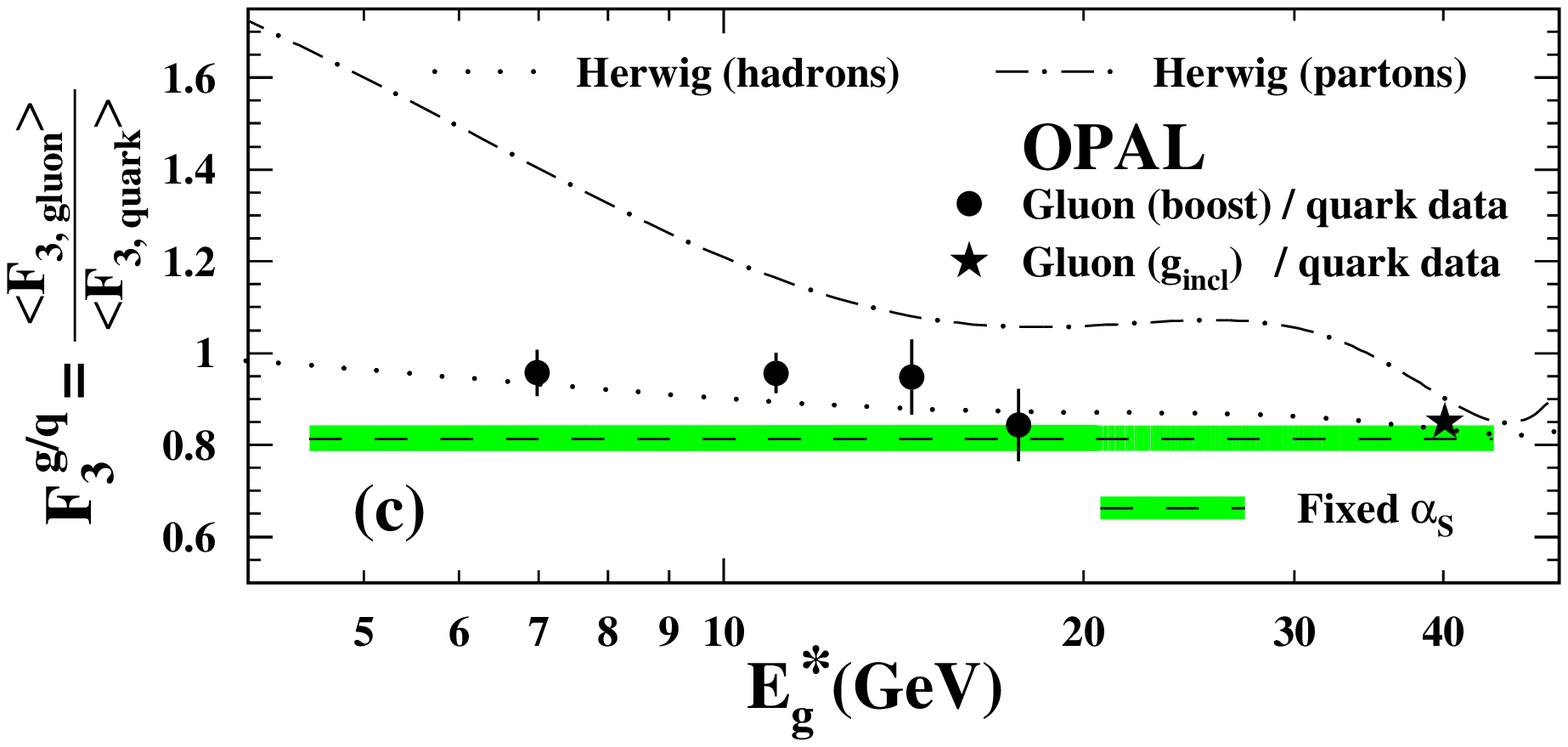}
 \end{center}
\vspace*{-6mm}
\caption{
(a)~The ratio between the mean charged particle multiplicities
of unbiased gluon and uds flavored quark jets, $\gqratio$, 
as a function of jet energy.
The data have been corrected for detector acceptance
and resolution,
for event selection,
and for gluon jet impurity.
The vertical lines indicate the total uncertainties.
Statistical uncertainties are too small to be visible.
(b,c)~The corresponding results for the factorial moments
of the multiplicity distributions, $\factwo$ and~$\facthree$.
The data are presented in comparison to the results of QCD calculations,
and to the Herwig Monte Carlo at the hadron and parton levels.
}
\label{fig-ratios}
\end{figure}

Fig.~\ref{fig-ratios}a shows the ratio of the 
mean charged particle multiplicities between gluon and quark jets,
$\gqratio$ (see eq.~(\ref{eq-gqratio})),
for the six energies for which the quark jet scales
correspond to our gluon data.
The analogous results for $\factwo$ and $\facthree$,
denoted $\rfactwo$ and $\rfacthree$,
are shown in Figs.~\ref{fig-ratios}b and~c.
The latter results are limited to energies of
6.98, 10.92, 14.24 and 17.72~GeV
because information about higher moments
of the quark jet (hemisphere) multiplicity distributions 
is not available for the other energies.
Figs.~\ref{fig-ratios}a--c include our previous measurements
at 40.1~GeV~\cite{bib-opalhemisphere-2,bib-opalhemisphere-3}.
Fig.~\ref{fig-ratios}a also includes the result for
$\gqratio$ we obtain by dividing the CLEO~\cite{bib-cleo92}
and ARGUS~\cite{bib-argus} measurements of unbiased 
gluon and quark multiplicities,
respectively.
The quark jet data have in all cases
been corrected for the small differences
in energy between the gluon and corresponding quark jet samples,
and for the presence of c and b flavored jets.
The reason for this latter correction is that the 
theoretical results for $\gqratio$ assume massless quarks.
The corrections were determined using
bin-by-bin factors derived from Herwig.
The total corrections for the mean multiplicities of quark jets
are about 10\% and are approximately independent of energy.
The corresponding corrections for the
$\factwo$ and $\facthree$ distributions of quark jets
are about 1\% and 3\%, respectively.
Very similar results are obtained using Jetset and Ariadne.

The dotted and dash-dotted
curves in Fig.~\ref{fig-ratios}a
show the Herwig predictions for $\gqratio$
at the hadron and parton levels.
It is seen that the parton and hadron level results are very similar,
even for small energies $\egstar$$\,\sim\,$5~GeV.
We conclude that hadronization effects are small for~$\gqratio$.
Comparing the dotted and dash-dotted curves in
Figs.~\ref{fig-ratios}b and~c,
it is seen that the hadronization corrections predicted
for $\rfactwo$ and $\rfacthree$ are fairly large and
have a significant dependence on energy.
The hadronization correction of
$\rfactwo$ is predicted to be about 20\% for $\egstar$$\,=\,$7~GeV,
decreasing to about 12\% at 14~GeV and 6\% at 40~GeV.
The corresponding values for $\rfacthree$
are 50\%, 25\% and 12\%.

A 3NLO analytic expression for $\gqratio$ 
is presented in~\cite{bib-capella}.
$\Lambda$ is the only free parameter in this expression.
The open band in Fig.~\ref{fig-ratios}a
shows the results we obtain by evaluating this expression
using $\nff$$\,=\,$5.
The lower edge of the band corresponds to $\Lambda$$\,=\,$0.296~GeV,
i.e.~the value from Sect.~\ref{sec-3nlo} from the
fit of the 3NLO expression for $\mnchgluon$.
The upper edge shows the result using the corresponding
value~\cite{bib-opaleden}
for unbiased quark jet multiplicity,
$\Lambda$$\,=\,$0.190~GeV.
The 3NLO prediction is seen to lie 15--20\% above the data.
%Similar results have been obtained previously
%(see e.g.~\cite{bib-opaleden,bib-dremingary-physrep}).
We also tried to fit the 3NLO expression for $\gqratio$
to the data in Fig.~\ref{fig-ratios}a.
We find that the theoretical expression is unable
to simultaneously provide a good description of the data 
at both low and high energies.
A fit of the three highest energy points
(14.24, 17.72 and 40.1~GeV)
yields $\Lambda$$\,=\,$$1.07\pm0.16\,$(stat.)~GeV,
%corresponding to $\alpha_S\,(\mzee)$$\,=\,$$0.155\pm0.004$,
with a $\chi^2$/d.o.f.\ of 6.1/2.
This value of $\Lambda$ is considerably larger than that
found from the fit to the $\mnchgluon$ data,
mentioned above.
In contrast to $\gqratio$,
3NLO perturbative expressions for $\rfactwo$
and $\rfacthree$ are not yet available.

The long-dashed line in Fig.~\ref{fig-ratios}a
shows the prediction of the fixed $\alpha_S$ calculation
for $\gqratio$,
assuming $\nff$$\,=\,$5
(see Table~\ref{tab-fixed-alphas}).
The shaded band corresponds to the one standard deviation total
uncertainty for this quantity
(Table~\ref{tab-fixed-alphas}).
The corresponding results for $\rfactwo$ and~$\rfacthree$
are shown in Figs.~\ref{fig-ratios}b and~c.
The fixed $\alpha_S$ result for $\rfactwo$ is determined 
by taking the ratio of the $\factwo$ expressions 
for gluon and quark jets~\cite{bib-dremin-hwa-f2f3},
using $\alpha_S$$\,=\,$0.301 from Table~\ref{tab-fixed-alphas}.
The fixed $\alpha_S$ result
for $\rfacthree$ is determined in an analogous manner.
The overall description of $\gqratio$ by the
fixed $\alpha_S$ calculation (Fig.~\ref{fig-ratios}a)
is seen to be similar to that of the 3NLO result,
being in somewhat better agreement with the data
at high energies ($\egstar$$\,\approx\,$40~GeV)
and in worse agreement at low energies
($\egstar$$\,\ltsim\,$10~GeV).
The fixed $\alpha_S$ prediction for $\rfactwo$
(Fig.~\ref{fig-ratios}b)
is in good agreement with the measurements,
while the prediction for $\rfacthree$
(Fig.~\ref{fig-ratios}c)
is in good agreement for $\egstar$$\,\gtsim\,$14~GeV.
Given the significant hadronization corrections predicted
for these last two distributions,
discussed above,
the good agreement between the data and fixed $\alpha_S$
results in Figs.~\ref{fig-ratios}b and~c
may be somewhat accidental.

A theoretical result for $\gqratio$ has also been
determined in the context of the dipole model~\cite{bib-eden-rgq}.
This result is shown by the short-dashed curve 
%labelled ``Dipole Model''
in Fig.~\ref{fig-ratios}a.
The dipole model prediction is seen to lie above the data,
but to be in somewhat better agreement with the measurements
than the 3NLO result.

Finally, we include in Fig.~\ref{fig-ratios}a
a theoretical result~\cite{bib-ochs} for $\gqratio$
based on a numerical, rather than an analytic,
solution of the QCD evolution equations for multiplicity.
This result is shown by the solid line.
Like the fixed $\alpha_S$ solution, 
the numerical solution is ``exact'' in the sense that it
is not based on a perturbative approximation.
The numerical result allows better accounting of
energy conservation effects and phase space limits
than the analytic results,
and incorporates a running value for $\alpha_S$
(see~\cite{bib-dremingary-physrep,bib-ochs}
for further discussion).
The value of $\Lambda$ used for the 
numerical calculation is~0.50~GeV,
determined from a fit to measurements of jet
rates in Z$^0$ decays~\cite{bib-ochs}.
%This value of $\Lambda$ corresponds to $\alpha_S\,(\mzee)$$\,=\,$0.13.
The numerical calculation is seen to provide a much 
improved description of the $\gqratio$ data
compared to the 3NLO or fixed $\alpha_S$ expressions.
This suggests that much of the discrepancy between the
data and analytic results in Fig.~\ref{fig-ratios}
is a consequence of technical difficulties in the calculations
(the inclusion of energy conservation, etc.),
rather than shortcomings of QCD.
Similar conclusions are presented in~\cite{bib-dremingary-physrep}.

\subsection{Fragmentation functions}
\label{sec-ff}

\begin{figure}[p]
 \begin{center}
   \epsfxsize=15.3cm
   \epsffile{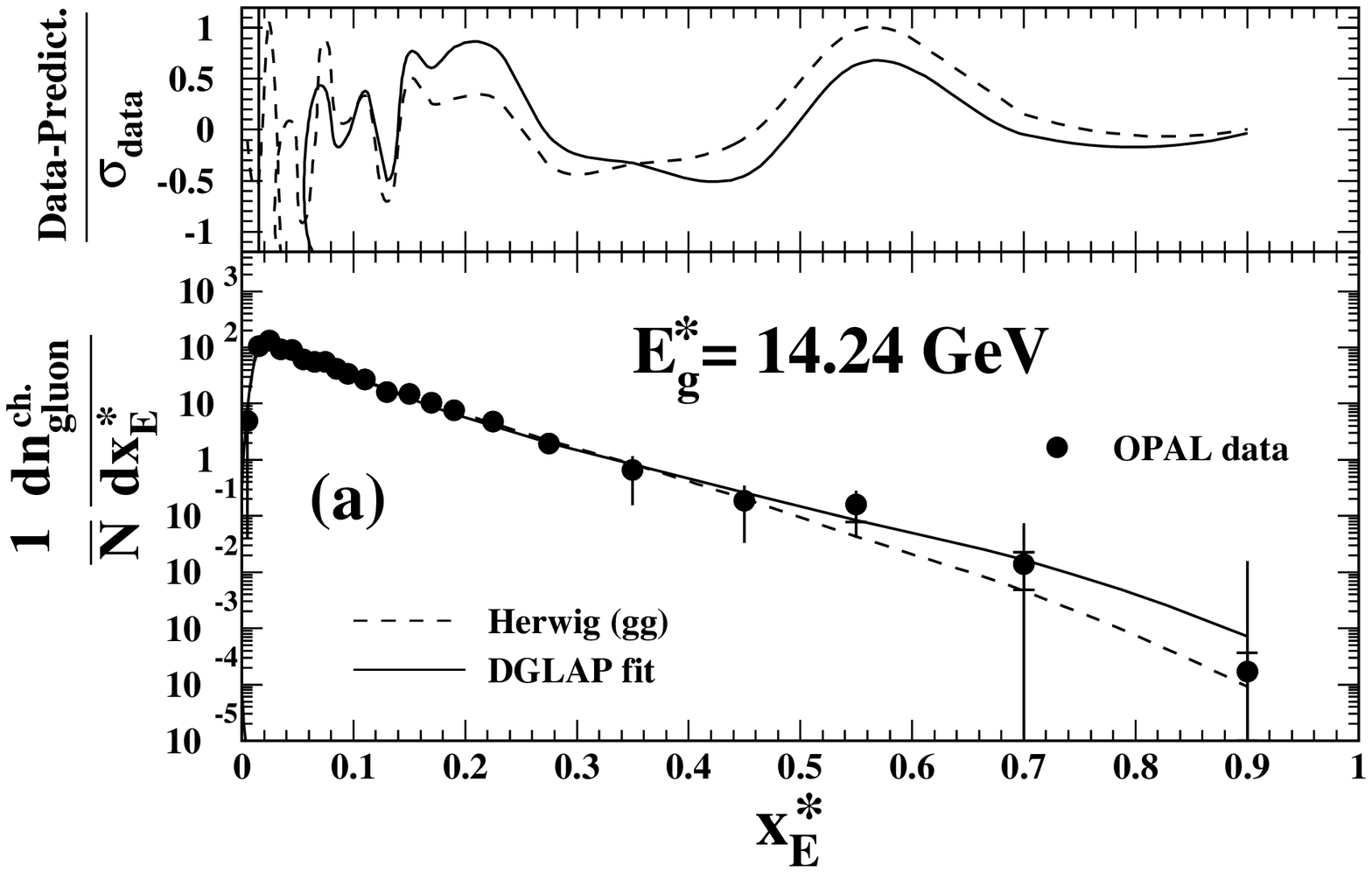} \\
   \epsfxsize=15.3cm
   \epsffile{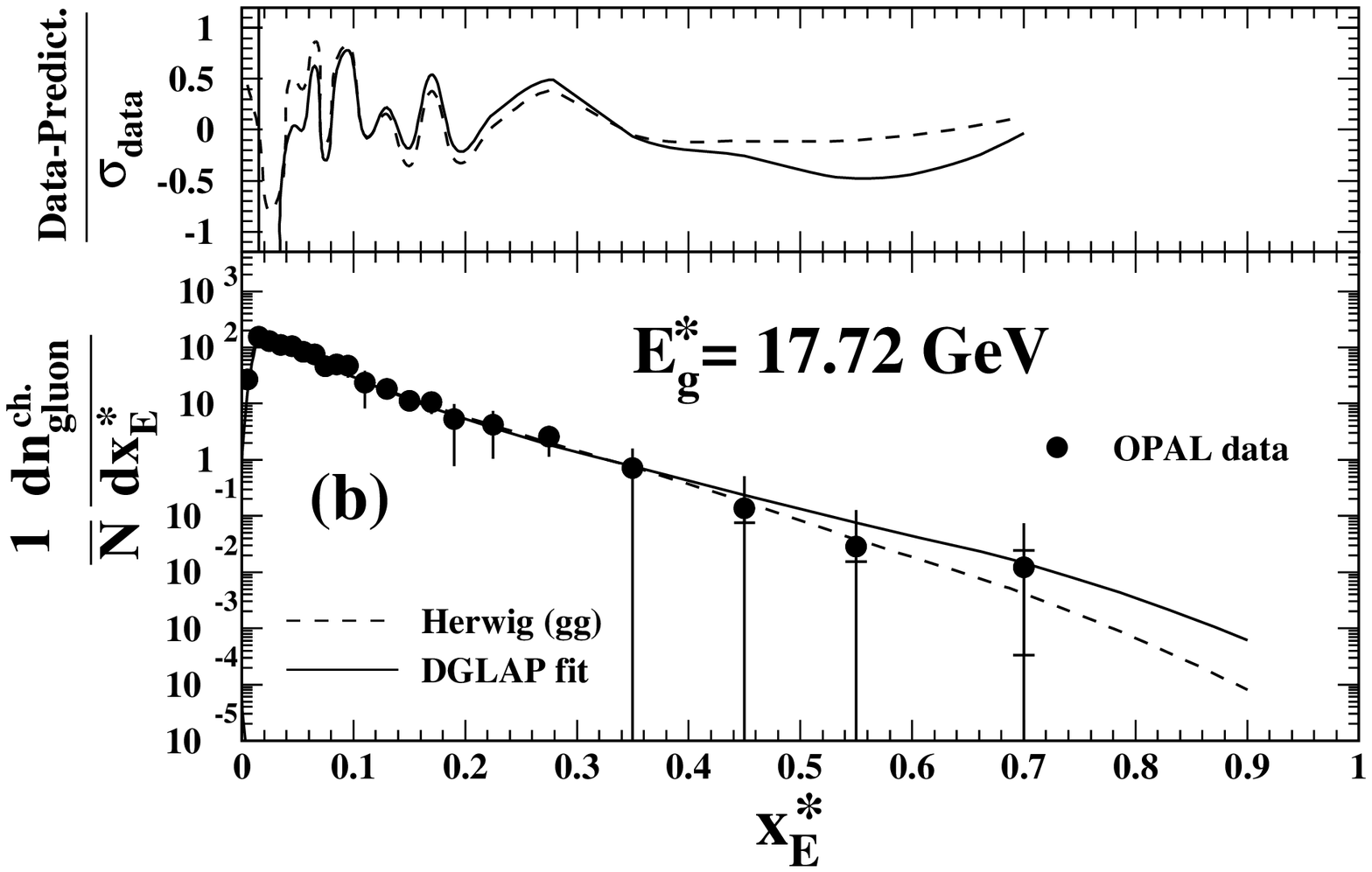}
 \end{center}
\vspace*{-6mm}
\caption{The charged particle fragmentation function of gluon jets,
$\fragfunc$, for (a)~$\egstar$$\,=\,$14.24 and (b)~17.72~GeV.
The data have been corrected for detector acceptance and resolution,
for event selection,
and for gluon jet impurity.
The total uncertainties are shown by the vertical lines, 
with the statistical component delimited by small horizontal lines.
The data are presented in comparison to a QCD prediction based on
DGLAP evolution of unbiased gluon and quark jet fragmentation functions
measured at 40.1 and 45.6~GeV, respectively,
and to the Herwig Monte Carlo at the hadron level.
The small figures above each distribution show the 
differences between the QCD and Herwig curves relative to the data,
in units of the total experimental uncertainties.
}
\label{fig-ffdata}
\end{figure}

We next turn to a discussion of
the gluon jet fragmentation function.
Our results for the corrected fragmentation functions 
of unbiased gluon jets
at $\egstar$$\,=\,$14.24 and 17.72~GeV are presented
in Figs.~\ref{fig-fftest}f and~g,
and again in Fig.~\ref{fig-ffdata}.
Numerical values for these data are given in
Table~\ref{tab-ff}.

\begin{table}[t]
\begin{center}
\begin{tabular}{|c|c|c|}
  \hline
  bin in $x_E^*$ & $\fragfunc$, $\egstar$$\,=\,$14.24~GeV &
    $\fragfunc$, $\egstar$$\,=\,$17.72~GeV  \\
  \hline
   0.00--0.01  & $4.9\pm   1.2\pm 4.7$   & $27.1\pm 5.3\pm 7.3$  \\
   0.01--0.02  & $104.7\pm 4.3\pm 4.9$   & $155\pm  13\pm 17$ \\
   0.02--0.03  & $131.4\pm 5.0\pm 7.2$   & $128\pm  11\pm 17$  \\
   0.03--0.04  & $93.4\pm  6.1\pm 5.8$   & $110\pm  9\pm 16$ \\
   0.04--0.05  & $89.0\pm  3.8\pm 6.6$   & $106\pm  8\pm 25$ \\
   0.05--0.06  & $61.8\pm  4.0\pm 8.3$   & $83\pm   8\pm 21$ \\
   0.06--0.07  & $55.4\pm  4.0\pm 7.1$   & $76\pm   7\pm 16$   \\
   0.07--0.08  & $54.7\pm  3.2\pm 5.6$   & $46\pm   6\pm 15$ \\
   0.08--0.09  & $41.1\pm  2.7\pm 4.7$   & $50\pm   6\pm 14$  \\
   0.09--0.10  & $33.9\pm  2.9\pm 3.5$   & $48\pm   5\pm 18$  \\
   0.10--0.12  & $27.0\pm  1.7\pm 3.7$   & $24\pm   3\pm 15$  \\
   0.12--0.14  & $15.9\pm  1.6\pm 2.8$   & $18.2\pm 2.4\pm 6.0$  \\
   0.14--0.16  & $14.7\pm  1.0\pm 3.0$   & $11.1\pm 2.1\pm 2.9$  \\
   0.16--0.18  & $10.2\pm  1.0\pm 2.0$   & $10.5\pm 2.0\pm 3.3$  \\
   0.18--0.20  & $7.6\pm   0.7\pm 1.2$   & $5.4\pm  1.5\pm 4.3$  \\
   0.20--0.25  & $4.75\pm  0.40\pm 0.88$ & $4.2\pm  0.7\pm 3.1$  \\
   0.25--0.30  & $1.96\pm  0.27\pm 0.66$ & $2.6\pm  0.6\pm 1.3$  \\
   0.30--0.40  & $0.66\pm  0.09\pm 0.50$ & $0.71\pm 0.17\pm 0.71$  \\
   0.40--0.50  & $0.19\pm  0.06\pm 0.14$ & $0.14\pm 0.06\pm 0.14$  \\
   0.50--0.60  & $0.160\pm 0.082\pm 0.086$    & $0.029\pm 0.013\pm 0.029$  \\
   0.60--0.80  & $0.014\pm 0.009\pm 0.014$    & $0.012\pm 0.012\pm 0.012$  \\
   0.80--1.00  & $0.0002\pm 0.0002\pm 0.0002$ & ---  \\
  \hline
\end{tabular}
\end{center}
  \caption{The charged particle fragmentation function of gluon jets,
$\fragfunc$, for $\egstar$$\,=\,$14.24 and 17.72~GeV.
The data have been corrected for detector acceptance and resolution,
for event selection,
and for gluon jet impurity.
The first uncertainty is statistical and the 
second systematic.}
  \label{tab-ff}
\end{table}

Unlike multiplicity,
the shape of fragmentation functions is not presently calculable.
If the shape of a fragmentation function is known 
at a particular scale,
the DGLAP~\cite{bib-dglap} evolution equations
can be used to predict the shape at a different scale, however.
Since gluon jets can evolve through splitting to a quark-antiquark pair,
as well as through gluon emission,
the evolution of the gluon jet fragmentation function depends on
the quark jet fragmentation function,
in addition to that of the gluon.
%the gluon jet fragmentation function.
In~\cite{bib-opalhemisphere-3},
we presented results for the unbiased gluon jet
fragmentation function at 40.1~GeV.
Measurements of unbiased,
flavor-separated (uds, c and b)
quark jet fragmentation functions at 45.6~GeV
are presented in~\cite{bib-opal-quark-ff}.
By applying the DGLAP equations to these measurements,
we can obtain QCD predictions for the
gluon jet fragmentation function at the scales of the present study.

We note that the quark jet data in~\cite{bib-opal-quark-ff}
are presented in terms of the scaled charged particle three-momenta
$x_p$$\,=\,$$2p$/$\ecm$ 
(with $p$ the particle three-momentum),
rather than~$x_E$$\,=\,$$2E$/$\ecm$ 
(with $E$ the particle energy),
and that $x_p$ and $x_E$ differ for small particle energies
(or momenta).
Using detector level events,
we find that fragmentation functions defined using
$x_p$ differ from those defined by $x_E$ by about 1\% for
$x_p$ (or $x_E)$$\,=\,$0.10,
and by about 2.5\% for $x_p$ (or $x_E$)$\,=\,$0.05,
for example.

The fragmentation functions of gluon and quark jets are
parametrized at a reference scale 
$\sqrt{s_0}$ using the empirical formula~\cite{bib-delphi-scaling}
\begin{equation}
  F^{i}(x_E;\sqrt{s_0}) = 
    a_i \, x_E^{b_i} \,(1-x_E)^{c_i}
    \exp\{ -d_i \, \ln^2 x_E \}
   \;\;\;\; ,
  \label{eq-empirical}
\end{equation}
where $i$$\,=\,$g, uds, c, or~b.
To determine the parameters $a$, $b$, $c$ and $d$ for quark jets,
we fit eq.~(\ref{eq-empirical})
to the measurements in~\cite{bib-opal-quark-ff},
i.e.\ we choose $\sqrt{s_0}$$\,=\,$45.6~GeV.
To determine the parameters for
gluon jets at this same scale,
%(the quark and gluon jets are required to be parametrized
%at the same scale $\sqrt{s_0}$ by the program Evolve, see below),
we first apply an energy correction to the 40.1~GeV data.
%The reason for this is that the scale $\sqrt{s_0}$ 
%in~(\ref{eq-empirical})
%is required be the same for all jet types~$i$.
The corrections are performed 
using bin-by-bin factors determined from Herwig and
have a typical size of about~5\%.
Eq.~(\ref{eq-empirical}) is then fitted 
to the corrected gluon jet data to determine
the parametrization of the gluon jet fragmentation function
at 45.6~GeV.
The fits are performed using the statistical uncertainties 
of the data
and provide good descriptions of the measurements to within
their overall uncertainties.
The results we obtain
are listed in Table~\ref{tab-ff-parametrizations}.

\begin{table}[t]
\begin{center}
\begin{tabular}{|c|cccc|}
\hline
     & $a$ & $b$ & $c$ & $d$  \\
\hline
uds & $0.2589\pm 0.0074$ & $-2.949\pm 0.016$ 
          & $0.859\pm 0.029$ & $0.2967\pm 0.0020$ \\
c   & $0.546\pm 0.035$ & $-2.67\pm 0.016$ 
          & $2.37\pm 0.21$ & $0.2651\pm 0.0029$ \\
b   & $0.3284\pm 0.0016$ & $-3.042\pm 0.0016$ 
          & $2.6055\pm 0.015$ & $0.31805\pm 0.00017$ \\
g   & $0.0891\pm0.0047$ & $-3.84\pm 0.012$ 
          & $2.62\pm 0.26$ & $0.4144\pm 0.0023$  \\
\hline
\end{tabular}
\caption{Parameter values used to describe the fragmentation
functions of unbiased quark and gluon jets at 45.6~GeV
(see eq.~(\ref{eq-empirical})).
The uncertainties are statistical.
}
\label{tab-ff-parametrizations}
\end{center}
\end{table}

We then use the program Evolve~\cite{bib-evolve}
to determine the QCD prediction for the gluon jet
fragmentation function at other scales.
Evolve is based on next-to-leading order expressions
(see~\cite{bib-nason-webber})
determined in the $\mathrm\overline{MS}$
renormalization scheme.
We determine the predictions
of the program for the gluon jet fragmentation functions
at 14.24 and 17.72~GeV,
and calculate the global $\chi^2$
with respect to our corresponding measurements.
The global $\chi^2$ is defined by the sum of
the $\chi^2$ from the two energies.
The $\chi^2$ are calculated using the statistical
uncertainties of the data.
To avoid the edges of the distribution where 
there are theoretical ambiguities~\cite{bib-evolve},
the global $\chi^2$ is evaluated in the $x_E$ range
from 0.10 to 0.80 only.
%0.10$\,\leq\,$$x_E$$\,\leq\,$0.80 only.
Note this excludes the small $x_E$ region where
fragmentation functions defined using $x_E$ or $x_p$ differ
by more than~1\%.

%the gluon jet fragmentation function
%determined using the boost method
%can differ from that found using gg hemispheres 
%by more than the experimental statistical uncertainties
%(see the difference plots in Figs.~\ref{fig-fftest}f and~g).
%This also excludes the region where
%fragmentation functions defined using $x_E$ or $x_p$ differ
%by more than~1\%.

We fit the value of $\alpha_S\,(\mzee)$ in Evolve
to minimize the global~$\chi^2$.
The result is 
$\alpha_S\,(\mzee)$$\,=\,$$\mathrm 0.128\pm0.008\,(stat.)\pm0.015\,(syst.)$.
The $\chi^2$/d.o.f.\ of the fit,
based on statistical uncertainties,
is~40.5/21.
The systematic uncertainty is defined
by adding the following contributions in quadrature:
(1)~the uncertainty returned by the
fitting routine when the total uncertainties of the data are used
to perform the fit,
rather than the statistical uncertainties only;
%
%(1)~the difference with respect to the standard result
%if the total uncertainties of the data are used,
%rather than the statistical uncertainties only,
%both for the quark and gluon jet measurements
%used to define the parametrizations at 45.6~GeV
%(see~(\ref{eq-empirical}))
%and for the gluon jet measurements at 14.24 and 17.72~GeV
%used to define the global~$\chi^2$;
%
(2)~the difference with respect to the standard result
if the range 0.05$\,\leq\,$$x_E$$\,\leq\,$0.80
is used to define the global $\chi^2$,
rather than 0.10$\,\leq\,$$x_E$$\,\leq\,$0.80.
The systematic uncertainty from the second term is 
about 50\% larger than that from the first term.
While our result for $\alpha_S\,(\mzee)$
is not competitive with other measurements
(see e.g.~\cite{bib-pdg}),
it does provide a unique consistency test of QCD
since it is the first determination of $\alpha_S\,(\mzee)$ in the
$\mathrm\overline{MS}$ scheme
based on unbiased gluon jets.
The result of the fit is shown in comparison to the data
in Figs.~\ref{fig-ffdata}a and~b.
The difference plots in the top portions of these figures
show the differences between the data and fit
in units of the total experimental uncertainties.
The fit is seen to provide a good description of the measurements.

\section{Summary}

In this paper,
we present the first experimental study to use
the jet boost algorithm,
a method based on the QCD dipole model to extract properties of
unbiased gluon jets from 
{\epem}$\,\rightarrow\,$$\mathrm q\overline{q}g$ events.
We test the jet boost algorithm using the Herwig Monte Carlo
QCD simulation program,
comparing the results of this method to
those derived from unbiased gluon jets defined by hemispheres 
of inclusive gg events from a color singlet point source.
We examine two distributions:
the distribution of charged particle multiplicity in the jets,
$\nchgluon$,
and the charged particle fragmentation functions,
$\fragfunc$.
We find that the results of the jet boost algorithm 
for the multiplicity distribution are in close
correspondence to those of the gg hemispheres for jet energies
$\egstar$ larger than about 5~GeV.
For the fragmentation functions,
the results of the two methods agree to good precision
for $\egstar$$\,\gtsim\,$14~GeV.

We use the jet boost algorithm to extract measurements
of the unbiased gluon jet multiplicity distribution for seven intervals
of energy between 5.25 and 17.72~GeV.
These are the first measurements of the $\nchgluon$
distribution in this energy range.
%Our results are much more precise than previous measurements
%in this energy range.
The distributions are analyzed to determine their means
$\mnchgluon$ and first two non-trivial factorial moments
$\factwogluon$ and~$\facthreegluon$.
The factorial moments are strongly correlated with the dispersion
and skew of the multiplicity distribution and thus characterize
its shape.

In conjunction with our previous results for unbiased gluon
jet multiplicity at 
40.1~GeV~\cite{bib-opalhemisphere-1}--\cite{bib-opalhemisphere-3},
we test two QCD analytic expressions for the energy evolution
of $\mnchgluon$, $\factwogluon$ and~$\facthreegluon$:
one based on the 
next-to-next-to-next-to-leading order
(3NLO) perturbative
approximation of QCD~\cite{bib-dremingary,bib-f2f3-3nlo}
and the other~\cite{bib-dremin-hwa-n,bib-dremin-hwa-f2f3}
utilizing a fixed
value of the strong coupling strength,~$\alpha_S$.
The 3NLO expression takes into account the running nature of
the coupling strength while the fixed $\alpha_S$ expression more accurately
incorporates higher order effects such as energy conservation.
%In this sense,
%the two expressions are complementary to each other.
%In conformity with common practice and 
To avoid the introduction of model dependent
hadronization correction factors,
the parton level analytic results are compared directly
to the hadron level measurements.

The 3NLO expression for $\mnchgluon$ is found to provide
a good description of the data using a value of the QCD scale
parameter $\Lambda$$\,=\,$$\mathrm0.296\pm0.037\,(stat.+syst.)$~GeV,
much more similar to the corresponding result for quark jets,
$\Lambda$$\,=\,$$\mathrm0.190\pm0.032\,(stat)$~GeV~\cite{bib-opaleden},
than found in previous studies.
Our results therefore provide a much improved
demonstration of the consistency
of the QCD expressions for gluon and quark jet multiplicity.
These results are found using $\nff$$\,=\,$5, 
with $\nff$ the number of active quark flavors.
Very similar descriptions of the data are found using
$\nff$$\,=\,$3 or~4.
Note that these $\Lambda$ values
are not defined in the context of a particular renormalization
scheme and so do not correspond e.g.\ to~$\mathrm\Lambda_{\overline{MS}}$.
The 3NLO expressions are found to provide a reasonable description
of the energy evolution of the $\factwogluon$ and~$\facthreegluon$ data
between about 14 and 40~GeV,
using values of $\Lambda$ which are globally similar
to that found from the fit to the $\mnchgluon$ data.
The fitted 3NLO curves lie below the 
$\factwogluon$ and~$\facthreegluon$ measurements 
at smaller energies, however.
These discrepancies at low energies may be a consequence
of hadronization effects,
which are predicted to be significant for the
$\factwogluon$ and~$\facthreegluon$ distributions.
%especially for $\egstar$$\,\ltsim\,$12~GeV.
The fixed $\alpha_S$ expressions are found to be in general 
agreement with the $\mnchgluon$ data,
and also with the $\factwogluon$ data within fairly
large theoretical uncertainties.
The fixed $\alpha_S$ result for $\facthreegluon$
lies above the data except for $\egstar$$\,\approx\,$40~GeV.

We also examine the ratio of the 
gluon to quark jet mean charged particle multiplicities,
$\gqratio$,
and the corresponding ratios for $\factwo$ and~$\facthree$.
We find that a numerical solution~\cite{bib-ochs}
of the QCD
evolution equations for particle multiplicity
provides a good description of the $\gqratio$ data,
while the 3NLO and fixed $\alpha_S$ calculations
with their fitted values of $\Lambda$ from the $\mnchgluon$ data
are 15--20\% too high.
This suggests that energy conservation and phase space limits,
which are more properly incorporated into the numerical solution than
into the analytic results,
are important considerations for the accurate
description of this quantity.
%Similar conclusions are presented in~zzzz.

We measure the fragmentation function of unbiased gluon jets
at 14.24 and 17.72~GeV.
%the only measurements of this quantity other than our
%previous result~\cite{bib-opalhemisphere-3} at 40.1~GeV.
In conjunction with our previous
measurements of unbiased gluon and quark jet fragmentation
functions at 40.1 and 45.6~GeV, respectively,
%unbiased, flavor-separated quark jet
%fragmentation functions at 45.6~GeV,
we fit these data using the DGLAP evolution
equations at next-to-leading-order in the $\mathrm\overline{MS}$ scheme.
This fit yields a result for the strong interaction
coupling strength
$\alpha_S\,(\mzee)$$\,=\,$$\mathrm 0.128\pm0.008\,(stat.)\pm0.015\,(syst.)$,
consistent with the world average.
While this result is not competitive in precision with
other measurements of $\alpha_S$,
it does provide a unique consistency test of QCD.

\section{Acknowledgments}

We thank Igor Dremin, Patrik Ed\'{e}n and Bryan Webber
for helpful comments and discussions.

We particularly wish to thank the SL Division for the efficient operation
of the LEP accelerator at all energies
and for their close cooperation with
our experimental group.  
In addition to the support staff at our own
institutions we are pleased to acknowledge the  \\
Department of Energy, USA, \\
National Science Foundation, USA, \\
Particle Physics and Astronomy Research Council, UK, \\
Natural Sciences and Engineering Research Council, Canada, \\
Israel Science Foundation, administered by the Israel
Academy of Science and Humanities, \\
Benoziyo Center for High Energy Physics,\\
Japanese Ministry of Education, Culture, Sports, Science and
Technology (MEXT) and a grant under the MEXT International
Science Research Program,\\
Japanese Society for the Promotion of Science (JSPS),\\
German Israeli Bi-national Science Foundation (GIF), \\
Bundesministerium f\"ur Bildung und Forschung, Germany, \\
National Research Council of Canada, \\
Hungarian Foundation for Scientific Research, OTKA T-038240, 
and T-042864,\\
The NWO/NATO Fund for Scientific Research, the Netherlands.

\newpage
\appendix

\section{Appendix: boost algebra}

\subsection{Boost to the back-to-back frame of a dipole}
\label{sec-boost1}

\begin{figure}[t]
 \begin{center}
    \epsfxsize=5cm
    \epsffile{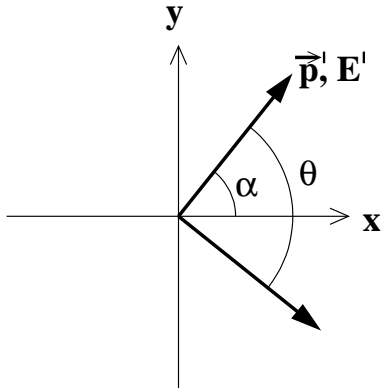} 
 \end{center}
\caption{
Schematic illustration of a two-jet system.
}
\label{fig-boost}
\end{figure}

Consider a massless jet with energy $E^\prime$
and 3-momentum of magnitude~$p^\prime$.
The jet lies in the $x$--$y$ plane and
makes angles $\theta$ and $\alpha$
with respect to another massless jet and 
the $x$ axis, respectively
(see Fig.~\ref{fig-boost}).
%see Fig.~\ref{fig-qqdipole}b).
We wish to boost the event to a Lorentz frame
in which the jet points along the $y$ axis.
The energy and momentum of the jet in the boosted
frame are $E^*$ and $p^*$.
A boost along the $x$ axis yields the condition
\begin{equation}
  p_x^* = \gamma\left(p_x^\prime -\beta\, E^\prime \right) = 0
   \;\;\;\; ,
  \label{eq-beta}
\end{equation}
with $\gamma$$\,=\,$$1/\sqrt{1-\beta^2}$ and 
$\beta$$\,=\,$$v$,
where $v$ is the relative speed between the boosted and
original frames
(note that the speed of light $c$ is set to unity).
Since $p_x^\prime$$\,=\,$$p^\prime\cos\alpha$ 
and $p^\prime$$\,=\,$$E^\prime$,
eq.~(\ref{eq-beta}) yields $\beta$$\,=\,$$\cos\alpha$.
The energy of the jet in the boosted frame is
\begin{equation}
   E^* = \gamma\left( E^\prime - \beta\, p_x^\prime \right) 
       = E^\prime\, \sin\alpha \;\;\;\; .
\end{equation}
%\begin{eqnarray}
%  E^* & = & \gamma\left( E^\prime - \beta\, p_x^\prime \right) \\
%           & = & \gamma\left
%    ( E^\prime - \beta\, E^\prime\cos\alpha \right) \nonumber \\
%           & = & \gamma E^\prime \left( 1 -\beta^2 \right) \nonumber \\
%           & = & E^\prime\, \sqrt{1-\beta^2} \nonumber \\
%           & = & E^\prime\, \sin\alpha \;\;\;\; . \nonumber
%\end{eqnarray}
If the $x$ axis corresponds to the bisector of the two jets,
so that $\alpha$$\,=\,$$\theta/2$,
%If the boost is performed along
%the bisector separating the two jets,
%i.e.\ $\alpha$$\,=\,$$\theta/2$,
then the same boost brings the other massless jet
to the $-y$ direction
so that the boosted frame corresponds to a frame in
which the two jets are back-to-back.
Then $\beta$$\,=\,$$\cos(\theta/2)$
and $E^*$$\,=\,$$E^\prime\,\sin(\theta/2)$.

\subsection{Boost from the c.m.\ frame to the symmetric 
frame of a three-jet {\boldmath$\mathrm q\overline{q}g$} event}
\label{sec-boost2}

In the c.m. frame of a three-jet $\mathrm q\overline{q}g$
event we define
\begin{equation} 
  \label{eq-xdef}
  x_{i} \equiv\frac{2\, E_{i}}{\sqrt{s}} \hspace{1cm} i={\mathrm q,\bar{q},g}
   \;\;\;\; ,
\end{equation} 
with $E_{i}$ the energy of jet $i$.
%We define $\theta_{ij}$ ($i,j={\mathrm q,\bar{q},g}$) 
%to be the angle between jets $i$ and~$j$.
$x^\prime_{i}$ is the corresponding quantity in the boosted
reference frame for which the event is symmetric,
%in the Lorentz frame for which 
i.e.~the frame in which the angle between the
gluon jet and the quark jet is the same as the angle
between the gluon jet and the antiquark jet,
$\theta_{\mathrm qg}^\prime$$\,=\,$$\theta_
{\mathrm\bar{q}g}^\prime$$\,\equiv\,$$\theta^\prime$
(cf.~Fig.~\ref{fig-qgdipole}a for which
$\theta^\prime$$\,=\,$$\theta$$\,=\,$$2\alpha$).

Under the assumption
the jets are massless, 
it is straightforward to show in the c.m. frame that 
\begin{equation}
  \label{eq-sij}
  s_{ij}=s\,(1-x_{k})
   \;\;\;\; ,
\end{equation} 
where $s_{ij}$$\,=\,$$(p_i+p_j)^2$ with $p_i$ the 4-momentum of object~$i$,
and where $i$, $j$ and $k$ are cyclic,
i.e.~$i$$\,=\,$q with $j$$\,=\,$$\mathrm\overline{q}$
means $k$$\,=\,$g, etc.
The virtuality scale (eq.~(\ref{eq-ptlu})) of the gluon jet
can then be written:
\begin{equation}
  \label{eq-plund2}
  \ptgluon = \frac{1}{2}
    \sqrt{ s(1-x_{\mathrm q})(1-x_{\mathrm\bar{q}}) }  \;\;\;\; .
\end{equation}
Setting $\ptgluon$ equal to the jet energy 
scale $\egstar$ (see eq.~(\ref{eq-ejetstar})),
as dictated by eq.~(\ref{eq-onescale}),
yields~\cite{bib-eden}
\begin{equation}
  \label{eq-constr2}
  (1-x_{\mathrm q})(1-x_{\mathrm\bar{q}})
  = (x^\prime_{\mathrm g})^2\sin^{2}{\frac{\theta^\prime}{2}}
   \;\;\;\; .
\end{equation}     
  
We can also express $s_{ij}$ using the
angle between partons $i$ and $j$, $\theta_{ij}$:
\begin{equation} 
  \label{eq-sijbis}
  s_{ij} = 4E_{i}E_{j}\sin^{2}{\frac{\theta_{ij}}{2}} 
   \;\;\;\; .
\end{equation}
This latter expression,
unlike eq.~(\ref{eq-sij}),
is valid in any frame.
Evaluating eq.~(\ref{eq-sijbis}) in the symmetric
frame and equating it to eq.~(\ref{eq-sij}) yields
\begin{equation}
  \label{eq-xk}
  1 - x_{k} =
  x'_{i}x'_{j}\sin^{2}{\frac{\theta_{ij}^\prime}{2} }
   \;\;\;\; ,
\end{equation}
which leads to the following expression:
\begin{equation}
  \label{eq-xkg}
  \frac{(1-x_{\mathrm q})(1-x_{\mathrm\bar{q}})}{1-x_{\mathrm g}} 
    = x^{\prime 2}_{\mathrm g}\frac{\sin^2{(\theta_{\mathrm qg}^\prime/2)}
           \sin^2{(\theta_{\mathrm\bar{q}g}^\prime/2)}}
       {\sin^{2}{(\theta_{\mathrm q\bar{q}}^\prime/2)}}
   \;\;\;\; .
\end{equation} 
Since $\theta^\prime_{\mathrm qg}=\theta^\prime_{\mathrm\bar{q}g}=\theta^\prime$
in the symmetric frame,
then $\theta^\prime_{\mathrm q\bar{q}}=2\pi-2\theta^\prime$,
so that
$\sin\,(\theta^\prime_{\mathrm q\bar{q} }/2) 
     = 2\sin\,(\theta^\prime/2)
        \cos\,(\theta^\prime/2)   
$.
Inserting these results into eq.~(\ref{eq-xkg}) yields~\cite{bib-eden}
\begin{equation}
  \label{eq-constr3}
  \frac{(1-x_{\mathrm q})(1-x_{\mathrm\bar{q}})}{1-x_{\mathrm g}}
    =\frac{ x^{\prime 2}_{\mathrm g} }{4}
      \frac{\sin^2{(\theta^\prime/2)}}{\cos^{2}{(\theta^\prime/2)}}
   \;\;\;\; .
\end{equation}

The expression for the angle $\theta^\prime$~\cite{bib-eden} is 
obtained by combining eqs.~(\ref{eq-constr2}) and (\ref{eq-constr3}):
\begin{equation}
  \label{eq-costp}
  \cos^{2}\frac{\theta^\prime}{2}=\frac{1-x_{\mathrm g}}{4}
   \;\;\;\; .
\end{equation} 
When inserted into eq.~(\ref{eq-constr2}),
this yields the expression for the gluon jet energy in
the symmetric frame:
\begin{equation} 
  \label{eq-xgcond}
  x^\prime_{\mathrm g} = \sqrt{\frac{4(1-x_{\mathrm q})
    (1-x_{\mathrm\bar{q}})}{3+x_{\mathrm g}}}
   \;\;\;\; .
\end{equation}
Similarly,
using eq.~(\ref{eq-xk}) to derive expressions
analogous to eq.~(\ref{eq-xkg})
for $(1-x_{\mathrm q})(1-x_{\mathrm g})/(1-x_{\mathrm\bar{q}})$ 
and
$(1-x_{\mathrm\bar{q}})(1-x_{\mathrm g})/(1-x_{\mathrm q})$ 
yields the results for the quark and antiquark energies
in the symmetric frame:
\begin{equation}
\label{eq-xq1cond}
  x'_{\mathrm q} = \frac{x'_{\mathrm g}}{1-x_{\mathrm q}} \hspace{0.5cm} , 
  \hspace{0.5cm} x'_{\bar{\mathrm q}} = \frac{x'_{\mathrm g}}{1-x_{\mathrm\bar{q}}}
   \;\;\;\; .
\end{equation} 

%The boosted frame is specified by~(\ref{eq-xgcond})
%and~(\ref{eq-xq1cond}).

Consider the c.m. frame of the event
to be described by a Cartesian coordinate 
system with the $z$ axis along the gluon jet direction
and the three-jet event in the $y$--$z$ plane.
The scaled three momenta of the jets are then 
\begin{eqnarray}
 2\,\vec{p}_{\mathrm g}/\ecm       & = 
     & (0,0,x_{\mathrm g}) \\
 2\,\vec{p}_{\mathrm q}/\ecm       & = 
     & (0,x_{{\mathrm q},y}, x_{{\mathrm q},z}) \\
 2\,\vec{p}_{\mathrm\bar{q}}/\ecm  & = 
     & (0,-x_{{\mathrm q},y}, x_{{\mathrm\bar{q}},z})
   \;\;\;\; .
\end{eqnarray}
with $|\vec{p}_i|$$\,=\,$$E_i$ such that
$x_{i,y}$ is the component of scaled momentum in 
the $y$ direction, etc.
Since we boost from the c.m.\ frame, 
the scaled energy in the symmetric frame is
\begin{equation}
  \label{eq-gamma}
  x^\prime_{\mathrm q}+x^\prime_{\mathrm g}+x^\prime_{\mathrm\bar{q}} 
    = \gamma(x_{\mathrm q}+x_{\mathrm g}+x_{\mathrm\bar{q}}) = 2\gamma
   \;\;\;\; ,
\end{equation}
where $\gamma=1/\sqrt{1-\beta_{y}^{2}-\beta_{z}^{2}}$, 
with $\beta_{y}$ and $\beta_{z}$ the Lorentz boost factors 
along the $y$ and $z$ directions.
%Obviously, with this specific choice of the reference frame, $\beta_{x}=0$.
Knowing all the $x_i$ (measured), 
$x^\prime_i$ (from eqs.~(\ref{eq-xgcond}) and~(\ref{eq-xq1cond}))
and $\gamma$ (from eq.~(\ref{eq-gamma})),
the transformation equations
\begin{equation}
  \label{eq-bz}
  x^\prime_{\mathrm g} = \gamma(x_{\mathrm g} + \beta_{z}x_{\mathrm g})
\end{equation}
\begin{equation}
  \label{eq-by}
  x^\prime_{\mathrm q} = \gamma(x_{\mathrm q} 
     + \beta_{z}x_{{\mathrm q},z} + \beta_{y}x_{{\mathrm q},y})
\end{equation}
can be solved to find the boost factors $\beta_{z}$ and~$\beta_{y}$.
With the Lorentz boost factors defined,
all the particles in the event can then be boosted
to the symmetric event frame to define the unbiased gluon
jets in the manner explained in Sect.~\ref{sec-boost}.

\end{document}